\documentclass[preprint, authoryear]{elsarticle}
\biboptions{sort&compress}
\makeatletter
\def\ps@pprintTitle{%
 \let\@oddhead\@empty
 \let\@evenhead\@empty
 \def\@oddfoot{}%
 \let\@evenfoot\@oddfoot}
\makeatother

\usepackage[utf8]{inputenc}
\usepackage{xcolor}
\usepackage{stix} 
\usepackage{dsfont}
\usepackage{multicol} 
\usepackage{color}
\usepackage{soul} 
\sethlcolor{yellow}
\usepackage{mathrsfs}
\usepackage{dirtytalk} 
\usepackage{graphicx}
\usepackage{mathtools}
\usepackage{amsfonts}
\usepackage{amsmath}
\allowdisplaybreaks[1] 
\usepackage{accents}
\usepackage{bbm}
\usepackage{chngcntr} 
\counterwithin{equation}{section}
\usepackage[colorlinks=true, linkcolor=blue]{hyperref}
\usepackage[capitalize, nameinlink]{cleveref} 
\usepackage[section]{placeins} 
\usepackage[normalem]{ulem} 
\usepackage{cancel}

\usepackage{geometry}
\begin{document}

\begin{frontmatter}

    \title{A finite deformation theory of dislocation thermomechanics}

    \author[fr_address]{Gabriel D. Lima-Chaves}

    \author[usa_address]{Amit Acharya}
    
    \author[fr_address]{Manas V. Upadhyay\corref{mycorrespondingauthor}}
    \cortext[mycorrespondingauthor]{Corresponding author}
    \ead{manas.upadhyay@polytechnique.edu}
    
    \address[fr_address]{
        Laboratoire de Mécanique des Solides (LMS), CNRS UMR 7649, École Polytechnique, Institut Polytechnique de Paris, Route de Saclay, 91128 Palaiseau, France
    }
    \address[usa_address]{
        Department of Civil and Environmental Engineering, and Center for Nonlinear Analysis, Carnegie Mellon University, Pittsburgh, PA 15213, United States
    }
    
    \begin{abstract}
        A geometrically nonlinear theory for field dislocation thermomechanics based entirely on measurable state variables is proposed. Instead of starting from an ordering-dependent multiplicative decomposition of the total deformation gradient tensor, the additive decomposition of the velocity gradient into elastic, plastic and thermal distortion rates is obtained as a natural consequence of the conservation of the Burgers vector. Based on this equation, the theory consistently captures the contribution of transient heterogeneous temperature fields on the evolution of the (polar) dislocation density. The governing equations of the model are obtained from the conservation of Burgers vector, mass, linear and angular momenta, and the First Law. The Second Law is used to deduce the hyperelastic constitutive equation for the Cauchy stress and the thermodynamical driving force for the dislocation velocity. An evolution equation for temperature is obtained from the First Law and the Helmholtz free energy density, which is taken as a function of the following measurable quantities: elastic distortion, temperature and the dislocation density (the theory allows prescribing additional measurable quantities as internal state variables if needed). Furthermore, the theory allows one to compute the Taylor-Quinney factor, which is material and strain rate dependent. Accounting for the polar dislocation density as a state variable in the Helmholtz free energy of the system allows for temperature solutions in the form of dispersive waves with finite propagation speed, i.e.~\emph{thermal waves}, despite using Fourier's law of heat conduction as the constitutive assumption for the heat flux vector.
    \end{abstract}
    
\end{frontmatter}

\section{Introduction}

We present a fully coupled finite-deformation thermomechanical theory of field dislocation mechanics, i.e., a theory based on partial differential equations (PDEs). The theory incorporates a two-way coupling between dislocation activity and temperature evolution, while accounting for unrestricted geometrical and material nonlinearities in a potentially anisotropic elastoplastic body. This theory is motivated by the finite deformation isothermal field dislocation mechanics theory (\cite{ACHARYA2001761}, \cite{acharya2004constitutive}) and the small-deformation thermomechanical framework of (\cite{upadhyay_thermomechanical_2020}). Specifically, the present theory is a generalisation of the thermomechanical theory in \cite{acharya_microcanonical_2011}, by accounting for a flux term in the dislocation density evolution that allows for capturing thermal strain effects in the stress response, and of the one in \cite{upadhyay_thermomechanical_2020} from a geometrical non-linearity perspective. \textcolor{black}{The theory deals with the thermomechanics of individually characterizable dislocations (hence, geometrically necessary); as shall be discussed at the end of section \ref{sec:conservationofB}, an extension to account for dislocation ensembles (both geometrically necessary and statistical) can be implemented using the procedure outlined in \cite{arora_unification_2020} and demonstrated in addressing critical experimental tests of models of (small-scale) plasticity in \cite{arora_mechanics_2022}, \cite{arora_interface-dominated_2023}, \cite{ARORA2020_FEM_FDM}.}

The motivation to develop this model arises from the need for a continuum framework in a geometrically nonlinear setting that has the minimum necessary tools to study the evolution of dislocations in bodies that undergo thermomechanical processes, such as additive manufacturing, welding, quenching, annealing, forming, forging, etc. The small deformation thermomechanical theory (\cite{upadhyay_thermomechanical_2020}) and its finite element implementation (\cite{2024_limachaves_upadhyay}) have shown how dislocation activity induces the evolution of the temperature field. The need to develop a geometrically nonlinear theory became evident when simulating dislocations moving at large velocities (at a considerable fraction of the material shear wave propagation velocity) (\cite{2024_limachaves_upadhyay}) or under rapidly evolving thermomechanical boundary conditions such as those occurring during additive manufacturing. 

The proposed theory relies only on measurable (observable) fields at any given instant of time, namely, the elastic distortion, the polar dislocation density, the temperature field, and the material velocity. The theory does not require a multiplicative decomposition of the deformation gradient tensor into elastic and plastic distortions ($\boldsymbol{F} = \boldsymbol{F}^e\boldsymbol{F}^p$) in the isothermal case, as introduced by \cite{bilby1957extrait},  \cite{kroner_allgemeine_1959}, \cite{lee_elastic-plastic_1969},  or into elastic, plastic and thermal parts in the thermomechanical case ($\boldsymbol{F} = \boldsymbol{F}^e\boldsymbol{F}^p\boldsymbol{F}^\theta$ or $\boldsymbol{F} = \boldsymbol{F}^e\boldsymbol{F}^\theta\boldsymbol{F}^p$), as considered in \cite{zeng_rate-dependent_2022}, \cite{bammann_kinematic_2010}, \cite{li_coupled_2022}, \cite{mcauliffe_unified_2015}, \cite{felder_thermo-mechanically_2022}, \cite{grilli_crystal_2022}, \cite{zhao_modeling_2013}, among others. \textcolor{black}{One of the earliest attempts at a continuum thermomechanical theory of dislocations (along with other dissipative phenomena) was proposed in \cite{berdichevskiiDynamicTheoryContinuously1967}. The structure of that model with respect to the evolution of geometrically necessary dislocation density is significantly different from ours as shall be discussed in section \ref{sec:smalldefexample}.} 

In the elastoplastic case, \cite{clifton_equivalence_1972} discussed the equivalence between the decompositions $\boldsymbol{F} = \boldsymbol{F}^e_{(1)}\boldsymbol{F}^p_{(1)}$ ("classical") and $\boldsymbol{F} = \boldsymbol{F}^p_{(2)}\boldsymbol{F}^e_{(2)}$ ("reverse"), stating that either can be used for the analysis of finite elastic and plastic deformations of isotropic solids, the choice being a matter of convenience for the study in question. \cite{lubarda_duality_1999} focused on the reverse decomposition $\boldsymbol{F} = \boldsymbol{F}^p_{(2)}\boldsymbol{F}^e_{(2)}$, and showed that for an isotropic solid, the same structure of constitutive equations is obtained as when using the classical decomposition, with $\boldsymbol{F}^e_{(1)} = \boldsymbol{F}^e_{(2)}$ if the material preserves its elastic properties during plastic deformation. More recently, \cite{yavari_direct_2023} proposed the equivalence of the classical and reverse decompositions for anisotropic solids, with the Cauchy stress computed by either being the same, ``when the direct and reverse decompositions represent the same anelastic deformation.'' Here, in a simple example of an elastoplastic \emph{evolution} considering $J_2$ plasticity and isotropic elasticity in a purely mechanical setting, we show in \cref{sec:ordering_example} that the Cauchy stress history corresponding to a simple shear or a combined loading depends on the chosen ordering of the multiplicative decomposition. The evolution of the plastic distortion in the two cases is different, although not directly comparable, being tensors with different invariance properties.

In the context of finite deformation thermoelasticity, the multiplicative decomposition of $\boldsymbol{F}$ into elastic and thermal components ($\boldsymbol{F} = \boldsymbol{F}^e\boldsymbol{F}^\theta$) was introduced in \cite{stojanovic_thermoel_1964} according to \cite{sadik_origins_2017}. As in elastoplasticity theories, an intermediate configuration is introduced, which is obtained from the current configuration upon isothermal elastic unloading (\cite{Vujosevic2002FinitestrainTB}). The non-uniqueness of the intermediate configuration is usually handled by considering a specific form for $\boldsymbol{F}^\theta$ according to the material to be modelled (\cite{Vujosevic2002FinitestrainTB}). Within a finite deformation thermo-elastoplastic theory, if one considers a material that thermally expands isotropically, then the ordering of the plastic and thermal distortion tensors in the decomposition $\boldsymbol{F} = \boldsymbol{F}^e\boldsymbol{F}^p\boldsymbol{F}^\theta$ is irrelevant, since, in this case, $\boldsymbol{F}^\theta = \alpha(\theta)\mathds{1}$, with $\alpha(\theta)$ being the thermal stretch ratio in an arbitrary direction (\cite{Vujosevic2002FinitestrainTB}). However, in a more general case that encompasses thermal anisotropy, the factors in the multiplicative decomposition do not commute, and there is little physical guidance as to what should be chosen, with each choice having an impact on the constitutive relations of the theory. 

Other researchers have also proposed finite deformation elastoplastic models that do not require the specification of a multiplicative decomposition of $\boldsymbol{F}$. \cite{rubin_eulerian_2023} proposed an Eulerian theory of size-dependent plasticity that does not rely on the multiplicative decomposition of $\boldsymbol{F}$, without including a detailed description of dislocation mechanics. In \cite{acharya2004constitutive}, a finite deformation time-dependent, isothermal dislocation mechanics theory is proposed, in which only the current configuration and a set of point-wise elastically unloaded configurations play a role in the theory, with the latter being defined through a kinematically fundamental statement of elastic incompatibility following the work of \cite{willis_second-order_1967} concerning dislocation statics. Following this line of work, \cite{acharya2015dislocation} show that, based on the conservation of the Burgers vector, in the isothermal case one recovers the well-accepted additive decomposition of the spatial velocity gradient into elastic and plastic parts, without the need to introduce the multiplicative decomposition $\boldsymbol{F}= \boldsymbol{F}^e\boldsymbol{F}^p$. This approach has seen significant development and validation against the experimental results in \cite{arora_dislocation_2020, ARORA2020_FEM_FDM, arora_unification_2020, arora_mechanics_2022, arora_interface-dominated_2023}.

The novelty of our work is the development of exact kinematics for thermo-elastoplastic problems based on dislocation mechanics in a finite-deformation setting within a transient heterogeneous temperature field which does not involve a reference configuration and deformation from it, and one that leads to a dynamical model whose dissipation is invariant to superposed rigid body motions. Similarly to \cite{upadhyay_thermomechanical_2020}, this kinematics naturally arises from the conservation of the Burgers vector. However, the extension to finite deformations under the required `design' constraints mentioned above is neither straightforward nor obvious; for instance, the present work differs fundamentally from \cite{upadhyay_thermomechanical_2020} by only involving observable fields along the lines of the finite deformation, thermomechanical model in \cite{acharya_microcanonical_2011}, the latter, however, not incorporating the effect of thermal strains in stress response. These differences impact the evolution statement of the dislocation density in comparison to \cite{acharya_microcanonical_2011}, \cite{upadhyay_thermomechanical_2020}, even apart from terms related to accounting for finite deformations. This difference is clearly shown in the comparison between the linearisation of the present model and the one in \cite{upadhyay_thermomechanical_2020} on \cref{sec:comparison_upadhyay}.

Building on the proposed kinematics and considering a constitutive assumption for the Helmholtz free energy that accounts for the (line-type) defect density in the body, the resulting structure of the PDE for temperature evolution is such that it allows for solutions in the form of dispersive temperature waves with \textit{finite speeds of propagation}. This is despite assuming Fourier's law as the constitutive equation for the heat flux vector, which results in the instantaneous propagation of temperature throughout a domain characteristic of a linear parabolic problem. It is shown that, at least on a linear level, the obtained PDE admits solutions with well-posed growth, allowing for the onset of spatial pattern formation from the amplification of wave components of specific wavenumbers. 

Our theory is well-suited for understanding the different sources of heat coming from thermomechanics and plastic work due to dislocation motion, with the latter being described by a geometrical argument of conservation of Burgers vector. The proposed framework allows us to evaluate the partition of plastic work into heat and stored energy in the material during thermomechanical processes. Understanding of this partition gained increased importance after the experiments conducted by \cite{taylorquinney_latent_1934}, which provided a measure of the latent energy stored in a material during cold work (see \cite{bever_stored_1973} for an extensive survey on the topic). Subsequently, many studies were directed towards computing the plastic work that remained stored in the material (and converted into heat) through modelling. Different techniques have been used, among them dislocation dynamics (\cite{benzerga2005stored}), molecular dynamics (\cite{kositski2021employing}, \cite{xiong2022atomistic}, \cite{stimac2022energy}) and continuum approaches (\cite{rosakis_thermodynamic_2000}, \cite{stainier_study_2010}, \cite{daehli2023experimental}, \cite{longere_thermodynamically_2023}), \cite{kunda2024study}, with the present fitting into the latter body of work.

This paper is divided as follows: \cref{sec:ordering_example} shows an example that illustrates the impact of the ordering of the elastoplastic multiplicative decomposition on the stress response of an isotropic material, which readers mainly interested in the development of the thermomechanical theory may safely skip. The kinematics based on the conservation of the Burgers vector is presented in \cref{sec:kinematics}. Then, the governing equations and the thermodynamical considerations of the theory are shown in \cref{sec:balance-constitutive}, where the temperature evolution PDE and the partition of plastic work in the model are also discussed. A geometrical linearisation of the proposed framework is shown in \cref{sec:small-def}. Considering a Saint-Venant-Kirchhoff material, a Helmholtz free energy expression is presented in \cref{sec:example_psi}, alongside the set of resulting equations of the model in the geometrically non-linear and linear cases, followed by a comparison with the theory developed in \cite{upadhyay_thermomechanical_2020}. The mathematical notation adopted in this paper can be consulted in \ref{sec:notation}.

\subsection{Ordering-dependence of the multiplicative decomposition in finite deformation elastoplasticity}\label{sec:ordering_example}
\setcounter{figure}{0}

The objective of this section is to study the impact of ordering on the multiplicative decomposition of the deformation gradient tensor ($\boldsymbol{F}$) into elastic ($\boldsymbol{F}^e$) and plastic ($\boldsymbol{F}^p$) distortions $\boldsymbol{F} = \boldsymbol{F}^e_{(1)}\boldsymbol{F}^p_{(1)}$ (denoted \textit{Case 1}) or $\boldsymbol{F} = \boldsymbol{F}^p_{(2)}\boldsymbol{F}^e_{(2)}$ (denoted \textit{Case 2}) in a simple and practical example within a purely mechanical setting, considering a given homogeneous deformation gradient history, constitutive relation for the stress response, and $J_2$ plasticity. More specifically, the evolution equations for $\dot{\boldsymbol{F}}^e_{(1)}$ (in \textit{Case 1}) and $\dot{\boldsymbol{F}}^e_{(2)}$ (in \textit{Case 2}) are solved for a given $\boldsymbol{F}(t)$ and $\boldsymbol{L}(t) = \dot{\boldsymbol{F}}\boldsymbol{F}^{-1}$, from which the Cauchy stress evolutions $\boldsymbol{\sigma}\big(\boldsymbol{F}^e_{(1)}(t)\big)$ and $\boldsymbol{\sigma}\big(\boldsymbol{F}^e_{(2)}(t)\big)$ are calculated and compared in simple shear and combined stretch-contraction-shear examples. The expression for $\boldsymbol{\sigma}$ is defined assuming hyperelasticity and following the frame-invariance requirements of $\boldsymbol{F}^e_{(1)}$ and $\boldsymbol{F}^e_{(2)}$. The plastic distortion rate $\boldsymbol{L}^p_{(i)}$, $i = 1,2$, is defined based on the simplest $J_2$ constitutive assumption and on the frame invariance requirements associated with $\text{sym}(\boldsymbol{L}^p_{(i)})$.  It is shown that $\boldsymbol{\sigma}\big(\boldsymbol{F}^e_{(1)}(t)\big)$ and $\boldsymbol{\sigma}\big(\boldsymbol{F}^e_{(2)}(t)\big)$ differ considerably in the examples considered here, so the choice of ordering in the multiplicative decomposition of $\boldsymbol{F}$ has a crucial impact on the evolution of stress in a body, without imposing any further conditions on the decompositions. As already mentioned, readers mainly interested in the theory presented in this work may skip to the end of this section. 

A central tenet in this study is that any mechanical theory, and hence the ones considered here based on either of the decompositions in \textit{Case 1} or \textit{Case 2}, should reduce to conventional nonlinear elasticity as a limiting case. That is, considering $\boldsymbol{F}^p_{(i)} = \mathds{1}$, $i = 1,2$, we have $\boldsymbol{F} = \boldsymbol{F}^e_{(i)}$ to recover elasticity, such that in both cases $\boldsymbol{F}^e_{(i)}$ is a two-point tensor with its co-domain being tangent spaces at points of the body in the current configuration.

The two cases are presented as follows:

\paragraph{Case 1: $\boldsymbol{F} = \boldsymbol{F}^e_{(1)}\boldsymbol{F}^p_{(1)}$} The spatial velocity gradient is written as
\begin{equation}
    \begin{split}
        &\boldsymbol{L} = \dot{\boldsymbol{F}}\boldsymbol{F}^{-1}  = \left(\dot{\boldsymbol{F}}^e_{(1)}\boldsymbol{F}^p_{(1)} + \boldsymbol{F}^e_{(1)}\dot{\boldsymbol{F}}^p_{(1)}\right)\left(\boldsymbol{F}^{p-1}_{(1)}\boldsymbol{F}^{e-1}_{(1)}\right) = \dot{\boldsymbol{F}}^e_{(1)}\boldsymbol{F}^{e-1}_{(1)} + \boldsymbol{F}^e_{(1)}\dot{\boldsymbol{F}}^p_{(1)}\boldsymbol{F}^{p-1}_{(1)}\boldsymbol{F}^{e-1}_{(1)}\\
        \implies& \boldsymbol{L} = \dot{\boldsymbol{F}}^e_{(1)}\boldsymbol{F}^{e-1}_{(1)} + \boldsymbol{L}^p_{(1)} \\
        \implies& \dot{\boldsymbol{F}}^e_{(1)} = \left(\boldsymbol{L} - \boldsymbol{L}^p_{(1)}\right) \boldsymbol{F}^e_{(1)},
    \end{split}
    \label{eqn:Fe_dot_FeFp}
\end{equation}
where we have defined $\boldsymbol{L}^p_{(1)} := \boldsymbol{F}^e_{(1)}\dot{\boldsymbol{F}}^p_{(1)}\boldsymbol{F}^{p-1}_{(1)}\boldsymbol{F}^{e-1}_{(1)}$. \vspace{0.2cm}

Under a rigid body motion as in \cref{eqn:app_RBM}, $\boldsymbol{F}$ transforms as $\boldsymbol{F}^* = \boldsymbol{QF}$, and we have 
\begin{equation}
    \boldsymbol{F}^* = \boldsymbol{F}_{(1)}^{e*}\boldsymbol{F}_{(1)}^{p*} \implies \boldsymbol{F}_{(1)}^{e*}\boldsymbol{F}_{(1)}^{p*} = \boldsymbol{Q}\left(\boldsymbol{F}_{(1)}^{e}\boldsymbol{F}_{(1)}^{p}\right). 
    \label{eqn:Fstar_Case1}
\end{equation}
To recover elasticity, we set $\boldsymbol{F}^p_{(1)} = \mathds{1}$ and $\boldsymbol{F}^{p*}_{(1)} = \mathds{1}$ which implies $\boldsymbol{F} = \boldsymbol{F}^e_{(1)}$ and
\begin{equation}
    \boldsymbol{F}^{e*}_{(1)} = \boldsymbol{Q}\boldsymbol{F}^e_{(1)}.
    \label{eqn:app_invariance_case1}
\end{equation}
We assume that the invariance of $\boldsymbol{F}^e_{(1)}$ under a superposed rigid body motion remains the same in the elastoplastic case; then, \mbox{\cref{eqn:app_invariance_case1,eqn:Fstar_Case1}} imply
\begin{equation}
    \boldsymbol{F}^{p*}_{(1)} = \boldsymbol{F}^p_{(1)}.
    \label{eqn:Fp1_rotation}
\end{equation}

\paragraph{Case 2: $\boldsymbol{F} = \boldsymbol{F}^p_{(2)}\boldsymbol{F}^e_{(2)}$} The velocity gradient now reads
\begin{equation}
    \begin{split}
        &\boldsymbol{L} = \dot{\boldsymbol{F}}\boldsymbol{F}^{-1} = \dot{\boldsymbol{F}}^p_{(2)}\boldsymbol{F}^{p-1}_{(2)} + \boldsymbol{F}^p_{(2)}\dot{\boldsymbol{F}}^e_{(2)}\boldsymbol{F}^{e-1}_{(2)}\boldsymbol{F}^{p-1}_{(2)}\\
        \implies&\dot{\boldsymbol{F}}^e_{(2)} = \boldsymbol{F}^{p-1}_{(2)}\left(\boldsymbol{L} - \boldsymbol{L}^p_{(2)}\right) \boldsymbol{F}^{p}_{(2)}\boldsymbol{F}^{e}_{(2)} = \boldsymbol{F}^{e}_{(2)}\boldsymbol{F}^{-1} \left(\boldsymbol{L} - \boldsymbol{L}^p_{(2)}\right) \boldsymbol{F}, 
    \end{split}
    \label{eqn:Fe_dot_FpFe}
\end{equation}
where $L^P_{(2)} := \dot{\boldsymbol{F}}^p_{(2)}\boldsymbol{F}^{p-1}_{(2)}$. \vspace{0.2cm}

Superposed rigid body motion in this case requires
\begin{equation}
    \boldsymbol{F}^* = \boldsymbol{F}_{(2)}^{p*}\boldsymbol{F}_{(2)}^{e*} \implies \boldsymbol{F}_{(2)}^{p*}\boldsymbol{F}_{(2)}^{e*} = \boldsymbol{Q}\left(\boldsymbol{F}_{(2)}^{p}\boldsymbol{F}_{(2)}^{e}\right). 
    \label{eqn:Fstar_Case2}
\end{equation}
As before, to recover elasticity we set  $\boldsymbol{F}^p_{(2)} = \mathds{1}$, such that $\boldsymbol{F} = \boldsymbol{F}^e_{(2)}$ and 
\begin{equation}
    \boldsymbol{F}^* = \boldsymbol{F}^{e*}_{(2)} \implies \boldsymbol{F}^{e*}_{(2)} = \boldsymbol{QF}^e_{(2)}.
    \label{eqn:app_invariance_case2}
\end{equation}
Assuming that this invariance requirement of $\boldsymbol{F}^e_{(2)}$ also applies in the elastoplastic case, \cref{eqn:app_invariance_case2,eqn:Fstar_Case2} gives
\begin{equation}
    \begin{split}
        &\boldsymbol{F}^{p*}_{(2)} = \boldsymbol{Q}\left(\boldsymbol{F}^p_{(2)}\boldsymbol{F}^e_{(2)}\right)\left(\boldsymbol{F}^{e*}_{(2)}\right)^{-1} = \boldsymbol{Q}\left(\boldsymbol{F}^p_{(2)}\boldsymbol{F}^e_{(2)}\right)\boldsymbol{F}^{e-1}\boldsymbol{Q}^T\\
        \implies& \boldsymbol{F}^{p*}_{(2)} = \boldsymbol{Q}\boldsymbol{F}^p_{(2)}\boldsymbol{Q}^T,
    \end{split}
    \label{eqn:Fp2_rotation}
\end{equation}
that is, $\boldsymbol{F}^p_{(2)}$ transforms as a tensor on the current configuration under a superposed rigid body motion.

To define the Cauchy stress tensor, we consider that the stress response given by $\boldsymbol{\sigma} = \hat{\boldsymbol{\sigma}}(\boldsymbol{F}^e)$. Given that $\boldsymbol{F}^e_{(i)}$ transforms as $\boldsymbol{F}^{e*}_{(i)} = \boldsymbol{QF}^e_{(i)}$ under a superposed rigid body motion for $i = 1,2$ (\cref{eqn:app_invariance_case1,eqn:app_invariance_case2}), the material frame indifference for the stress tensor (c.f., \cite{truesdell_non-linear_2004}) requires the reduced constitutive equation $\hat{\boldsymbol{\sigma}}(\boldsymbol{F}^e_{(i)}) = \boldsymbol{F}^e_{(i)}\tilde{\boldsymbol{\sigma}}(C^e_{(i)}) \boldsymbol{F}^{eT}_{(i)}$, where $\boldsymbol{C}^e_{(i)} = \boldsymbol{F}^{eT}_{(i)}\boldsymbol{F}^e_{(i)}$ is the right Cauchy-Green tensor. For this example, we choose $\tilde{\boldsymbol{\sigma}}(\boldsymbol{C}^e_{(i)}) = \mathds{C}:\boldsymbol{E}_{(i)}$ with $\boldsymbol{E}_{(i)} = \frac{1}{2}(\boldsymbol{C}^e_{(i)} - \mathds{1})$ and $\mathds{C}$ the fourth order isotropic stiffness tensor. Hence, we write the Cauchy stress tensor as
\begin{equation}
    \boldsymbol{\sigma}_{(i)} = \boldsymbol{F}^{e}_{(i)} \left[\frac{1}{2}\mathds{C} : \left(\boldsymbol{F}^{eT}_{(i)}\boldsymbol{F}^{e}_{(i)} - \mathds{1}\right)\right] \boldsymbol{F}^{eT}_{(i)}.
    \label{eqn:app_sigma_constitutive_i}
\end{equation}

Next, we choose an expression for $\boldsymbol{L}^p_{(i)}$ based on $J_2$ plasticity theory and the invariance requirements of $\text{sym}(\boldsymbol{L}^p_{(i)})$. For a superposed rigid body motion, in \textit{Case 1} we write \cref{eqn:Fe_dot_FeFp} as
\begin{equation}
    \begin{split}
        &\text{sym}(\boldsymbol{L}^*) = \text{sym}\big(\dot{\boldsymbol{F}}^{e*}_{(1)}(\boldsymbol{F}^{e*}_{(1)})^{-1}\big) + \text{sym}(\boldsymbol{L}^{p*}_{(1)}) \\
        \implies& \boldsymbol{Q}\text{sym}(\boldsymbol{L})\boldsymbol{Q}^T = \boldsymbol{Q}\text{sym}\big(\dot{\boldsymbol{F}}^{e}_{(1)}\boldsymbol{F}^{e-1}_{(1)}\big)\boldsymbol{Q}^T + \text{sym}(\boldsymbol{L}^{p*}_{(1)}) \\
        \implies& \text{sym}(\boldsymbol{L}^{p*}_{(1)}) = \boldsymbol{Q}\text{sym}(\boldsymbol{L}^{p}_{(1)})\boldsymbol{Q}^T,
    \end{split}
    \label{eqn:app_LpRequirement_case1}
\end{equation}
where \cref{eqn:app_invariance_case1} was used. For \textit{Case 2} we write \cref{eqn:Fe_dot_FpFe} as  
\begin{equation}
    \begin{split}
        &\text{sym}(\boldsymbol{L}^*) = \text{sym}(\boldsymbol{L}^{p*}_{(2)}) + \text{sym}\big(\boldsymbol{F}^{p*}\dot{\boldsymbol{F}}^{e*}_{(2)}(\boldsymbol{F}^{e*}_{(2)})^{-1}(\boldsymbol{F}^{p*}_{(2)})^{-1}\big) \\
        \implies& \boldsymbol{Q}\text{sym}(\boldsymbol{L})\boldsymbol{Q}^T = \text{sym}(\boldsymbol{L}^{p*}_{(2)}) + \boldsymbol{Q}\text{sym}\big(\boldsymbol{F}^{p}\dot{\boldsymbol{F}}^{e}_{(2)}\boldsymbol{F}^{e-1}_{(2)}\boldsymbol{F}^{p-1}_{(2)}\big)\boldsymbol{Q}^T\\
        \implies& \text{sym}(\boldsymbol{L}^{p*}_{(2)}) = \boldsymbol{Q}\text{sym}(\boldsymbol{L}^{p}_{(2)})\boldsymbol{Q}^T.
    \end{split}
    \label{eqn:app_LpRequirement_case2}
\end{equation}
Hence, for both \textit{Case 1} and \textit{Case 2} we use
\begin{equation}
    \begin{split}
        \boldsymbol{L}^p_{(i)} &= a_{(i)}\frac{\text{dev}(\boldsymbol{\sigma}_{(i)})}{\lVert\text{dev}(\boldsymbol{\sigma}_{(i)}\rVert}\\
        a_{(i)} &= \hat{\gamma}_0 \left(\frac{\lVert\text{dev}(\boldsymbol{\sigma}_{(i)})\rVert}{\sqrt{2}g}\right)^{\frac{1}{m}},
    \end{split}
    \label{eqn:Lp_J2}
\end{equation}
where $\hat{\gamma}_0$ is the reference strain rate, $g$ is the material strength and $m$ is the material rate sensitivity coefficient, and $\text{dev}(\boldsymbol{\sigma}) = \boldsymbol{\sigma} - \frac{1}{3}\text{tr}(\boldsymbol{\sigma})\mathds{1}$ is the stress deviator. This allows a ``most'' unbiased choice allowing for minimum deviations between the predictions of the two cases.

\subsubsection{Example: simple shear}

Consider a homogeneous, time-dependent deformation gradient tensor whose components with respect to an orthonormal basis $\{\hat{\boldsymbol{e}}_1, \hat{\boldsymbol{e}}_2, \hat{\boldsymbol{e}}_3\}$ are expressed in matrix form as
\begin{equation}
    \boldsymbol{F}(t) = 
    \begin{pmatrix}
        1  & \gamma_0 t & 0 \\
        0 & 1  & 0 \\
        0 & 0 & 1
    \end{pmatrix},
    \label{eqn:given_F_history_shear}
\end{equation}
which corresponds to a simple shear along $\hat{\boldsymbol{e}}_1 \otimes \hat{\boldsymbol{e}}_2$. The corresponding components of the velocity gradient $\boldsymbol{L} = \dot{\boldsymbol{F}}\boldsymbol{F}^{-1}$ are given by
\begin{equation}
    \boldsymbol{L}(t) =     
    \begin{pmatrix}
        0 & \gamma_0 & 0 \\
        0 & 0 & 0 \\
        0 & 0 & 0
    \end{pmatrix}.
    \label{eqn:given_L_history_shear}
\end{equation}
The example consists of solving the evolution equations for $\dot{\boldsymbol{F}}^e_{(i)}$, $i = 1,2$ given by \cref{eqn:Fe_dot_FeFp,eqn:Fe_dot_FpFe}, with $\boldsymbol{F}$ and $\boldsymbol{L}$ given by \cref{eqn:given_F_history_shear,eqn:given_L_history_shear}, respectively, $\boldsymbol{L}^p_{(i)}$ computed by \cref{eqn:Lp_J2}, and $\boldsymbol{\sigma}_{(i)}$ expressed as in \cref{eqn:app_sigma_constitutive_i}. The parameters used in the calculations are $E = 100$ GPa, $\nu = 0.3$, $\gamma_0 = 1$ s$^{-1}$, $g = 50$ MPa, $m = 0.01$. The numerical integration of \cref{eqn:Fe_dot_FeFp,eqn:Fe_dot_FpFe} is carried out at a single material point up to $t_F = 1$ s, with a time step of $\Delta t = 10^{-5}$ s, resulting in a $\gamma_0 t_F = 1$ shear strain. 

\begin{figure}[!ht]
    \centering
    \includegraphics[width=.8\textwidth]{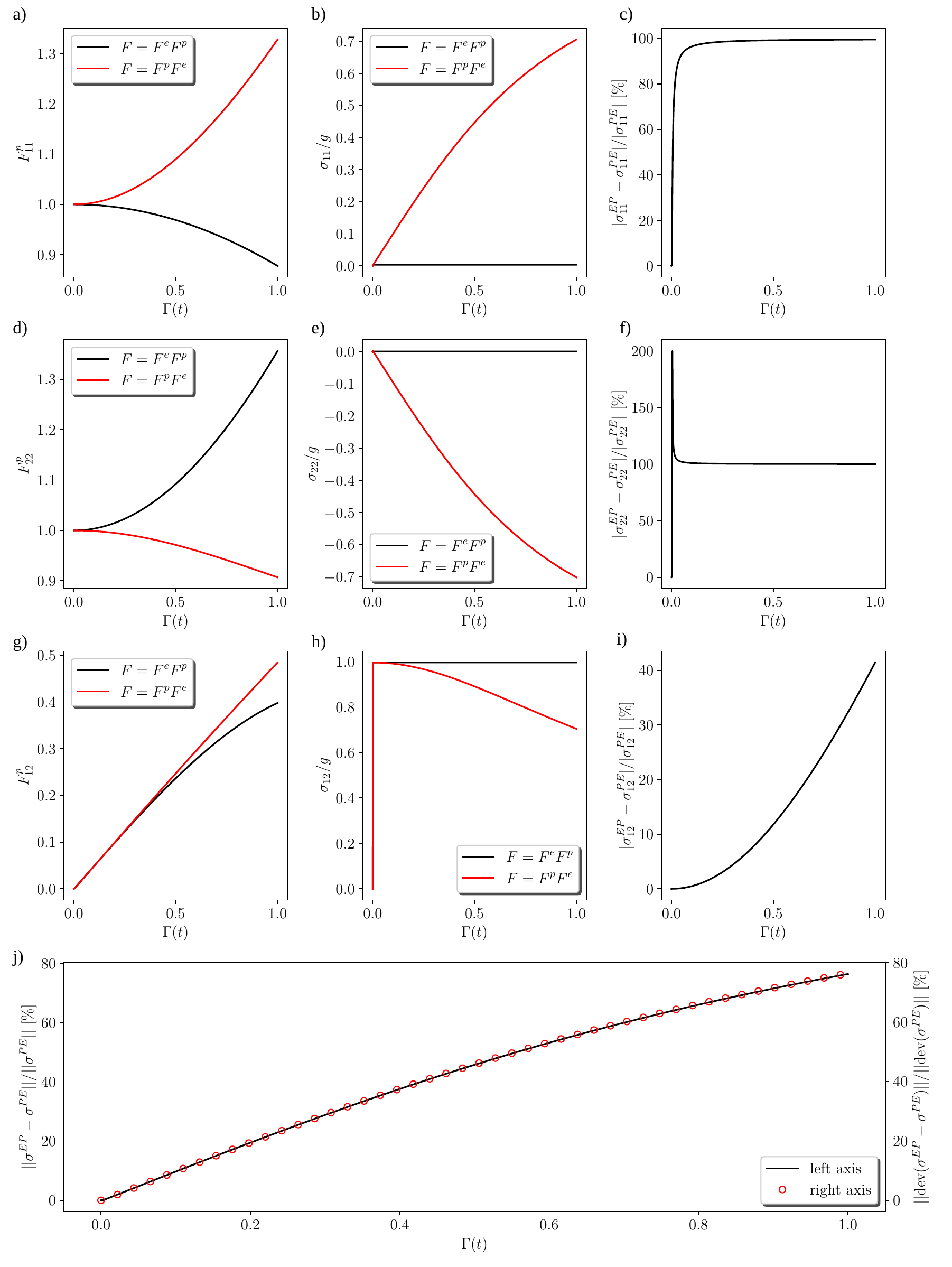}
    \caption[Stress and plastic distortion evolution during simple shear, considering different orderings of the multiplicative decomposition of the deformation gradient tensor.]{Stress and plastic distortion evolution during simple shear considering different orderings of the multiplicative decomposition of the deformation gradient tensor, plotted against $\Gamma(t) = \lVert\int \boldsymbol{L}\, \text{d}t\rVert$. a), d), g): Evolution of the 11, 12 and 22 components of the plastic distortion tensor for each case; the $EP$ and $PE$ superscripts indicate variables resulting from the decompositions $\boldsymbol{F} = \boldsymbol{F}^e\boldsymbol{F}^p$ and $\boldsymbol{F}=\boldsymbol{F}^p\boldsymbol{F}^e$, respectively. b), e), h): Evolution of the 11, 12 and 22 components of the Cauchy stress tensor for each case, with the relative differences shown in c), f), and i). The curve in f) is capped at 200 \% to suppress the large peak due to the red curve in e) crossing 0. j): Evolution of the relative differences in the Frobenius norm of the Cauchy stress deviator and full stress tensor for each case.}
    \label{fig:Shear_FeFp}
\end{figure}

For this simple example, considerable differences emerge in the stress response according to the ordering of the multiplicative decomposition, as shown in \cref{fig:Shear_FeFp}. The evolutions depicted in \mbox{\cref{fig:Shear_FeFp}}a, d, and g cannot be directly compared, as $\boldsymbol{F}^p_{(1)}$ and $\boldsymbol{F}^p_{(2)}$ are not tensors with the same domain and co-domain. The normal Cauchy stress components $\sigma_{11}$ and $\sigma_{22}$ remain constant and close to zero in \textit{Case 1}, whereas in \textit{Case 2}, they evolve to positive and negative values, respectively, producing a significant deviation (\cref{fig:Shear_FeFp}b, c, e, and f). The shear component $\sigma_{12}$ behaves differently according to the ordering, with a constant plateau after yielding in \textit{Case 1}, and softening in \textit{Case 2}, such that the relative difference constantly increases to reach above 40 \% at $\Gamma = 1$ (\cref{fig:Shear_FeFp}h and i). Finally, $\lVert\text{dev}(\boldsymbol{\sigma})\rVert$ and $\lVert\boldsymbol{\sigma}\rVert$ present a similar evolution in both cases, but with a considerable relative difference that reaches a maximum of 80 \% for both (\cref{fig:Shear_FeFp}j).

\subsubsection{Example: combined stretch, contraction and shear}

In this example, we write the deformation gradient tensor components as
\begin{equation}
    \boldsymbol{F}(t) = 
    \begin{pmatrix}
        1 + d\gamma_0 t & \gamma_0 t & 0 \\
        0 & 1 - d\gamma_0 t & 0 \\
        0 & 0 & 1
    \end{pmatrix},
    \label{eqn:given_F_history}
\end{equation}
where $d$ is a constant factor. This deformation gradient corresponds to a combined stretch along $\hat{\boldsymbol{e}}_1$, contraction along $\hat{\boldsymbol{e}}_2$, and shear along $\hat{\boldsymbol{e}}_1 \otimes \hat{\boldsymbol{e}}_2$. The corresponding components of the velocity gradient are given by 
\begin{equation}
    \boldsymbol{L}(t) =     
    \begin{pmatrix}
        \frac{d\gamma_0}{1 + d\gamma_0 t} & \frac{\gamma_0}{1 - (d\gamma_0 t)^2} & 0 \\
        0 & \frac{d\gamma_0}{-1 + d\gamma_0 t} & 0 \\
        0 & 0 & 0
    \end{pmatrix}.
    \label{eqn:given_L_history}
\end{equation}
The example consists of solving the evolution equations for $\dot{\boldsymbol{F}}^e_{(i)}$, $i = 1,2$ given by \cref{eqn:Fe_dot_FeFp,eqn:Fe_dot_FpFe}, with $\boldsymbol{F}$ and $\boldsymbol{L}$ given by \cref{eqn:given_F_history,eqn:given_L_history}, respectively, $\boldsymbol{L}^p_{(i)}$ computed by \cref{eqn:Lp_J2}, and $\boldsymbol{\sigma}_{(i)}$ expressed as in \cref{eqn:app_sigma_constitutive_i}. The parameters used in the calculations are $E = 100$ GPa, $\nu = 0.3$, $d = 0.05$, $\gamma_0 = 1$ s$^{-1}$, $g = 50$ MPa, $m = 0.01$. The numerical integration of \cref{eqn:Fe_dot_FeFp,eqn:Fe_dot_FpFe} is also carried out at a single material point and up to $t_F = 1$ s, with a time step of $\Delta t = 10^{-5}$ s, resulting in a $\gamma_0 t_F = 1$  strain in shear and $d \gamma_0 t_F = 0.05$ strain in stretch and contraction. 

\begin{figure}[!ht]
    \centering
    \includegraphics[width=.8\textwidth]{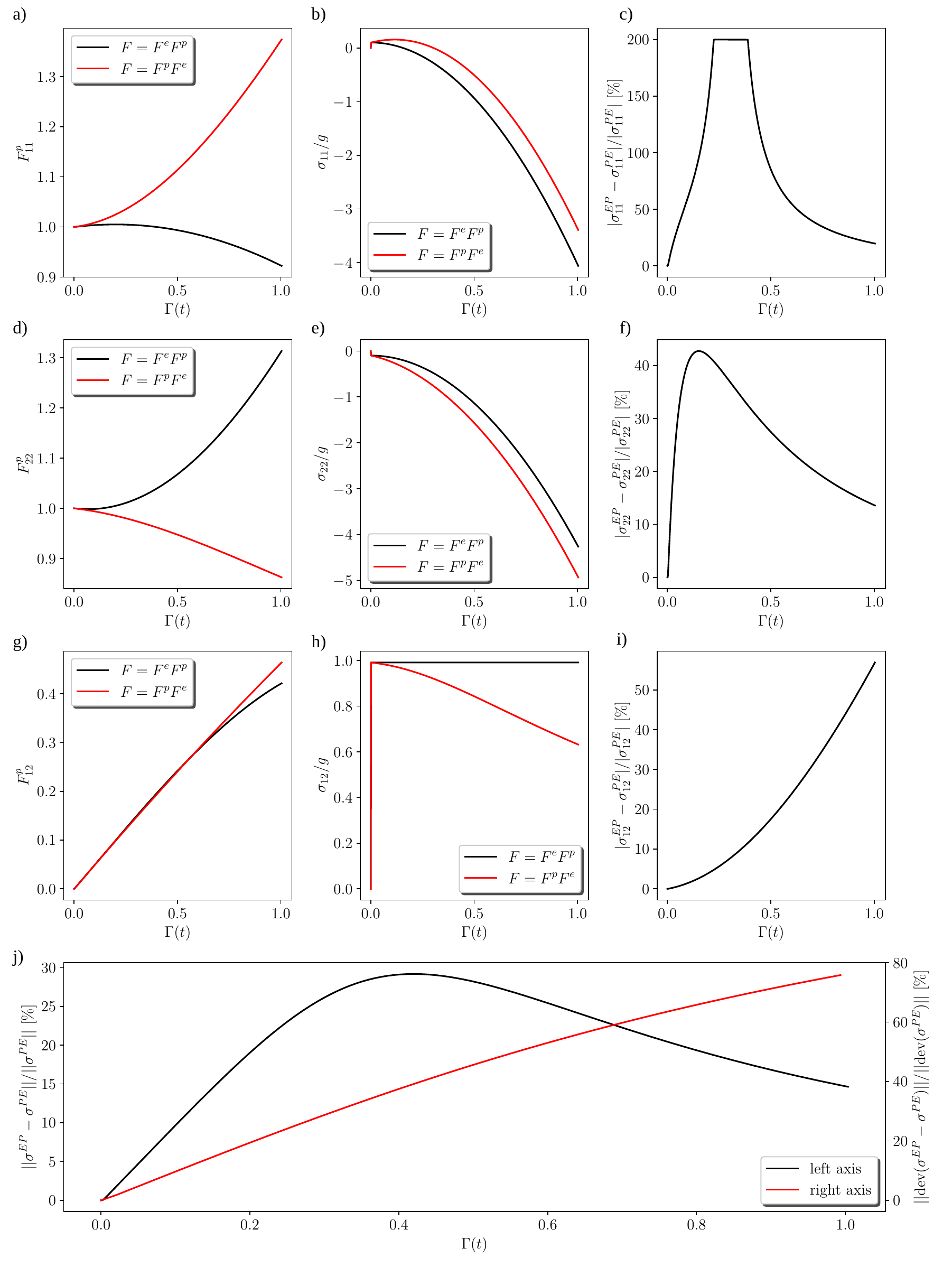}
    \caption[Stress and plastic distortion evolution during a combined stretch, contraction and shear, considering different orderings of the multiplicative decomposition of the deformation gradient tensor.]{Stress and plastic distortion evolution for a combined stretch, contraction and shear considering different orderings of the multiplicative decomposition of the deformation gradient tensor, plotted against $\Gamma(t) = \lVert\int \boldsymbol{L}\, \text{d}t\rVert$. a), d), g): Evolution of the 11, 12 and 22 components of the plastic distortion tensor for each case; the $EP$ and $PE$ superscripts indicate variables resulting from the decompositions $\boldsymbol{F} = \boldsymbol{F}^e\boldsymbol{F}^p$ and $\boldsymbol{F}=\boldsymbol{F}^p\boldsymbol{F}^e$, respectively. b), e), h): Evolution of the 11, 12 and 22 components of the Cauchy stress tensor for each case, with the relative differences shown in c), f), and i); the curve in c) is capped at 200 \% to avoid having a large peak due to the red curve in b) crossing 0. j): Evolution of the relative differences in the Frobenius norm of the Cauchy stress deviator and full stress tensor for each case.}
    \label{fig:stretchContracShear_FeFp}
\end{figure}

In this combined loading, considerable differences also arise due to the ordering of the multiplicative decomposition. As in the previous case, the components of $\boldsymbol{F}^p$ cannot be directly compared. The normal Cauchy stress component $\sigma_{11}$ is positive in the beginning due to the stretching in the $\hat{\boldsymbol{e}}_1$ direction, but becomes negative due to the combined and nonlinear interactions between the contraction and shear in the other directions. Similar behaviour is obtained for both orderings of the multiplicative decomposition but with a considerable relative difference of around 25 \% at $\Gamma = 1$ (\cref{fig:stretchContracShear_FeFp}b and c). The component $\sigma_{22}$ is purely negative due to contraction along the $\hat{\boldsymbol{e}}_2$, and the relative difference between the stress evolutions reaches around 40 \% before decreasing to 20 \% at $\Gamma = 1$ (\cref{fig:stretchContracShear_FeFp}e, f). The shear component $\sigma_{12}$ presents a behaviour similar to the previous simple shear example (\cref{fig:stretchContracShear_FeFp}h and i). Finally, $\lVert\text{dev}(\boldsymbol{\sigma})\rVert$ and $\lVert\boldsymbol{\sigma}\rVert$ present a similar evolution for both orderings, but with a considerable relative difference that reaches a maximum of 80 \% and 30 \% in each case, respectively (\cref{fig:stretchContracShear_FeFp}j).

\subsubsection{Summary and implications}

 The ordering-dependence of the decompositions becomes evident in the presence of rotation of the material, that is when shear is involved. Pure stretch/contraction simulations result in no difference in stress evolution due to the decomposition ordering. The present discussion also extends to the thermo-elastoplastic case, in which a third thermal distortion tensor $\boldsymbol{F}^\theta$ can be included in the multiplicative decomposition, thus yielding six possible multiplicative decompositions of the deformation gradient into elastic, plastic, and thermal components. Of course, in the thermally isotropic case, the position $\boldsymbol{F}^\theta$ in the multiplicative decomposition would be irrelevant, since it would be expressed as a multiple of the identity tensor. However, in the more general anisotropic case, we expect that the six possible orderings would result in even more differences in the stress response of the body. 

Considering a given body in its as-received state, it is impossible to uniquely define its plastic/thermal history, its current stress state and temperature distribution being the only accessible internal variables relevant in the context of this work. Therefore, both the definition of $\boldsymbol{F}^p$ and $\boldsymbol{F}^\theta$, and the order in which they appear in the decomposition, are arbitrary. In this sense, a unique multiplicative decomposition of the deformation gradient would require assuming knowledge of the precise deformation history of the body, which is not available to us. The only deformation history that we can follow starts from the first "current" configuration, that is, the as-received body.

These are some of the reasons why in our theory we avoid relying on such decompositions, working instead on the current configuration with kinematics based on the conservation of the Burgers vector, presented in the next section, which does not require the introduction of a global reference configuration and a plastic distortion from it. 

\section{Kinematics}\label{sec:kinematics}

\subsection{Distortion fields and configurations}\label{sec:distortion}

\begin{figure}
    \centering\includegraphics[width=.8\textwidth]{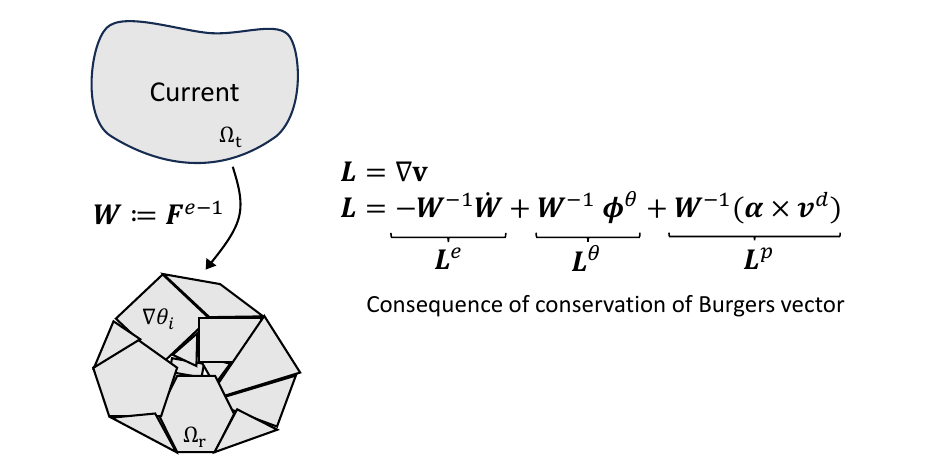}
    \caption{The transformation of $\Omega_t$ by $\boldsymbol{W}$, the only distortion tensor involved in this work. The $\nabla\theta_i$ represents that each polygon $i$ is allowed to have a temperature gradient, as long as $\Omega_r$ remains stress-free. The additive decomposition of the velocity gradient into elastic, thermal and plastic parts is also shown and further discussed in \cref{sec:conservationofB}.}
    \label{fig:configurations}
\end{figure}

Consider a body $\Omega$ that contains a distribution of dislocation lines and a temperature gradient at a given moment in time $t$ due to some combination of constant mechanical and thermal boundary conditions, as well as internal forces and heat sources within $\Omega$. We shall assume that the local temperature $\theta(\boldsymbol{x}, t)$ is below the solidus temperature everywhere in $\Omega$ at any given instant in time, that is, $\Omega$ always remains in the solid state. These configurations of $\Omega$ parametrized by time shall be called its current configurations and is denoted as $\Omega_t$. Next, suppose that the body, at each fixed instant of time, can be relaxed pointwise to a set ($\Omega_r$) of local stress-free configurations through the inverse elastic distortion tensor $\boldsymbol{W}:= \boldsymbol{F}^{e-1}$ (\mbox{\cref{fig:configurations}}).\hspace{1cm}  

Despite not adopting the multiplicative decomposition, as shall be shown, our model yields the well-known and accepted additive decomposition of the spatial velocity gradient ($\boldsymbol{L}$) into elastic ($\boldsymbol{L^e}$), thermal ($\boldsymbol{L^\theta}$) and plastic ($\boldsymbol{L^p}$) distortion rates (see, e.g. \cite{NEMATNASSER_1982} and the references therein). Furthermore, and crucially, this decomposition arises as a natural consequence of the conservation of the Burgers vector. We note that one could indeed build (non-unique) plastic and thermal distortion tensors out of this information by invoking an arbitrarily fixed reference configuration; however, this consideration is merely a consequence of the theory, if so desired, and not a necessary physical element.

\subsection{The thermomechanical line defect -- a consequence of the definition of the Burgers vector}

In a uniform $\theta$ field and the absence of dislocations, $\boldsymbol{W}$ is compatible (curl-free) i.e., $\nabla\times\boldsymbol{W} = 0$, in $\Omega_t$, and could thus be represented as the gradient of a vector field (in a simply-connected domain). In the presence of dislocations and/or temperature gradients, an incompatibility might be introduced in $\boldsymbol{W}$, that is, $\nabla\times\boldsymbol{W} \neq 0$, at one or more points in the body. 

If we are to draw a closed circuit (a Burgers circuit) in $\Omega_t$ and take the line integral of $\boldsymbol{W}$ over that circuit, and if this line integral is non-vanishing, then it characterises the vector topological charge (Burgers vector) carried by a line-type defect as
\begin{equation}\label{eqn:burgersdef}
    \begin{aligned}
        \boldsymbol{b}_r := -\oint_{c_t} \boldsymbol{W}  \textbf{d}\boldsymbol{x} = - \int_{s_t} (\nabla\times\boldsymbol{W})  \hat{\boldsymbol{n}} \text{ d}s = \int_{s_t} \boldsymbol{\alpha}  \hat{\boldsymbol{n}} \text{ d}s,
    \end{aligned}
\end{equation}
where $c_t$ denotes a closed curve in $\Omega_t$ with line element $\boldsymbol{\textbf{d}x}$, and $s_t$ is an arbitrary surface enclosed by $c_t$, whose normal is $\hat{\boldsymbol{n}}$. The two-point tensor $\boldsymbol{\alpha}$ is an areal density defined as
\begin{equation}\label{eqn:alpha_curlW}
    \begin{aligned}
        \boldsymbol{\alpha} := - \nabla\times\boldsymbol{W}.
    \end{aligned}
\end{equation}

In regions where only dislocations contribute to the incompatibility in $\boldsymbol{W}$, $\boldsymbol{\alpha}$ and $\boldsymbol{b}$ only characterise dislocations. This association is very well known and accepted since the pioneering works of \cite{nye_geometrical_1953}, \cite{kroner_kontinuumstheorie_1958}, \cite{de_wit_continuum_1960}, and \cite{mura_continuous_1963}. In fact, over the years, the definition of Burgers vector has become synonymous with the character of a dislocation.

However, the very definition of the Burgers vector in \cref{eqn:burgersdef} allows one to capture the incompatibility in $\boldsymbol{W}$ irrespective of the source of that incompatibility; in the context of this work, this incompatibility could arise from either dislocations or heterogeneous $\theta$ fields or both (\cite{kroner_kontinuumstheorie_1958,upadhyay_thermomechanical_2020}). Furthermore, the definition of the Burgers vector makes it difficult to distinguish between the different contributors to the incompatibility of $\boldsymbol{W}$, especially in a geometrically non-linear setting. We note that this point is made from a measurement point of view (e.g.\@ direct measurement of $\boldsymbol{W}$ or computing Burgers vectors from molecular statics simulations).

The consequences of different contributors to the incompatibility of elastic distortion were first explored by \cite{kroner_kontinuumstheorie_1958} in a small deformation stationary thermomechanics setting and later in a kinematic and dynamic setting by \cite{upadhyay_thermomechanical_2020}. Kröner argued that the incompatibility in $\boldsymbol{W}$ induced by a heterogeneous $\theta$ arises from a defect that also has a line-type nature. Kröner named this defect as the thermal quasi-dislocation, where \say{quasi} indicates that while this incompatibility is captured by the same type of areal density and vector used to characterise dislocations, it does not manifest itself as a line-type defect. However, both \cite{kroner_kontinuumstheorie_1958} and \cite{upadhyay_thermomechanical_2020} relied on the definition of plastic and thermal strain tensors. This definition can be applied in the small-deformation case, where each contribution (elastic, plastic, thermal) to the strain tensor can be separately added to the total strain tensor (from a reference configuration that is necessarily introduced for geometric linearisation). In the geometrically non-linear case, the introduction of these tensors would require assuming full knowledge of the thermomechanical history of a body by introducing a global reference configuration and a plastic distortion tensor from it. In this work, the thermomechanics of field dislocations is treated in the dynamic, finite-deformation setting without such a requirement.

Following the arguments presented by \cite{kroner_kontinuumstheorie_1958}, in regions where dislocations are not present but where the heterogeneous $\theta$ field induces an incompatibility in $\boldsymbol{W}$, $\boldsymbol{\alpha}$ and $\boldsymbol{b}$ can be non-vanishing. There is no evidence that a line-type thermal defect manifests itself in such situations, but non-vanishing $\boldsymbol{\alpha}$ and $\boldsymbol{b}$ do not have to arise from a line-type defect. They will be non-zero whenever $\nabla\times\boldsymbol{W} \neq 0$ is respected. As demonstrated in the small deformation setting in \cite{upadhyay_thermomechanical_2020}, in a domain containing a constant temperature gradient but no dislocations, constant non-zero $\boldsymbol{\alpha}$ and $\boldsymbol{b}$ are generated everywhere, and these quantities are measurable.

In the more general case, where both dislocations and heterogeneous $\theta$ field are present in the domain, $\boldsymbol{\alpha}$ and $\boldsymbol{b}$ can be non-vanishing and we assert that it is not possible to uniquely separate their contributions to the incompatibility in $\boldsymbol{W}$. In this situation, we postulate that the line defect has a thermomechanical character and henceforth we shall call it the \textit{thermomechanical defect}, with $\boldsymbol{\alpha}$ representing its density. In the case where only dislocations contribute to incompatibility in $\boldsymbol{W}$, this thermomechanical defect manifests itself simply as a dislocation line, but not necessarily, when the dislocation lines may form dense distribution on the scale of observation. Similarly, in a dislocation-free medium with a heterogeneous $\theta$ that contributes to incompatibilities in $\boldsymbol{W}$, the thermomechanical defect does not manifest itself as a line, but has the character of the thermal defect as postulated by \cite{kroner_kontinuumstheorie_1958}. This feature becomes important when dealing with the kinematics of this line defect (\cref{sec:conservationofB}).

Note that in the case where only dislocations contribute to $\boldsymbol{\alpha}$ and $\boldsymbol{b}$, their magnitudes will depend on $\theta$ regardless of whether $\theta$ contributes to the incompatibility in $\boldsymbol{W}$, through the temperature dependence of the crystal lattice spacing.

\subsection{Conservation of Burgers vector}\label{sec:conservationofB}

Let the thermomechanical boundary conditions evolve with time, resulting in the motion of the thermomechanical line defect. Then, the conservation of the Burgers vector of this line defect can be written as follows (see \cite{acharya_microcanonical_2011}, Appendix B for a detailed derivation):
\begin{equation}\label{eqn:conservationofb}
    \begin{aligned}
        \frac{\text{d}}{\text{d}t} \boldsymbol{b}_r &= - \frac{\text{d}}{\text{d}t} \oint_{c_t} \boldsymbol{W}  \textbf{d}\boldsymbol{x}  = - \oint_{c_t} \left( \boldsymbol{\alpha} \times \boldsymbol{v}^d + \boldsymbol{\phi}^\theta\right)  \textbf{d}\boldsymbol{x} \\
        & =  \frac{\text{d}}{\text{d}t} \int_{s_t} \boldsymbol{\alpha}  \hat{\boldsymbol{n}} \text{ d}s = - \int_{s_t} \nabla\times\left( \boldsymbol{\alpha} \times \boldsymbol{v}^d + \boldsymbol{\phi}^\theta\right)\hat{\boldsymbol{n}} \text{ d}s,\end{aligned}
\end{equation}
where $\boldsymbol{v}^d$ is the velocity \textit{relative to the material} of the thermomechanical defect, whose expression is constitutively prescribed guided by the condition of non-negativeness of the global thermomechanical dissipation (see \cref{subsec:helmholtz_constitutive_relations}). $\boldsymbol{\phi}^\theta$ represents a crucial source term that arises from the transient heterogeneous $\theta$. The argument to support this comes from the idea proposed in the work of \mbox{\cite{kroner_kontinuumstheorie_1958}} and substantiated in \mbox{\cite{upadhyay_thermomechanical_2020}}. In the presence of large transient temperature gradients, the thermomechanical defect evolution can have a contribution from the incompatibility in thermal strains induced by the evolving heterogeneous temperature field, and that contribution is accounted for in $\boldsymbol{\phi}^\theta$. Alternatively, such a contribution leads to the additive decomposition of the velocity gradient into thermal and plastic parts (\cref{eqn:velgradadddecomp}), resulting, in the small deformation theory, in the familiar expression for the elastic strain rate affecting the stress rate in classical thermoelasticity. From \cref{eqn:conservationofb} we obtain the following evolution statement of $\boldsymbol{\alpha}$:
\begin{equation}
    \accentset{\circ}{\boldsymbol{\alpha}} = - \nabla\times\left( \boldsymbol{\alpha} \times \boldsymbol{v}^d + \boldsymbol{\phi}^\theta \right),
    \label{eqn:transport}
\end{equation}
where $\accentset{\circ}{\boldsymbol{\alpha}} := \text{tr}(\boldsymbol{L}) \boldsymbol{\alpha} + \dot{\boldsymbol{\alpha}} - \boldsymbol{\alpha}\boldsymbol{L}^T$, is the convected derivative of $\boldsymbol{\alpha}$ ({\cite{ACHARYA2001761}}), $\boldsymbol{L} = \nabla \boldsymbol{v}$ is the velocity gradient, and $\boldsymbol{v}$ is the material velocity in $\Omega_t$. In \cref{eqn:conservationofb,eqn:transport}, the term $(\boldsymbol{\alpha}\times\boldsymbol{v}^d)$ represents the flux of Burgers vector carried by thermomechanical defect lines across the curve $c_t$ with a velocity $\boldsymbol{v}^d$ (\cite{acharya_microcanonical_2011}), and $\nabla\times\boldsymbol{\phi}^\theta$ represents a source of elastic incompatibility (areal density) due to the transient heterogeneous $\theta$ in $s_t$.

In the small deformation case (\cite{upadhyay_thermomechanical_2020}), it was shown that $\boldsymbol{\phi}^\theta$ is directly related to the evolution of thermal strains as $\boldsymbol{\phi}^\theta = \partial_t\boldsymbol{\varepsilon}^\theta = \partial_t\big[\boldsymbol{\gamma}(\theta - \theta_0)\big]$, where $\partial_t$ denotes the partial derivative with respect to time, $\boldsymbol{\varepsilon}^\theta$ is the thermal strain, $\boldsymbol{\gamma}$ is a positive-definite second-order tensor of thermal expansion coefficients, and $\theta_0$ is a reference temperature value. In our work, we introduce the contribution of the incompatibility in thermal strains to the evolution of $\boldsymbol{\alpha}$ through the flux term $\nabla\times\boldsymbol{\boldsymbol{\phi}}^\theta$ and define 
\begin{equation}
    \boldsymbol{\phi}^\theta = \boldsymbol{Y}\dot\theta
    \label{phitheta_Ythetadot}
\end{equation}
where $\boldsymbol{Y}$ is a two-point second-order tensor of coefficients of thermal expansion that transforms vectors from the current to the relaxed lattice state at any given point.

To comply with frame invariance requirements (see \ref{sec:invariance_Y}), the simplest choice for $\boldsymbol{Y}$ is 
\begin{equation}
    \boldsymbol{Y} = \boldsymbol{W \gamma}
\end{equation}
where $\boldsymbol{\gamma}$ is the second-order tensor of thermal expansion coefficients defined in $\Omega_t$, thus being a measurable quantity on the current configuration. Frame invariance requires that, under a rigid body motion, the thermal expansion tensor transform as $\boldsymbol{\gamma}^* = \boldsymbol{Q\gamma Q}^T$ (see \ref{sec:invariance_Y}), where $\boldsymbol{\gamma}^*$ is the rotated tensor, and $\boldsymbol{Q}$ is a proper rotation tensor. Complying with this requirement, we define 
\begin{equation}
    \boldsymbol{\gamma} = \sum_{i=1}^3 d_i(\theta) \,\boldsymbol{l}_i \otimes \boldsymbol{l}_i
\end{equation}
where $d_i(\theta)$ are the thermal expansion coefficients along the directions $\boldsymbol{l}_i$, which correspond to the eigenvectors of the left Cauchy-Green deformation tensor $\boldsymbol{B}^e = \boldsymbol{F}^e\boldsymbol{F}^{eT}$. In the thermally isotropic case, we have that $d_i(\theta) = d(\theta)$, such that $\boldsymbol{Y} = d(\theta)\boldsymbol{W}$. 

\textcolor{black}{Based on the aforementioned arguments, at minimum we have $\boldsymbol{Y} \equiv \boldsymbol{Y}\left( \theta, \boldsymbol{l}, \boldsymbol{W} \right)$.} \textcolor{black}{Therefore, even though the time derivative of $\boldsymbol{Y}$ is not entering into \cref{phitheta_Ythetadot}, $\boldsymbol{Y}$ nevertheless evolves in time through its dependence on $\theta$ and $\boldsymbol{l}$; the contribution of $\dot{\boldsymbol{Y}}$ to $\boldsymbol{\phi}^\theta$ is expected to be negligible in comparison to $\dot{\theta}$; this can be deduced from the expression of coefficient of thermal expansion as a function of temperature shown for the case of a stainless steel in \cite{mohananIntergranularStressPlastic2024}. And yet, as shall be shown in section \ref{sec:temperature_evolution_LD}, this form of $\boldsymbol{\phi}^\theta$ accounts for the contribution of the temporal evolution of coefficient of thermal expansion tensor to temperature evolution, which may be non-negligible in the form in which it enters that equation.}

The thermal flux from \cref{phitheta_Ythetadot} can be rewritten as 
\begin{equation}
    \boldsymbol{\phi}^\theta = \boldsymbol{W\gamma}\dot\theta
\end{equation}

\cref{eqn:conservationofb} also imposes a specific expression for the evolution of $\boldsymbol{W}$ up to a gradient term. This gradient term is ignored as a physically motivated constitutive choice (\cite{acharya2015dislocation}) for which plastic strain rate arises only where dislocations are present and a transient temperature gradient generates incompatible thermal strains. The evolution of $\boldsymbol{W}$ then is
\begin{equation}\label{eqn:Wrate}
    \begin{aligned}
        \dot{\boldsymbol{W}} + \boldsymbol{W}\boldsymbol{L} = \boldsymbol{\alpha} \times \boldsymbol{v}^d + \boldsymbol{W\gamma}\dot\theta.
    \end{aligned}
\end{equation}
By left-multiplying by $\boldsymbol{W}^{-1}$, we obtain the following additive decomposition of $\boldsymbol{L}$ as a natural consequence of the statement of conservation of topological charge \cref{eqn:conservationofb}: 
\begin{equation}\label{eqn:velgradadddecomp}
    \begin{aligned}
        \boldsymbol{L} = {\underbrace{\vphantom{\boldsymbol{W}^{-1}\left(\boldsymbol{\alpha} \times \boldsymbol{v}^d + \boldsymbol{S}\right)}-\boldsymbol{W}^{-1}\dot{\boldsymbol{W}}}_{\mbox{\normalsize\(\boldsymbol{L}^e\)}}} + 
        {\underbrace{\boldsymbol{W}^{-1}\left(\boldsymbol{\alpha} \times \boldsymbol{v}^d\right)}_{\mbox{\normalsize\(\boldsymbol{L}^p\)}}} + {\underbrace{\vphantom{\boldsymbol{W}^{-1}\left(\boldsymbol{\alpha} \times \boldsymbol{v}^d + \boldsymbol{S}\right)}\boldsymbol{\gamma}\dot\theta}_{\mbox{\normalsize\(\boldsymbol{L}^\theta\)}}}.
    \end{aligned}
\end{equation}
where $\boldsymbol{L}^e$, $\boldsymbol{L}^p$ and $\boldsymbol{L}^\theta$ correspond to the elastic, plastic and thermal distortion rates, respectively. Crucially, we have arrived at this result without the need to assume a multiplicative decomposition of the deformation gradient. We note here that, just like $\boldsymbol{\phi}^\theta$, other mechanisms of inelastic strain rate, such as arising from phase transformations and twinning, can be incorporated as an additive term $\boldsymbol{S}^i$ inside the curl in \cref{eqn:transport} as well as in \mbox{\cref{eqn:Wrate}}  leading to 
\begin{equation}\label{eqn:velgradadddecomp_sources}
    \begin{aligned}
    \boldsymbol{L} = {\underbrace{\vphantom{\boldsymbol{W}^{-1}\left(\boldsymbol{\alpha} \times \boldsymbol{v}^d + \boldsymbol{S}\right)}-\boldsymbol{W}^{-1}\dot{\boldsymbol{W}}}_{\mbox{\normalsize\(\boldsymbol{L}^e\)}}} + 
        {\underbrace{\boldsymbol{W}^{-1}\left(\boldsymbol{\alpha} \times \boldsymbol{v}^d\right)}_{\mbox{\normalsize\(\boldsymbol{L}^p\)}}} + {\underbrace{\vphantom{\boldsymbol{W}^{-1}\left(\boldsymbol{\alpha} \times \boldsymbol{v}^d + \boldsymbol{S}\right)}\boldsymbol{\gamma}\dot\theta}_{\mbox{\normalsize\(\boldsymbol{L}^\theta\)}}} +
        {\underbrace{\vphantom{\boldsymbol{W}^{-1}\left(\boldsymbol{\alpha} \times \boldsymbol{v}^d + \boldsymbol{S}\right)}\boldsymbol{W}^{-1}\boldsymbol{S}^i}_{\mbox{\normalsize\(\boldsymbol{L}^i\)}}}.
    \end{aligned}
\end{equation}
provided $\boldsymbol{S}^i$ is a measurable quantity in the current configuration, with $\boldsymbol{L}^i$ being the corresponding inelastic strain rate. In the event that $\boldsymbol{S}^i$ has a non-zero curl, then it would appear in \cref{eqn:transport} as an additive source term with $\boldsymbol{\phi}^\theta$. \textcolor{black}{The equivalence between the first term and the classical form of the elastic distortion rate can be seen by considering
\begin{equation}
    -\boldsymbol{W}^{-1}\dot{\boldsymbol{W}} = - \boldsymbol{F}^e\dot{\overline{\boldsymbol{F}^{e-1}}} = \boldsymbol{F}^e\boldsymbol{F}^{e-1}\boldsymbol{L}^e = \boldsymbol{L}^e,
\end{equation}
where the second equality can be shown by an argument similar to \cref{eqn:F_inv_dot} with $\boldsymbol{L}^e = \dot{\boldsymbol{F}}^e\boldsymbol{F}^{e-1}$.}

\textcolor{black}{As described in the introduction, this theory deals with individually characterizable dislocations. An extension to account for dislocation ensembles (both geometrically necessary and statistical) can be done by performing a space-time running average as outlined in the works of \cite{acharya_size_2006, arora_dislocation_2020}. Then, the only changes to the form of the equations will occur in \cref{eqn:transport} and \cref{eqn:velgradadddecomp_sources} through the introduction of a source term akin to $\boldsymbol{\phi}^\theta$ and $\boldsymbol{S}^i$ that accounts for statistical dislocations; in this case, $\boldsymbol{\alpha}$ will represent the polar density of the ensemble of thermomechanical dislocations and $\boldsymbol{v}^d$ their velocity.}

\textcolor{black}{As shown in \cite{acharyaJumpConditionGND2007} in the isothermal mechanics context, the conservation of Burgers vector \cref{eqn:conservationofb} is equipped to impose continuity conditions across grain boundary interfaces on the dislocation evolution that also involves $\boldsymbol{\phi}^\theta$ and $\boldsymbol{S}^i$. Therefore, the framework is directly applicable to single- or poly-crystalline single-phase materials. For treating multi-phase materials, one could straightforwardly introduce a phenomenological distortion rate through the term $\boldsymbol{S}^i$ that is associated with phase transformations, e.g., similar to \cite{thamburajaPolycrystallineShapememoryMaterials2001}. For a more fundamental (bottom up) approach, the concept of g-disclination dynamics (\cite{acharyaCoupledPhaseTransformations2012,hirth2021straight,hirth2024role}) can be utilized at the level of microscopic defects and whose homogenization can provide guidance to obtain the form of $\boldsymbol{S}^i$, similar to the homogenization of isothermal FDM giving guidance for $\boldsymbol{L}^p$ (\cite{arora_dislocation_2020, chatterjee2020plasticity}).}

\section{Balance laws, dissipation analysis and constitutive equations}\label{sec:balance-constitutive}

\subsection{Mass balance}

If $\rho$ is a space and time-dependent mass density field, then the conservation of mass statement is
\begin{equation}
    \begin{aligned}
        \frac{\text{d}}{\text{d}t} \int_{\Omega_t} \rho \ \text{d}v = 0 \quad\Rightarrow\quad \dot{\rho} + \nabla\cdot \left( \rho \boldsymbol{v} \right) = 0.
    \end{aligned}
\end{equation}

\subsection{Balance of linear and angular momentum}

The balance of linear momentum reads
\begin{equation}\label{eqn:dynamics}
    \begin{aligned}
        \nabla\cdot\boldsymbol{\sigma}  + \rho \boldsymbol{b}_f = \rho \dot{\boldsymbol{v}},
    \end{aligned}
\end{equation}
where $\boldsymbol{\sigma}$ is the Cauchy stress tensor and $\boldsymbol{b}_f$ is the body force density per unit mass.

Angular momentum balance implies that the Cauchy stress tensor is symmetric i.e.,
\begin{equation}\label{eqn:angularmomentum}
    \begin{aligned}
        \boldsymbol{\sigma} = \boldsymbol{\sigma}^T.
    \end{aligned}
\end{equation}

\subsection{First law of thermodynamics}

The first law of thermodynamics for the continuum thermomechanical problem is written as
\begin{equation}
    \begin{aligned}
        \frac{\text{d}}{\text{d}t} \left(\int_{\Omega_t} \rho {\varepsilon} \text{ d}v + \frac{1}{2} \int_{\Omega_t} \rho \boldsymbol{v} \cdot \boldsymbol{v} \text{ d}v \right) = - \int_{\partial \Omega_t} \boldsymbol{q} \cdot \hat{\boldsymbol{n}} \text{ d}s + \int_{\partial \Omega_t} \boldsymbol{t} \cdot {\boldsymbol{v}} \text{ d}s + \int_{\Omega_t} \rho r \text{ d}v + \int_{\Omega_t} \rho \boldsymbol{b}_f \cdot \boldsymbol{v} \text{ d}v,
    \end{aligned}
    \label{eqn:firstlaw_boundary_ints}
\end{equation}
where $\varepsilon$ is the internal energy density, $\boldsymbol{q}$ is the heat flux vector, $\boldsymbol{t}$ is the traction vector and $\rho r$ are internal heat sources.

Considering Cauchy's theorem, we have that $\boldsymbol{t} = \boldsymbol{\sigma}\hat{\boldsymbol{n}}$, such that using \cref{eqn:dynamics} in \cref{eqn:firstlaw_boundary_ints} and rearranging terms gives
\begin{equation}\label{eqn:firstlaw}
    \begin{aligned}
        \int_{\Omega_t}\rho \dot{\varepsilon} \ \text{d}v = - \int_{\Omega_t} \nabla\cdot\boldsymbol{q}\ \text{d}v + \int_{\Omega_t} \boldsymbol{\sigma} : \boldsymbol{L} \ \text{d}v + \int_{\Omega_t} \rho r \ \text{d}v.
    \end{aligned}
\end{equation}
In the local form, it is written as
\begin{equation}\label{eqn:localfirstlaw}
    \begin{aligned}
        \rho \dot{\varepsilon} = -  \nabla\cdot \boldsymbol{q} + \boldsymbol{\sigma} : \boldsymbol{L} + \rho r.
    \end{aligned}
\end{equation}

\subsection{Second law of thermodynamics}

We consider the second law of thermodynamics for a continuum body as
\begin{equation}\label{eqn:secondlaw_integral}
    \begin{aligned}
        \frac{\text{d}}{\text{d}t} \int_{\Omega_t} \rho \eta \ \text{d}v \geq - \int_{\partial\Omega_t} \frac{\boldsymbol{q}}{\theta} \cdot \hat{\boldsymbol{n}} \text{ d}s + \int_{\Omega_t} \frac{\rho r}{\theta} \text{ d}v,
    \end{aligned}
\end{equation}

\noindent where $\eta$ is the entropy density of the body. Eliminating $\rho r$ from \cref{eqn:secondlaw_integral} by using \cref{eqn:localfirstlaw}, and using the divergence theorem on the boundary term, we get
\begin{equation}\label{eqn:secondlaw_with_eps}
    \begin{aligned}
        \int_{\Omega_t} \rho (\theta\dot{\eta} - \dot\varepsilon) \ \text{d}v 
        - \int_{\Omega_t} \frac{1}{\theta}\boldsymbol{q}\cdot\nabla\theta \,\text{d}v
        + \int_{\Omega_t} \boldsymbol{\sigma} : \boldsymbol{L} \,\text{d}v
        \geq 0.
    \end{aligned}
\end{equation}

Then, we define the Helmholtz free energy density $\Psi$ as 
\begin{equation}
    \Psi = \varepsilon - \eta\theta \implies \dot\varepsilon = \dot\Psi + \dot\eta\theta + \eta\dot\theta.
    \label{eqn:legendre_transform&rate}
\end{equation}

Inserting \cref{eqn:legendre_transform&rate} in \cref{eqn:secondlaw_with_eps} gives the global dissipation inequality 
\begin{equation}\label{eqn:localsecondlaw2}
    \begin{aligned}
        D := \int_{\Omega_t} \left[ - \rho \left( \dot\Psi + \eta \dot\theta \right) - \frac{1}{\theta} \boldsymbol{q} \cdot \nabla \theta + \boldsymbol{\sigma}:\boldsymbol{L} \right] \text{d}v \geq 0.
    \end{aligned}
\end{equation}  
We use this form of the second law to provide guidance on the possible constitutive assumptions that guarantee non-negative dissipation.

\subsection{Helmholtz free energy density and constitutive relations}\label{subsec:helmholtz_constitutive_relations}

Let us assume that the Helmholtz free energy is a function of the (measurable) state variables $\left( \boldsymbol{W}, \theta, \boldsymbol{\alpha} \right)$ such that $\Psi \equiv \Psi\left( \boldsymbol{W}, \theta, \boldsymbol{\alpha} \right)$. Then
\begin{equation}\label{eqn:psirate}
    \begin{aligned}
        \dot\Psi = \partial_{\boldsymbol{W}} \Psi: \dot{\boldsymbol{W}} + \partial_{\theta}\Psi\, \dot{\theta} + \partial_{\boldsymbol{\alpha}}\Psi: \dot{\boldsymbol{\alpha}}.
    \end{aligned}
\end{equation}

\textcolor{black}{Note that in the upscaled (homogenized) version of the theory, one could also introduce other measurable state variables into the free energy/entropy such as the statistical dislocation density (see for example \cite[Sec.~6]{acharya_microcanonical_2011})}.

Next, substitute \cref{eqn:Wrate} for $\dot{\boldsymbol{W}}$ and plug \cref{eqn:psirate} into \cref{eqn:localsecondlaw2} to arrive at
\begin{equation}
    \begin{aligned}
        D = &\int_{\Omega_t} \left(\boldsymbol{\sigma} : \boldsymbol{L} - \frac{1}{\theta}\boldsymbol{q}\cdot\nabla\theta\right)\, \text{d}v + \int_{\Omega_t} \rho\boldsymbol{W}^T\partial_{\boldsymbol{W}}\Psi : \boldsymbol{L}\, \text{d}v - \int_{\Omega_t} \rho\partial_{\boldsymbol{W}}\Psi : \left(\boldsymbol{\alpha}\times\boldsymbol{v}^d + \boldsymbol{W\gamma}\dot\theta\right)\, \text{d}v \\
        &- \int_{\Omega_t} \rho\partial_{\boldsymbol{\alpha}}\Psi : \left[-\text{tr}(\boldsymbol{L})\boldsymbol{\alpha} + \boldsymbol{\alpha L}^T - \nabla\times\left( \boldsymbol{\alpha} \times \boldsymbol{v}^d + \boldsymbol{W\gamma}\dot\theta \right)\right]\, \text{d}v - \int_{\Omega_t}\rho \big(\partial_\theta\Psi + \eta \big)\dot\theta\, \text{d}v \geq 0,
    \end{aligned}
    \label{eqn:D_after_replacing_dotf}
\end{equation}
where \cref{eqn:transport} was used for $\dot{\boldsymbol{\alpha}}$. After rearranging the terms, we get (see \ref{app:derivation_dissipation})

\begin{equation}\label{eqn:dissipation}
    \begin{aligned}
        D = &\int_{\Omega_t}  \left[ \boldsymbol{\sigma} +\text{sym}( \rho\boldsymbol{W}^T \partial_{\boldsymbol{W}}\Psi -  \rho(\partial_{\boldsymbol{\alpha}}\Psi)^T\boldsymbol{\alpha} + \rho(\partial_{\boldsymbol{\alpha}}\Psi: \boldsymbol{\alpha}) \mathds{1})\right]:\text{sym}(\boldsymbol{L}) \ \text{d}v\\
        &+\int_{\Omega_t}  \text{skew}( \rho\boldsymbol{W}^T \partial_{\boldsymbol{W}}\Psi -  \rho(\partial_{\boldsymbol{\alpha}}\Psi)^T\boldsymbol{\alpha} + \rho(\partial_{\boldsymbol{\alpha}}\Psi: \boldsymbol{\alpha}) \mathds{1}):\text{skew}(\boldsymbol{L}) \ \text{d}v\\
        &- \int_{\Omega_t} \bigg[\rho\eta + \rho \partial_\theta\Psi + \big( \rho\partial_{\boldsymbol{W}}\Psi - \nabla\times\rho \partial_{\boldsymbol{\alpha}}\Psi \big) :  \boldsymbol{W\gamma}  \bigg] \dot\theta  \ \text{d}v \\
        & - \int_{\Omega_t}  \left\{ \left[ \left( \rho\partial_{\boldsymbol{W}}\Psi  - \nabla\times \rho\partial_{\boldsymbol{\alpha}}\Psi   \right)^T\boldsymbol{\alpha} \right] : \textbf{X} \right\} \cdot \boldsymbol{v}^d\ \text{d}v \\
        & - \int_{\Omega_t} \frac{1}{\theta} \boldsymbol{q} \cdot \nabla \theta \ \text{d}v
        \\ 
        &- \int_{\partial \Omega_t} \rho \partial_{\boldsymbol{\alpha}}\Psi : \left[ \left( \boldsymbol{\alpha} \times \boldsymbol{v}^d + \boldsymbol{W\gamma} \dot{\theta}\right) \times \hat{\boldsymbol{n}} \right] \ \text{d}s \geq 0.
    \end{aligned}
\end{equation}
where we have considered the balance of angular momentum (\cref{eqn:angularmomentum}). Since $\text{skew}(\boldsymbol{L})$ is related to a rigid rotation of the body, the second integral in \cref{eqn:dissipation} would indicate a dissipation associated with this rigid motion. Hence, the objectivity of dissipation requires this integral to vanish, which is shown to be the case in \ref{sec:ericksen_equality} as a stringent test of the statements of kinematic evolution of the theory. Thus, the second term in \mbox{\cref{eqn:dissipation}} vanishes.

Now, consider a motion where dislocations do not move relative to the material, i.e., $\boldsymbol{v}^d = 0$, and $\theta$ remains uniform and constant in $\Omega_t$. Such a process should not result in any dissipation, and this is only possible if the following constitutive relationship is assumed for the Cauchy stress:
\begin{equation}    
    \boldsymbol{\sigma} = -\rho\boldsymbol{W}^T\partial_{\boldsymbol{W}}\Psi + \rho(\partial_{\boldsymbol{\alpha}}\Psi)^T\boldsymbol{\alpha} - \rho(\partial_{\boldsymbol{\alpha}}\Psi: \boldsymbol{\alpha}) \mathds{1} 
    \label{eqn:elasticlaw}
\end{equation}
\noindent\cref{eqn:elasticlaw} is a nonclassical form of a hyperelastic law (\cite{chaboche_cyclic_1993}), with $-\rho\boldsymbol{W}^T\partial_{\boldsymbol{W}}\Psi$ the hyperelastic part of $\boldsymbol{\sigma}$, and the two last terms being a direct consequence of the geometrical non-linearity and the chosen dependence of $\Psi$ on $\boldsymbol{\alpha}$. Indeed, considering the expression for $\accentset{\circ}{\boldsymbol{\alpha}}$ in \cref{eqn:transport}, a reversible process (e.g., a quasi-static elastic loading with negligible temperature changes) would produce an instantaneous change in $\boldsymbol{\alpha}$ through the terms in $\boldsymbol{L}$ (which are present only in the finite strain case and required for the Burgers vector conservation), as also shown in \cite{ARORA2020_FEM_FDM}, thus requiring the presence of the last two terms in $\boldsymbol{\sigma}$ to ensure no dissipation during such process. The validity of \cref{eqn:elasticlaw} as a constitutive statement for $\boldsymbol{\sigma}$ with respect to the balance of angular momentum is discussed in \ref{sec:ericksen_equality}. With the expression in \cref{eqn:elasticlaw} for the Cauchy stress, the first term in \cref{eqn:dissipation} vanishes. 

Next, we consider the case where no mechanical loading is applied, the dislocations are not moving, and the body temperature is uniform but can undergo homogeneous heating/cooling. In this situation, the only remaining term in the dissipation inequality \cref{eqn:dissipation} is the third term involving entropy $\eta$. Since $\dot{\theta}$ can be arbitrary in this case, the term in square brackets is set to zero to ensure non-negative dissipation, which yields
\begin{equation}\label{eqn:entropy}
    \begin{aligned}
        \eta = - \partial_\theta\Psi - \left( \partial_{\boldsymbol{W}}\Psi - \frac{1}{\rho }\nabla\times\rho\partial_{\boldsymbol{\alpha}}\Psi \right) :  \boldsymbol{W\gamma}.
    \end{aligned}
\end{equation}

Note that the second term in \cref{eqn:entropy} is a direct consequence of considering the contribution of the incompatibility induced by the transient heterogeneous $\theta$ as a source term in the thermomechanical defect density evolution (\cref{eqn:transport}). With these relationships, the dissipation inequality is reduced to 
\begin{equation}\label{eqn:dissipation2}
    \begin{aligned}
        D = & \int_{\Omega_t}  \left\{ \left[ \left( -\rho\partial_{\boldsymbol{W}}\Psi  + \nabla\times \rho\partial_{\boldsymbol{\alpha}}\Psi    \right)^T\boldsymbol{\alpha} \right] : \textbf{X} \right\} \cdot \boldsymbol{v}^d\ \text{d}v 
        - \int_{\Omega_t} \frac{1}{\theta} \boldsymbol{q} \cdot \nabla \theta \ \text{d}v\\
        &- \int_{\partial \Omega_t} \rho \partial_{\boldsymbol{\alpha}}\Psi : \left[ \left( \boldsymbol{\alpha} \times \boldsymbol{v}^d + \boldsymbol{W\gamma} \dot{\theta} \right) \times \hat{\boldsymbol{n}} \right] \ \text{d}s
        \geq 0.
    \end{aligned}
\end{equation}

To ensure the non-negativeness of the heat flux term, we consider the generalized Fourier's law of heat conduction
\begin{equation}
    \begin{aligned}
        \boldsymbol{q} := -\boldsymbol{K}\nabla \theta,
    \end{aligned}
    \label{eqn:fourier_law}
\end{equation}

\noindent where $\boldsymbol{K}$ is the positive definite second-order thermal heat conductivity tensor which, in the general case, could be $\theta$ and/or $\boldsymbol{W}$-dependent. We assume that the driving force $\boldsymbol{f}_v$ for the dislocation velocity takes the form 
\begin{equation}
    \boldsymbol{f}_v = \left[ \left( -\rho\partial_{\boldsymbol{W}}\Psi  + \nabla\times \rho\partial_{\boldsymbol{\alpha}}\Psi\right)^T\boldsymbol{\alpha} \right] : \textbf{X}.
    \label{eqn:drive_force_disl_vel}
\end{equation}
In the specific case of a single dislocation line and neglecting the dependence of $\Psi$ in $\boldsymbol{\alpha}$, \cref{eqn:drive_force_disl_vel} reduces to the form of the well-known Peach-Koehler force acting on the dislocation line (\cite{peach_forces_1950}).

Note that, in \cref{eqn:dissipation2},
we neglect the contribution of the boundary term to the enforcement of non-negative entropy production in the body. With these considerations, the global dissipation of the model is written as
\begin{equation}
    D = \int_{\Omega_t}  \boldsymbol{f}_v \cdot \boldsymbol{v}^d\ \text{d}v 
    - \int_{\Omega_t} \frac{1}{\theta} \boldsymbol{q} \cdot \nabla \theta \ \text{d}v. 
\end{equation}

This enables us to consider a simple kinetic assumption on the dislocation velocity expression that ensures the non-negativeness of $D$ such as 
\begin{equation}
    \boldsymbol{v}^d = \frac{1}{B}\boldsymbol{f}_v,\quad B > 0,
\end{equation}
where $B$ is a material parameter corresponding to the dislocation drag coefficient.

\subsection{Temperature evolution}\label{sec:temperature_evolution_LD}

Inserting the rate form of \cref{eqn:legendre_transform&rate} into \cref{eqn:localfirstlaw} for $\dot\varepsilon$, and using \cref{eqn:psirate,eqn:entropy} for $\dot\Psi$, and $\eta$, respectively, gives
\begin{equation}
    \begin{split}
        \rho\bigg[\partial_{\boldsymbol{W}}\Psi:\dot{\boldsymbol{W}}
        + \partial_{\boldsymbol{\alpha}}\Psi:\dot{\boldsymbol{\alpha}}
        - \dot\theta\left( \partial_{        \boldsymbol{W}}\Psi - \frac{1}{\rho }\nabla\times\rho\partial_{\boldsymbol{\alpha}}\Psi \right) :  \boldsymbol{W\gamma} + \theta\dot\eta\bigg] 
        = -\nabla\cdot \boldsymbol{q} + \boldsymbol{\sigma} : \boldsymbol{L} + \rho r.
    \end{split}
    \label{eqn:theta_derivation1}
\end{equation}
Next, taking the material time derivative of \cref{eqn:entropy}, we have 
\begin{equation}
    \begin{split}
        \dot\eta &= 
        -\dot{\overline{\partial_{\theta}\Psi}} 
        - \dot{\overline{\left(\partial_{\boldsymbol{W}}\Psi - \frac{1}{\rho }\nabla\times\rho \partial_{\boldsymbol{\alpha}}\Psi \right) :  \boldsymbol{W\gamma}}} \\
        &= -\partial^2_{\theta\boldsymbol{W}}\Psi:\dot{\boldsymbol{W}} - \partial^2_{\theta\theta}\Psi\dot\theta -\partial^2_{\theta\boldsymbol{\alpha}}\Psi:\dot{\boldsymbol{\alpha}} 
        -\Bigg[\partial^2_{\boldsymbol{W}\boldsymbol{W}}\Psi:\dot{\boldsymbol{W}} + \partial^2_{\boldsymbol{W}\theta}\Psi\dot\theta + \partial^2_{\boldsymbol{W}\boldsymbol{\alpha}}\Psi:\dot{\boldsymbol{\alpha}} \\
        &\qquad\qquad\qquad + \frac{\dot\rho}{\rho^2} \nabla\times\rho\partial_{\boldsymbol{\alpha}}\Psi -\frac{1}{\rho}\dot{\overline{\nabla\times\left(\rho\partial_{\boldsymbol{\alpha}}\Psi \right)}}\Bigg] : \boldsymbol{W\gamma } - \left(\partial_{\boldsymbol{W}}\Psi - \frac{1}{\rho }\nabla\times\rho \partial_{\boldsymbol{\alpha}}\Psi \right) :  \dot{\overline{\boldsymbol{W}\boldsymbol{\gamma}}}.
    \end{split}
    \label{eqn:eta_dot}
\end{equation}
Consider a tensor field $\boldsymbol{A}(\boldsymbol{x}, t)$, and an arbitrary reference configuration $\Omega_0$ (e.g. the as-received body), with $\boldsymbol{x}_0 \in \Omega_0$ denoting the position vector in $\Omega_0$. Then, we have 
\begin{equation}
    \frac{\dot{\overline{\partial \boldsymbol{A}}}}{\partial \boldsymbol{x}} = \dot{\overline{\frac{\partial\boldsymbol{A}}{\partial\boldsymbol{x}_0}\frac{\partial\boldsymbol{x}_0}{\partial\boldsymbol{x}}}} = \frac{\partial\dot{\boldsymbol{A}}}{\partial\boldsymbol{x}_0}\boldsymbol{F}^{-1} + \frac{\partial\boldsymbol{A}}{\partial\boldsymbol{x}_0}\dot{\overline{\boldsymbol{F}^{-1}}}.
    \label{eqn:mat_time_der_def_1}
\end{equation}
with $\boldsymbol{F}$ a two-point tensor between $\Omega_0$ and $\Omega_t$. Now,
\begin{equation}
    \boldsymbol{F}\boldsymbol{F}^{-1} = \mathds{1} \implies \boldsymbol{F}\dot{\overline{\boldsymbol{F}^{-1}}} = -\dot{\boldsymbol{F}}\boldsymbol{F}^{-1} \implies \dot{\overline{\boldsymbol{F}^{-1}}} = -\boldsymbol{F}^{-1}\boldsymbol{L},
    \label{eqn:F_inv_dot}
\end{equation}
such that, by inserting into \cref{eqn:mat_time_der_def_1}, we have
\begin{equation}
    \frac{\dot{\overline{\partial \boldsymbol{A}}}}{\partial \boldsymbol{x}} = \frac{\partial\dot{\boldsymbol{A}}}{\partial\boldsymbol{x}} - \frac{\partial\boldsymbol{A}}{\partial\boldsymbol{x}}\boldsymbol{L}.
    \label{eqn:mat_time_der_def_final}
\end{equation}
regardless of the choice of $\Omega_0$. With that, we can write
\begin{equation}
    \begin{split}
        \big[\dot{\overline{\nabla\times\boldsymbol{A}}}\big]_{ij} &= \epsilon_{jkl}\dot{\overline{A_{il,k}}} = \epsilon_{jkl}(\dot A_{il,k} - A_{il,m} L_{mk}) = \epsilon_{jkl}\dot A_{il,k} + A_{il,m} L_{mk}\epsilon_{jlk}\\
        \implies \dot{\overline{\nabla\times\boldsymbol{A}}} &= \nabla\times\dot{\boldsymbol{A}} + \left[(\nabla\boldsymbol{A})\boldsymbol{L}\right]\, :\,\mathbf{X},
    \end{split}
\end{equation}
such that the second term in the last line of \cref{eqn:eta_dot} becomes
\begin{equation}
    \dot{\overline{\nabla\times\left(\rho\partial_{\boldsymbol{\alpha}}\Psi \right)}} = \nabla\times\left(\dot\rho\partial_{\boldsymbol{\alpha}}\Psi + \rho\partial^2_{\boldsymbol{\alpha}\boldsymbol{W}}\Psi:\dot{\boldsymbol{W}}+ \rho\partial^2_{\boldsymbol{\alpha}\theta}\Psi\dot\theta +\rho\partial^2_{\boldsymbol{\alpha}\boldsymbol{\alpha}}\Psi:\dot{\boldsymbol{\alpha}}\right) + \left[\nabla(\rho\partial_{\boldsymbol{\alpha}}\Psi)\boldsymbol{L}\right]\, :\,\mathbf{X} 
\end{equation}
and \cref{eqn:eta_dot} can be written as
\begin{equation}
    \begin{split}
        \dot\eta  &= -\partial^2_{\theta\boldsymbol{W}}\Psi:\dot{\boldsymbol{W}} - \partial^2_{\theta\theta}\Psi\dot\theta -\partial^2_{\theta\boldsymbol{\alpha}}\Psi:\dot{\boldsymbol{\alpha}} 
        -\Bigg[\partial^2_{\boldsymbol{W}\boldsymbol{W}}\Psi:\dot{\boldsymbol{W}} + \partial^2_{\boldsymbol{W}\theta}\Psi\dot\theta + \partial^2_{\boldsymbol{W}\boldsymbol{\alpha}}\Psi:\dot{\boldsymbol{\alpha}} \\
        &\qquad + \frac{\dot\rho}{\rho^2} \nabla\times\rho\partial_{\boldsymbol{\alpha}}\Psi
        -\frac{1}{\rho}\nabla\times\left(\dot\rho\partial_{\boldsymbol{\alpha}}\Psi + \rho\partial^2_{\boldsymbol{\alpha}\boldsymbol{W}}\Psi:\dot{\boldsymbol{W}}+ \rho\partial^2_{\boldsymbol{\alpha}\theta}\Psi\dot\theta +\rho\partial^2_{\boldsymbol{\alpha}\boldsymbol{\alpha}}\Psi:\dot{\boldsymbol{\alpha}}\right)\\
        &\qquad - \frac{1}{\rho}\left(\nabla(\rho\partial_{\boldsymbol{\alpha}}\Psi)\boldsymbol{L}\right)\, :\,\mathbf{X}\Bigg] : \boldsymbol{\gamma W} - \left( \partial_{\boldsymbol{W}}\Psi - \frac{1}{\rho }\nabla\times\rho\partial_{\boldsymbol{\alpha}}\Psi \right) :  \dot{\overline{\boldsymbol{W}\boldsymbol{\gamma}}}.
    \end{split}
    \label{eqn:eta_dot2}
\end{equation}
Inserting \cref{eqn:eta_dot2} in \cref{eqn:theta_derivation1}, using \cref{eqn:elasticlaw} for $\boldsymbol{\sigma}$ and rearranging terms results in the following equation for the temperature evolution:
\begin{equation}
    \begin{split}
        \Big[\big( -\rho&\partial_{\boldsymbol{W}}\Psi + \nabla\times\rho\partial_{\boldsymbol{\alpha}}\Psi -\rho\theta\partial^2_{\boldsymbol{W}\theta}\Psi\big) :  \boldsymbol{W\gamma} -\rho\theta\partial^2_{\theta\theta}\Psi \Big]\dot\theta  + \theta\nabla\times\left(\rho\partial^2_{\boldsymbol{\alpha}\theta}\Psi\dot\theta\right) : \boldsymbol{W\gamma}\\
        = &-\nabla\cdot \boldsymbol{q} + \rho r -\rho\partial_{\boldsymbol{W}}\Psi:\left(\dot{\boldsymbol{W}} + \boldsymbol{WL}\right)
        - \rho\partial_{\boldsymbol{\alpha}}\Psi:\left(\dot{\boldsymbol{\alpha}} + \text{tr}(\boldsymbol{L})\boldsymbol{\alpha} - \boldsymbol{\alpha L}^T\right) + \rho\theta\partial^2_{\theta\boldsymbol{W}}\Psi:\dot{\boldsymbol{W}} + \rho\theta\partial^2_{\theta\boldsymbol{\alpha}}\Psi:\dot{\boldsymbol{\alpha}}\\
        &+ \rho\theta\Bigg[\partial^2_{\boldsymbol{W}\boldsymbol{W}}\Psi:\dot{\boldsymbol{W}} + \partial^2_{\boldsymbol{W}\boldsymbol{\alpha}}\Psi:\dot{\boldsymbol{\alpha}} + \frac{\dot\rho}{\rho^2} \nabla\times\rho\partial_{\boldsymbol{\alpha}}\Psi-\frac{1}{\rho}\nabla\times\big(\dot\rho\partial_{\boldsymbol{\alpha}}\Psi + \rho\partial^2_{\boldsymbol{\alpha}\boldsymbol{W}}\Psi:\dot{\boldsymbol{W}} \\
        &\quad+ \rho\partial^2_{\boldsymbol{\alpha}\boldsymbol{\alpha}}\Psi:\dot{\boldsymbol{\alpha}} \big) - \frac{1}{\rho}\left(\nabla(\rho\partial_{\boldsymbol{\alpha}}\Psi)\boldsymbol{L}\right)\, :\,\mathbf{X}\Bigg] : \boldsymbol{W\gamma} + \theta\left(\rho\partial_{\boldsymbol{W}}\Psi - \nabla\times\rho\partial_{\boldsymbol{\alpha}}\Psi \right) :  \left(\dot{\boldsymbol{W}}\boldsymbol{\gamma} + \boldsymbol{W}\mathds{G}:\dot{\boldsymbol{W}}\right),
    \end{split}
    \label{eqn:temperature_evolution}
\end{equation}
where we consider
\begin{equation}
    \dot{\boldsymbol{\gamma}} = \left(\frac{\partial\boldsymbol{\gamma}}{\partial \boldsymbol{B}^e}: \frac{\partial\boldsymbol{B}^e}{\partial \boldsymbol{W}} \right): \dot{\boldsymbol{W}} = \mathds{G}:\dot{\boldsymbol{W}}.
\end{equation}
Using \cref{eqn:Wrate,eqn:transport}  for $\dot{\boldsymbol{\alpha}}$ and $\dot{\boldsymbol{W}}$ gives
\begin{equation}
    \begin{split}
        \Bigg\{\bigg[ \nabla&\times\rho\partial_{\boldsymbol{\alpha}}\Psi(\mathds{1} + \theta\boldsymbol{\gamma}^T) -2\rho\theta\partial^2_{\boldsymbol{W}\theta}\Psi - \rho\theta\boldsymbol{W\gamma}:\partial^2_{\boldsymbol{WW}}\Psi - \rho\theta\partial_{\boldsymbol{W}}\Psi\boldsymbol{\gamma}^T - \theta(\rho\partial_{\boldsymbol{W}}\Psi - \nabla\times\rho\partial_{\boldsymbol{\alpha}}\Psi):\boldsymbol{W}\mathds{G}\bigg] :  \boldsymbol{W\gamma} \\
        &-\rho\theta\partial^2_{\theta\theta}\Psi \Bigg\}\dot\theta + \theta\nabla\times\left(\rho\partial^2_{\boldsymbol{\alpha}\theta}\Psi\dot\theta\right) : \boldsymbol{W\gamma} - \big[\rho\partial_{\boldsymbol{\alpha}}\Psi - \rho\theta\left(\partial^2_{\theta\boldsymbol{\alpha}}\Psi + \boldsymbol{W\gamma}:\partial^2_{\boldsymbol{W}\boldsymbol{\alpha}}\Psi\right)\big]:\nabla\times\left(\boldsymbol{W\gamma}\dot\theta\right) \\
        &\hspace{-0.5cm}+\theta\nabla\times\bigg[\rho\partial^2_{\boldsymbol{\alpha}\boldsymbol{W}}\Psi:\boldsymbol{W\gamma}\dot\theta -\rho\partial^2_{\boldsymbol{\alpha}\boldsymbol{\alpha}}\Psi:\nabla\times\left(\boldsymbol{W\gamma}\dot\theta\right) \bigg] : \boldsymbol{W\gamma}\\
        = &\\
        &- \nabla\cdot \boldsymbol{q} + \rho r +\bigg[-\rho\partial_{\boldsymbol{W}}\Psi\left(\mathds{1} - \theta\boldsymbol{\gamma}^T\right) +\theta\left(\rho\partial_{\boldsymbol{W}}\Psi - \nabla\times\rho\partial_{\boldsymbol{\alpha}}\Psi\right):\boldsymbol{W}\mathds{G} - \theta\nabla\times(\rho\partial_{\boldsymbol{\alpha}}\Psi)\boldsymbol{\gamma}^T \\
        &\quad+ \rho\theta\left(\partial^2_{\theta\boldsymbol{W}}\Psi + \boldsymbol{W\gamma}:\partial^2_{\boldsymbol{W}\boldsymbol{W}}\Psi\right)\bigg]:\left(\boldsymbol{\alpha}\times\boldsymbol{v}^d\right) + \Bigg\{- \rho\theta\boldsymbol{W}^T\left(\partial^2_{\theta\boldsymbol{W}}\Psi + \boldsymbol{W\gamma}:\partial^2_{\boldsymbol{W}\boldsymbol{W}}\Psi\right)
        \\
        &\quad-\left[\rho\theta\left(\partial^2_{\theta\boldsymbol{\alpha}}\Psi + \boldsymbol{W\gamma}:\partial^2_{\boldsymbol{W}\boldsymbol{\alpha}}\Psi\right):\boldsymbol{\alpha}\right]\mathds{1}  + \left[\rho\theta\left(\partial^2_{\theta\boldsymbol{\alpha}}\Psi + \boldsymbol{W\gamma}:\partial^2_{\boldsymbol{W}\boldsymbol{\alpha}}\Psi\right)^T\boldsymbol{\alpha}\right] \\
        &\quad+ \theta\left[-\rho(\boldsymbol{W}^T\partial_{\boldsymbol{W}}\Psi)\boldsymbol{\gamma}^T +\boldsymbol{W}^{T}\nabla\times(\rho\partial_{\boldsymbol{\alpha}}\Psi)\boldsymbol{\gamma}^T + \left(-\rho\boldsymbol{W}^T\partial_{\boldsymbol{W}}\Psi + \boldsymbol{W}^{T}\nabla\times\rho\partial_{\boldsymbol{\alpha}}\Psi \right):\boldsymbol{W}\mathds{G} \right]\Bigg\}:\boldsymbol{L}\\
        &+ \left[\rho\partial_{\boldsymbol{\alpha}}\Psi - \rho\theta\left(\partial^2_{\theta\boldsymbol{\alpha}}\Psi + \boldsymbol{W\gamma}:\partial^2_{\boldsymbol{W}\boldsymbol{\alpha}}\Psi\right)\right]:\nabla\times\left(\boldsymbol{\alpha}\times\boldsymbol{v}^d\right) \\
        &+ \Bigg\{ \frac{\theta\dot\rho}{\rho} \nabla\times\rho\partial_{\boldsymbol{\alpha}}\Psi-\theta\nabla\times\bigg[\dot\rho\partial_{\boldsymbol{\alpha}}\Psi + \rho\partial^2_{\boldsymbol{\alpha}\boldsymbol{W}}\Psi:\big(-\boldsymbol{WL}+\boldsymbol{\alpha}\times\boldsymbol{v}^d\big) \\
        &\quad+ \rho\partial^2_{\boldsymbol{\alpha}\boldsymbol{\alpha}}\Psi:\big(-\text{tr}(\boldsymbol{L})\boldsymbol{\alpha} + \boldsymbol{\alpha L}^T - \nabla\times(\boldsymbol{\alpha}\times\boldsymbol{v}^d)\big) \bigg] - \theta\left(\nabla(\rho\partial_{\boldsymbol{\alpha}}\Psi)\boldsymbol{L}\right)\, :\,\mathbf{X}\Bigg\} : \boldsymbol{W\gamma}.
    \end{split}
    \label{eqn:temperature_evolution_2}
\end{equation}

\cref{eqn:temperature_evolution_2} covers the most general case, in which $\Psi$ could have coupled terms between the state variables $\boldsymbol{W}$, $\theta$ and $\boldsymbol{\alpha}$. We recall here that the present model is applicable in the solid state of the body, such that \cref{eqn:temperature_evolution_2} is valid for $\theta$ below the solidus temperature.

To address the main implications of the new physical coupling between the evolution of the general thermomechanical defect density $\boldsymbol{\alpha}$ and the flux of thermal strains (\cref{eqn:transport}), an analysis of the structure of \cref{eqn:temperature_evolution_2} is carried out in \ref{sec:linear_stability} in a simplified linear, one-dimensional case. The following aspects stand out:
\begin{itemize}
    \item In the adiabatic case, neglecting the heat diffusion term, the temperature evolution remains governed by a PDE, owing to the presence of spatial derivatives of $\theta$ on the left-hand side of \cref{eqn:temperature_evolution_2}.
    \item The behaviour of the solutions of \cref{eqn:temperature_evolution_2} is in the form of \textit{dispersive temperature waves} i.e., with finite propagation speed (that varies with wave number). It arises from the mixed temporal and spatial derivatives in \cref{eqn:temperature_evolution_2}, which in turn occur because $\boldsymbol{\alpha}$ is introduced as an internal state variable.
    \item The well-posedness of \cref{eqn:temperature_evolution_2} is shown in a linearised setting. Along with the expected decay for a range of wavenumbers, the solution also admits well-posed growth. This can be the source of spatial patterning resulting from the growth of some Fourier components in the temperature.  
\end{itemize}

Generally, when considering Fourier's law of heat conduction as the constitutive statement for the heat conduction vector (\cref{eqn:fourier_law}), the resulting temperature evolution is such that a temperature change in the body is immediately felt throughout the entirety of it, giving rise to an infinite temperature propagation speed. To overcome this limitation, multiple strategies have been employed in the literature, usually involving the use of an extended constitutive relation for the heat flux vector that considers its rate associated with a certain relaxation time. This results in a hyperbolic evolution law for the temperature evolution, thus giving rise to wave-like solutions of the temperature field, which propagate at finite speeds (see \cite{joseph_heat_1989} and the references therein for a broad overview, including that of the pioneering work of \cite{Cattaneo1948conduzione}).

More recently, \cite{mariano_finite-speed_2017,mariano_sources_2022} analysed a rigid thermal conductor and showed that, by considering the existence of a geometrical descriptor of the microstructure of a body, the energy balance statement leads to a finite speed of temperature propagation governed by a hyperbolic PDE. Their result is said to be independent of the microstructure type, provided that the latter is sensitive to temperature changes. Moreover, their approach did not require any changes to Fourier's law of heat conduction. The theory presented here seems to align well with these results. In our case, the thermomechanical defect density $\boldsymbol{\alpha}$ can be seen as the descriptor that provides information about low-dimensional structures (dislocation lines and incompatibilities in thermal strains) on a larger scale. The fundamental statement of the evolution of $\boldsymbol{\alpha}$ (\cref{eqn:transport}) comprises the effect of a transient temperature field, which in turn results in convection-diffusion-dispersion effects in the PDE that governs the temperature evolution in the body (\cref{eqn:temperature_evolution_2}) due to the accounting of the energetic contribution of $\boldsymbol{\alpha}$ to Helmholtz free energy (\cref{eqn:psirate}). Similarly to \cite{mariano_sources_2022}, the use here of Fourier's law (\cref{eqn:fourier_law}) did not impede this result, and the obtained temperature field propagates at finite speeds.

\subsection{Taylor-Quinney coefficient}\label{sec:taylor-quinney}
In the pioneering work of \cite{taylorquinney_latent_1934}, the authors measured the amount of energy that remained stored in metallic rods after severe plastic deformation and introduced a coefficient, later called the Taylor-Quinney coefficient (TQC), defined as 
\begin{equation}
    \beta_{\text{int}} = \frac{W - Q}{W},
    \label{eqn:beta_TQC}
\end{equation}
where $W$ corresponds to the total work done on a rod, and $Q$ is the heat dissipated by it during the deformation. Subsequently, different definitions were used that fall under the same denomination of TQC (\cite{rittel2017dependence}). In particular, the so-called \textit{differential} TQC $\beta_{\text{diff}}$ is used as a measure of the instantaneous partition of plastic work into heat and stored energy during deformation, while the \textit{integral} TQC (\cref{eqn:beta_TQC}) expresses the amount of plastic work that is stored as latent energy in the body after deformation (\cite{rittel1999conversion}, \cite{rittel2017dependence}, \cite{stimac2022energy}).

Adopting \cite{rosakis_thermodynamic_2000} for our work, from \cref{eqn:temperature_evolution_2} we define 
\begin{equation}
    \begin{aligned}
        &\dot Q^p = \bigg[-\rho\partial_{\boldsymbol{W}}\Psi\left(\mathds{1} - \theta\boldsymbol{\gamma}^T\right) +\theta\left(\rho\partial_{\boldsymbol{W}}\Psi - \nabla\times\rho\partial_{\boldsymbol{\alpha}}\Psi\right):\boldsymbol{W}\mathds{G} - \theta\nabla\times(\rho\partial_{\boldsymbol{\alpha}}\Psi)\boldsymbol{\gamma}^T \\
        &\hspace{.9cm}+ \rho\theta\left(\partial^2_{\theta\boldsymbol{W}}\Psi + \boldsymbol{W\gamma}\!:\!\partial^2_{\boldsymbol{W}\boldsymbol{W}}\Psi\right)\bigg]\!:\!\left(\boldsymbol{\alpha}\times\boldsymbol{v}^d\right) + \left[\rho\partial_{\boldsymbol{\alpha}}\Psi - \rho\theta\left(\partial^2_{\theta\boldsymbol{\alpha}}\Psi + \boldsymbol{W\gamma}\!:\!\partial^2_{\boldsymbol{W}\boldsymbol{\alpha}}\Psi\right)\right]\!:\!\nabla\times\left(\boldsymbol{\alpha}\times\boldsymbol{v}^d\right)\\
        &\hspace{.5cm}- \Bigg\{\theta\nabla\times\bigg[ \rho\partial^2_{\boldsymbol{\alpha}\boldsymbol{W}}\Psi:\big(\boldsymbol{\alpha}\times\boldsymbol{v}^d\big) -\rho\partial^2_{\boldsymbol{\alpha}\boldsymbol{\alpha}}\Psi:\nabla\times(\boldsymbol{\alpha}\times\boldsymbol{v}^d) \bigg]\Bigg\} : \boldsymbol{W\gamma} \\
        & \dot Q^e = -\Bigg\{ \rho\theta\boldsymbol{W}^T\left(\partial^2_{\theta\boldsymbol{W}}\Psi + \boldsymbol{W\gamma}\!:\!\partial^2_{\boldsymbol{W}\boldsymbol{W}}\Psi\right)
        +\left[ \rho\theta\left(\partial^2_{\theta\boldsymbol{\alpha}}\Psi + \boldsymbol{W\gamma}\!:\!\partial^2_{\boldsymbol{W}\boldsymbol{\alpha}}\Psi\right)\!:\!\boldsymbol{\alpha}\right]\mathds{1} - \left[\rho\theta\left(\partial^2_{\theta\boldsymbol{\alpha}}\Psi + \boldsymbol{W\gamma}\!:\!\partial^2_{\boldsymbol{W}\boldsymbol{\alpha}}\Psi\right)^T\boldsymbol{\alpha}\right] \\
        &\hspace{2cm}- \theta\left[-\rho(\boldsymbol{W}^T\partial_{\boldsymbol{W}}\Psi)\boldsymbol{\gamma}^T +\boldsymbol{W}^{T}\nabla\times(\rho\partial_{\boldsymbol{\alpha}}\Psi)\boldsymbol{\gamma}^T + \left(-\rho\boldsymbol{W}^T\partial_{\boldsymbol{W}}\Psi + \boldsymbol{W}^{T}\nabla\times\rho\partial_{\boldsymbol{\alpha}}\Psi \right):\boldsymbol{W}\mathds{G} \right]\Bigg\}:\boldsymbol{L}\\
        &\hspace{.5cm}- \Bigg\{\theta\nabla\times\bigg[\dot\rho\partial_{\boldsymbol{\alpha}}\Psi - \rho\partial^2_{\boldsymbol{\alpha}\boldsymbol{W}}\Psi:\boldsymbol{WL} + \rho\partial^2_{\boldsymbol{\alpha}\boldsymbol{\alpha}}\Psi:\big(-\text{tr}(\boldsymbol{L})\boldsymbol{\alpha} + \boldsymbol{\alpha L}^T\big) \bigg] \\
        &\hspace{5cm}+ \theta\left(\nabla(\rho\partial_{\boldsymbol{\alpha}}\Psi)\boldsymbol{L}\right)\, :\,\mathbf{X} -\frac{\theta\dot\rho}{\rho} \nabla\times\rho\partial_{\boldsymbol{\alpha}}\Psi \Bigg\} : \boldsymbol{W\gamma},
    \end{aligned}
\end{equation}
where $\dot Q^p$ is the heating due to inelastic effects governed by the thermomechanical defect density evolution and $\dot Q^e$ is the thermoelastic contribution to heating. We introduce the plastic work rate as $\dot W^p = -\rho\partial_{\boldsymbol{W}}\Psi : (\boldsymbol{\alpha}\times\boldsymbol{v}^d)$, so that the fraction of $\dot W^p$ converted into $\dot Q^p$ can be defined as 
\begin{equation}
    \beta_{\text{diff}} = \frac{\dot Q^p}{\dot W^p}.
    \label{eqn:beta_rate}
\end{equation}

Denoting the left-hand side of \cref{eqn:temperature_evolution_2} as $\mathcal{L}[\dot\theta]$, and considering \cref{eqn:beta_rate}, we can rewrite the temperature evolution as
\begin{equation}
    \begin{aligned}
        \mathcal{L}[\dot\theta] = - \nabla\cdot\boldsymbol{q} + \rho r + \beta_{\text{diff}}\dot W^p + \dot Q^e,
    \end{aligned}
    \label{eqn:temperature_eqn_with_beta}
\end{equation}
which shows that it is governed by heat diffusion, internal heat sources, plastic work due to the evolution of the thermomechanical defect density, and thermoelastic effects. 

In our model, the rate of conversion of plastic work into heat (\cref{eqn:beta_rate}) is influenced by the strain rates. This influence is manifested through the stress dependence of the line defect velocity $\boldsymbol{v}^d$ and defect source $\boldsymbol{S}$. The dependence of $\beta$ on the loading conditions is well-established in the literature (see \cite{rittel2017dependence} and the references therein). The key aspect of our approach is that the rather straightforward argument of conservation of Burgers vector in \cref{eqn:conservationofb} yields the structure of \cref{eqn:temperature_eqn_with_beta} that allows for studying temperature evolution during plastic work. In other approaches, a similar result is obtained with the introduction of phenomenological expressions for the accumulated plastic strains or other conventional plasticity-related variables (\cite{stainier_study_2010}, \cite{nieto-fuentes_dislocation-based_2018}, \cite{longere_thermodynamically_2023}, \cite{zeng_rate-dependent_2022}, \cite{daehli2023experimental}). Note that $\beta_{\text{diff}}$ depends on the choice of $\Psi$, and the key point here is that our description of plasticity allows for studying plastic work repartition into heat directly from the evolution of thermomechanical defects while accounting for temperature effects and involving unambiguously definable quantities, measurable from the current state (at least in principle).

\subsection{Initial boundary value problem of finite deformation field dislocations thermomechanics}\label{sec:IBVP_large_def}

In this section, we summarize the set of governing equations and constitutive relations of the model. The initial and boundary conditions are also shown. For an approach to solving for $\boldsymbol{W}$ using the Stokes-Helmholtz decomposition, as detailed in \cite{acharya2004constitutive}, we direct the reader to \ref{sec:decompose_W}.
\begin{subequations}
    \def\myquad{\hskip7\fontdimen5\font}
    \def\myrulesize{0.9\textwidth}
    \begin{align}
        &\rule{\myrulesize}{0.4pt}\nonumber\\
        &\text{Kinematics and dislocation density evolution}\nonumber\\
        \label{eqn:setGen_W_rate}
        &\dot{\boldsymbol{W}} + \boldsymbol{WL} = \boldsymbol{\alpha} \times \boldsymbol{v}^d + \boldsymbol{W\gamma}\dot\theta \\ 
        \label{eqn:setGen_alpha_circ}
        &\accentset{\circ}{\boldsymbol{\alpha}} = - \nabla\times\big( \boldsymbol{\alpha} \times \boldsymbol{v}^d + \boldsymbol{W\gamma}\dot\theta \big), \quad \accentset{\circ}{\boldsymbol{\alpha}} = \text{tr}(\boldsymbol{L}) \boldsymbol{\alpha} + \dot{\boldsymbol{\alpha}} - \boldsymbol{\alpha}\boldsymbol{L}^T\\
        \label{eqn:setGen_V_dislocation}
        &\boldsymbol{v}^d = \frac{1}{B}\left[ \left( -\rho\partial_{\boldsymbol{W}}\Psi  + \nabla\times \rho\partial_{\boldsymbol{\alpha}}\Psi\right)^T\boldsymbol{\alpha} \right] : \textbf{X}\\
        &\rule{\myrulesize}{0.4pt}\nonumber\\
        &\text{Mass density evolution}\nonumber\\
        \label{eqn:setGen_rho_evolution}
        &\dot{\rho} + \nabla\cdot \left( \rho \boldsymbol{v} \right) = 0\\
        &\rule{\myrulesize}{0.4pt}\nonumber\\
        &\text{Dynamics}\nonumber\\
        \label{eqn:setGen_divsig}
        &\nabla\cdot\boldsymbol{\sigma} +\rho\boldsymbol{b}_f = \rho \dot{\boldsymbol{v}}\\
        \label{eqn:setGen_sig_constitutive}
        &\boldsymbol{\sigma} = - \rho\boldsymbol{W}^T \partial_{\boldsymbol{W}}\Psi + \rho(\partial_{\boldsymbol{\alpha}}\Psi)^T\boldsymbol{\alpha} - \rho(\partial_{\boldsymbol{\alpha}}\Psi: \boldsymbol{\alpha}) \mathds{1}\\
        &\rule{\myrulesize}{0.4pt}\nonumber\\
        &\text{Temperature evolution}\nonumber\\
        \label{eqn:setGen_temperature_evol}
        \begin{split}
        &\Bigg\{\bigg[ \nabla\times\rho\partial_{\boldsymbol{\alpha}}\Psi(\mathds{1} + \theta\boldsymbol{\gamma}^T) -2\rho\theta\partial^2_{\boldsymbol{W}\theta}\Psi - \rho\theta\boldsymbol{W\gamma}:\partial^2_{\boldsymbol{WW}}\Psi - \rho\theta\partial_{\boldsymbol{W}}\Psi\boldsymbol{\gamma}^T - \theta(\rho\partial_{\boldsymbol{W}}\Psi - \nabla\times\rho\partial_{\boldsymbol{\alpha}}\Psi):\boldsymbol{W}\mathds{G}\bigg] :  \boldsymbol{W\gamma} \\
        &\quad-\rho\theta\partial^2_{\theta\theta}\Psi \Bigg\}\dot\theta  + \theta\nabla\times\left(\rho\partial^2_{\boldsymbol{\alpha}\theta}\Psi\dot\theta\right) : \boldsymbol{W\gamma} - \big[\rho\partial_{\boldsymbol{\alpha}}\Psi - \rho\theta\left(\partial^2_{\theta\boldsymbol{\alpha}}\Psi + \boldsymbol{W\gamma}:\partial^2_{\boldsymbol{W}\boldsymbol{\alpha}}\Psi\right)\big]:\nabla\times\left(\boldsymbol{W\gamma}\dot\theta\right) \\
        &+\theta\nabla\times\bigg[\rho\partial^2_{\boldsymbol{\alpha}\boldsymbol{W}}\Psi:\boldsymbol{W\gamma}\dot\theta -\rho\partial^2_{\boldsymbol{\alpha}\boldsymbol{\alpha}}\Psi:\nabla\times\left(\boldsymbol{W\gamma}\dot\theta\right) \bigg] : \boldsymbol{W\gamma}\\
        &\quad= - \nabla\cdot \boldsymbol{q} + \rho r +\bigg[-\rho\partial_{\boldsymbol{W}}\Psi\left(\mathds{1} - \theta\boldsymbol{\gamma}^T\right) +\theta\left(\rho\partial_{\boldsymbol{W}}\Psi - \nabla\times\rho\partial_{\boldsymbol{\alpha}}\Psi\right):\boldsymbol{W}\mathds{G} - \theta\nabla\times(\rho\partial_{\boldsymbol{\alpha}}\Psi)\boldsymbol{\gamma}^T \\
        &\myquad+ \rho\theta\left(\partial^2_{\theta\boldsymbol{W}}\Psi + \boldsymbol{W\gamma}:\partial^2_{\boldsymbol{W}\boldsymbol{W}}\Psi\right)\bigg]:\left(\boldsymbol{\alpha}\times\boldsymbol{v}^d\right) + \Bigg\{- \rho\theta\boldsymbol{W}^T\left(\partial^2_{\theta\boldsymbol{W}}\Psi + \boldsymbol{W\gamma}:\partial^2_{\boldsymbol{W}\boldsymbol{W}}\Psi\right)
        \\
        &\myquad-\left[\rho\theta\left(\partial^2_{\theta\boldsymbol{\alpha}}\Psi + \boldsymbol{W\gamma}:\partial^2_{\boldsymbol{W}\boldsymbol{\alpha}}\Psi\right):\boldsymbol{\alpha}\right]\mathds{1}  + \left[\rho\theta\left(\partial^2_{\theta\boldsymbol{\alpha}}\Psi + \boldsymbol{W\gamma}:\partial^2_{\boldsymbol{W}\boldsymbol{\alpha}}\Psi\right)^T\boldsymbol{\alpha}\right] \\
        &\myquad+ \theta\left[-\rho(\boldsymbol{W}^T\partial_{\boldsymbol{W}}\Psi)\boldsymbol{\gamma}^T +\boldsymbol{W}^{T}\nabla\times(\rho\partial_{\boldsymbol{\alpha}}\Psi)\boldsymbol{\gamma}^T + \left(-\rho\boldsymbol{W}^T\partial_{\boldsymbol{W}}\Psi + \boldsymbol{W}^{T}\nabla\times\rho\partial_{\boldsymbol{\alpha}}\Psi \right):\boldsymbol{W}\mathds{G} \right]\Bigg\}:\boldsymbol{L}\\
        &\qquad+ \left[\rho\partial_{\boldsymbol{\alpha}}\Psi - \rho\theta\left(\partial^2_{\theta\boldsymbol{\alpha}}\Psi + \boldsymbol{W\gamma}:\partial^2_{\boldsymbol{W}\boldsymbol{\alpha}}\Psi\right)\right]:\nabla\times\left(\boldsymbol{\alpha}\times\boldsymbol{v}^d\right) \\
        &\qquad+ \Bigg\{ \frac{\theta\dot\rho}{\rho} \nabla\times\rho\partial_{\boldsymbol{\alpha}}\Psi-\theta\nabla\times\bigg[\dot\rho\partial_{\boldsymbol{\alpha}}\Psi + \rho\partial^2_{\boldsymbol{\alpha}\boldsymbol{W}}\Psi:\big(-\boldsymbol{WL}+\boldsymbol{\alpha}\times\boldsymbol{v}^d\big) \\
        &\myquad+ \rho\partial^2_{\boldsymbol{\alpha}\boldsymbol{\alpha}}\Psi:\big(-\text{tr}(\boldsymbol{L})\boldsymbol{\alpha} + \boldsymbol{\alpha L}^T - \nabla\times(\boldsymbol{\alpha}\times\boldsymbol{v}^d)\big) \bigg] - \theta\left(\nabla(\rho\partial_{\boldsymbol{\alpha}}\Psi)\boldsymbol{L}\right)\, :\,\mathbf{X}\Bigg\} : \boldsymbol{W\gamma}
        \end{split}\\ 
        \label{eqn:setGen_q_constitutive}
        &\boldsymbol{q} = -\boldsymbol{K}\nabla\theta.\\
        &\rule{\myrulesize}{0.4pt}\nonumber
    \end{align}
    \label{eqn:complete_eq_set_general}
\end{subequations}

\subsubsection{Boundary conditions}\label{sec:BCs_set_large_def}

Naturally, the closure of \cref{eqn:complete_eq_set_general} requires the specification of the initial and boundary conditions, presented in what follows.

The thermomechanical defect density transport, \cref{eqn:setGen_alpha_circ}, requires the specification

\begin{equation}
    (\boldsymbol{v}^d\cdot\hat{\boldsymbol{n}})\boldsymbol{\alpha} = \overline{\boldsymbol{F}}_\alpha\quad \text{on} \quad \partial\Omega_t^-,
    \label{eqn:BC_alpha}
\end{equation}
which is enough to ensure uniqueness, where $\overline{\boldsymbol{F}}_\alpha$ is a prescribed dislocation flux and $\partial\Omega_t^-$ corresponds to the part of $\partial\Omega_t$ where $\boldsymbol{v}^d\cdot\hat{\boldsymbol{n}} < 0$, i.e., only the inflow of defects into the body needs to be prescribed (\cite{acharya2003driving}). 

The balance of linear momentum in \cref{eqn:setGen_divsig} requires the specification of standard velocity and traction rate boundary conditions on complementary parts of $\partial\Omega_t$ (\cite{ARORA2020_FEM_FDM}). 

Finally, the temperature evolution \cref{eqn:setGen_temperature_evol} is completed with 
\begin{equation}
    \begin{aligned}
        &\theta = \bar\theta \quad &&\text{on} \quad \partial\Omega_t^\theta \\
        &\boldsymbol{q}\cdot\hat{\boldsymbol{n}} = \overline{\boldsymbol{q}}\quad &&\text{on} \quad \partial\Omega_t^q,
    \end{aligned}
    \label{eqn:BC_theta}
\end{equation}
where $\bar\theta$ and $\overline{\boldsymbol{q}}$ are prescribed quantities, $\partial\Omega^q_t \cap \partial\Omega^\theta_t = \emptyset$, and $\partial\Omega^q_t \cup \partial\Omega^\theta_t = \partial\Omega_t$.

\subsubsection{Initial conditions}\label{sec:IC_large_def}

The defect transport \cref{eqn:setGen_alpha_circ} is solved considering a given initial defect density $\boldsymbol{\alpha}_0$, i.e. $\boldsymbol{\alpha}(\boldsymbol{x}, 0) = \boldsymbol{\alpha}_0(\boldsymbol{x})$. 

The balance of linear momentum \mbox{\cref{eqn:setGen_divsig}} requires the specification of an initial material velocity profile $\boldsymbol{v}_0$, i.e. \\ \mbox{$\boldsymbol{v}(\boldsymbol{x}, 0) = \boldsymbol{v}_0(\boldsymbol{x})$}.

Furthermore, the temperature evolution \cref{eqn:setGen_temperature_evol} requires the specification of an initial temperature profile in the body, that is, $\theta(\boldsymbol{x}, 0) = \theta_0(\boldsymbol{x})$ for a given $\theta_0$.

\section{Geometric linearisation}\label{sec:small-def}

In this section, we geometrically linearise the model, i.e. develop it in a small deformation framework (but large temperature changes are allowed). In the small deformation case, $\boldsymbol{W}$  can be approximated as 
\begin{equation}
     \boldsymbol{W} = \boldsymbol{F}^{e-1} = \left(\mathds{1} + \boldsymbol{U}^e \right)^{-1} \approx \mathds{1} - \boldsymbol{U}^e,
    \label{eqn:Fe_W_small_def}
\end{equation}

\noindent where $\boldsymbol{U}^e$ is the elastic distortion tensor, with $\lVert \boldsymbol{U}^e \rVert \ll 1$. In the present small deformation setting, the superposed dot denotes a partial derivative with respect to $t$, and we assume that $\dot\rho = 0$ and $\rho/\rho_0 \approx 1$. 

We also assume $\boldsymbol{v} = \dot{\boldsymbol{u}}$, where $\boldsymbol{u}$ is the displacement field. Then, using \cref{eqn:Fe_W_small_def}, the velocity gradient $\boldsymbol{L}$ given by \cref{eqn:velgradadddecomp} can be approximated as
\begin{equation}
    \boldsymbol{L} = \nabla \dot{\boldsymbol{u}} = \dot{\boldsymbol{U}}^e + \boldsymbol{\alpha}\times\boldsymbol{v}^d +\boldsymbol{\gamma}\dot\theta.
    \label{eqn:velGradient_smallStrain}
\end{equation}
where in this case $\boldsymbol{\gamma} = \sum_{i=1}^3 d_i(\theta) \hat{\boldsymbol{e}}_i \otimes \hat{\boldsymbol{e}}_i$. Without loss of generality, we can define $\dot{\boldsymbol{U}}^p := \boldsymbol{\alpha} \times \boldsymbol{v}^d$ as the plastic distortion rate and $\dot{\boldsymbol{\varepsilon}}^\theta := \boldsymbol{\gamma} \dot{\theta}$ as the thermal strain rate in the small deformation formulation. Integrating \cref{eqn:velGradient_smallStrain} in time, we get
\begin{equation}
    \nabla \boldsymbol{u} = \boldsymbol{U}^e + \boldsymbol{U}^p + \boldsymbol{\varepsilon}^\theta,
    \label{eqn:additiv_decomp_grad_u}
\end{equation}
where a time-independent tensor field is ignored. This expression is the well-known additive decomposition of the displacement gradient tensor that is relevant in the small deformation case. 

Considering \cref{eqn:Fe_W_small_def}, the definition of the Burgers vector in \cref{eqn:burgersdef} becomes
\begin{equation}
    \boldsymbol{b} = \oint_{c} \boldsymbol{U}^e\boldsymbol{\textbf{d}x} = \int_{s}(\nabla\times\boldsymbol{U}^e)\hat{\boldsymbol{n}}\, \text{d}s = \int_{s}\boldsymbol{\alpha}\hat{\boldsymbol{n}}\, \text{d}s,
\end{equation}
where $s$ in an arbitrary closed surface in $\Omega$, whose boundary and unit normal are $c$ and $\hat{\boldsymbol{n}}$, respectively, and the second equality is obtained through Stokes' theorem. The thermomechanical defect density is now defined as
\begin{equation}
    \boldsymbol{\alpha} = \nabla\times\boldsymbol{U}^e,
    \label{eqn:alpha_curl_Ue}
\end{equation}
for which we obtain the evolution statement (\cite{acharya_microcanonical_2011}, \cite{upadhyay_thermomechanical_2020}) 
\begin{equation}
    \begin{aligned}
        \dot{\boldsymbol{\alpha}} &= \nabla\times\dot{\boldsymbol{U}}^e \\
        \implies \dot{\boldsymbol{\alpha}} &= -\nabla\times(\boldsymbol{\alpha}\times\boldsymbol{v}^d) - \nabla\times(\boldsymbol{\gamma}\dot\theta)
    \end{aligned}
    \label{eqn:alpha_dot_smallstrains}
\end{equation}
where \cref{eqn:velGradient_smallStrain} was used. These expressions are very similar to the ones derived in \cite{upadhyay_thermomechanical_2020}, except for one important detail, which is addressed in section \ref{sec:comparison_upadhyay}.

In the geometrically linear setting, the additive decomposition of the displacement gradient (\cref{eqn:additiv_decomp_grad_u}) allows for considering the Helmholtz free energy density as  $\Psi \equiv \Psi(\boldsymbol{\varepsilon}^e,\boldsymbol{\alpha}) = \Psi(\boldsymbol{\varepsilon} - \boldsymbol{\varepsilon}^p - \boldsymbol{\varepsilon}^\theta, \boldsymbol{\alpha}) = \Psi(\boldsymbol{\varepsilon} - \boldsymbol{\varepsilon}^p, \theta, \boldsymbol{\alpha})$, with $\boldsymbol{\varepsilon} = \text{sym}(\nabla\boldsymbol{u})$ and $\boldsymbol{\varepsilon}^{e,p} = \text{sym}(\boldsymbol{U}^{e,p})$, such that 
\begin{equation}
    \dot\Psi = \partial_{(\boldsymbol{\varepsilon} - \boldsymbol{\varepsilon}^p)} \Psi: (\dot{\boldsymbol{\varepsilon}} - \dot{\boldsymbol{\varepsilon}}^p) + \partial_{\theta}\Psi\, \dot{\theta} + \partial_{\boldsymbol{\alpha}}\Psi: \dot{\boldsymbol{\alpha}}.
    \label{eqn:psirate_smalldef}
\end{equation}

Considering $\boldsymbol{\sigma}:\boldsymbol{L} = \boldsymbol{\sigma}:\boldsymbol{\varepsilon}$ in the global dissipation (\cref{eqn:localsecondlaw2}), as well as replacing \cref{eqn:psirate_smalldef} and \cref{eqn:alpha_dot_smallstrains} for $\dot\Psi$ and $\dot{\boldsymbol{\alpha}}$, respectively, and using \cref{eqn:appB_int_A_curlB,eqn:appB_A_alphaTimesV} allows for the definition of the following constitutive relations: \vspace{-0.5cm}

\begin{subequations}
    \begin{align}
        \label{eqn:constit_sig_smalldef}
        &\boldsymbol{\sigma} = \rho\partial_{(\boldsymbol{\varepsilon}-\boldsymbol{\varepsilon}^p)}\Psi\\
        \label{eqn:constit_eta_smalldef}
        &\eta = -\partial_\theta\Psi +  \frac{1}{\rho} \nabla\times\left(\rho\partial_{\boldsymbol{\alpha}}\Psi\right):\boldsymbol{\gamma}\\
        \label{eqn:constit_dislVel_smalldef}
        &\boldsymbol{v}^d = \frac{1}{B}\left[(\boldsymbol{\sigma} + \nabla\times\rho\partial_{\boldsymbol{\alpha}}\Psi)^T\boldsymbol{\alpha}\right]:\mathbf{X}\\
        \label{eqn:constit_q_smalldef}
        &\boldsymbol{q} = \boldsymbol{K}\nabla\theta.
    \end{align}
\end{subequations}

For the temperature evolution equation, in the small-strains case, \cref{eqn:theta_derivation1} becomes 
\begin{equation}
    \rho\left[\partial_{(\boldsymbol{\varepsilon}-\boldsymbol{\varepsilon}^p)}\Psi : (\dot{\boldsymbol{\varepsilon}}-\dot{\boldsymbol{\varepsilon}}^p) + \partial_{\theta}\Psi\dot\theta + \partial_{\boldsymbol{\alpha}}\Psi:\dot{\boldsymbol{\alpha}} + \dot\theta\eta + \theta\dot\eta\right] = \boldsymbol{\sigma}: \dot{\boldsymbol{\varepsilon}} - \nabla\cdot\boldsymbol{q} + \rho r.
    \label{eqn:temperature_eq_1_smallstrains}
\end{equation}
Now, by taking the derivative of \cref{eqn:constit_eta_smalldef} with respect to time, we get
\begin{equation}
    \begin{aligned}
        \dot\eta = -\partial^2_{\theta(\boldsymbol{\varepsilon}-\boldsymbol{\varepsilon}^p)}\Psi:(\dot{\boldsymbol{\varepsilon}}-\dot{\boldsymbol{\varepsilon}}^p) - \partial^2_{\theta\theta}\Psi\dot\theta -\partial^2_{\theta\boldsymbol{\alpha}}\Psi:\dot{\boldsymbol{\alpha}}
        +\frac{1}{\rho}\nabla\times\left(\rho\partial^2_{\boldsymbol{\alpha}(\boldsymbol{\varepsilon}-\boldsymbol{\varepsilon}^p)}\Psi:(\dot{\boldsymbol{\varepsilon}}-\dot{\boldsymbol{\varepsilon}}^p)+ \rho\partial^2_{\boldsymbol{\alpha}\theta}\Psi\dot\theta +\rho\partial^2_{\boldsymbol{\alpha}\boldsymbol{\alpha}}\Psi:\dot{\boldsymbol{\alpha}}\right) : \boldsymbol{\gamma}.
    \end{aligned}
    \label{eqn:etadot_smallstrains}
\end{equation}
Inserting \cref{eqn:etadot_smallstrains} into \cref{eqn:temperature_eq_1_smallstrains}, considering \cref{eqn:constit_sig_smalldef,eqn:alpha_dot_smallstrains,eqn:velGradient_smallStrain}, and rearranging terms, results in the temperature evolution equation
\begin{equation}
    \begin{aligned}
        &\left[\left(\nabla\times\rho\partial_{\boldsymbol{\alpha}}\Psi - \rho\theta\partial^2_{\theta(\boldsymbol{\varepsilon} - \boldsymbol{\varepsilon}^p)}\Psi\right):\boldsymbol{\gamma} - \rho\theta\partial^2_{\theta\theta}\Psi\right]\dot\theta + \theta\nabla\times\left(\rho\partial^2_{\boldsymbol{\alpha}\theta}\Psi\dot\theta\right):\boldsymbol{\gamma} - \rho\left(\partial_{\boldsymbol{\alpha}}\Psi + \theta\partial^2_{\theta\boldsymbol{\alpha}}\Psi\right):\nabla\times(\boldsymbol{\gamma}\dot\theta) \\
        &\hspace{.4cm}+ \theta\nabla\times\left[\rho\partial^2_{\boldsymbol{\alpha}(\boldsymbol{\varepsilon} - \boldsymbol{\varepsilon}^p)}\Psi: \boldsymbol{\gamma}\dot\theta - \rho\partial^2_{\boldsymbol{\alpha\alpha}}\Psi:\nabla\times(\boldsymbol{\gamma}\dot\theta) \right] : \boldsymbol{\gamma} = \boldsymbol{\sigma}:\left(\boldsymbol{\alpha}\times\boldsymbol{v}^d\right) + \rho\theta\partial^2_{\theta(\boldsymbol{\varepsilon} - \boldsymbol{\varepsilon}^p)}\Psi : \dot{\boldsymbol{\varepsilon}}^e  -\nabla\cdot\boldsymbol{q} + \rho r\\
        &\hspace{.4cm}+ \bigg\{\rho(\partial_{\boldsymbol{\alpha}}\Psi - \theta\partial^2_{\theta\boldsymbol{\alpha}}\Psi):\nabla\times(\boldsymbol{\alpha}\times\boldsymbol{v}^d) + \theta\nabla\times\left[ \rho\partial^2_{\boldsymbol{\alpha\alpha}}\Psi:\nabla\times(\boldsymbol{\alpha}\times\boldsymbol{v}^d) \right] : \boldsymbol{\gamma}\bigg\} - \theta\nabla\times\left(\rho\partial^2_{\boldsymbol{\alpha}(\boldsymbol{\varepsilon} - \boldsymbol{\varepsilon}^p)}\Psi:\dot{\boldsymbol{\varepsilon}}^e \right) : \boldsymbol{\gamma},
    \end{aligned}
    \label{eqn:temp_evol_smallstrains}
\end{equation}
where the left-hand side represents the heat storage, with coefficients depending on the thermodynamic driving force of $\boldsymbol{\alpha}$, thermoelastic coupling,  heat capacity and coupling between $\boldsymbol{\alpha}$ and $\theta$ or $(\boldsymbol{\varepsilon} - \boldsymbol{\varepsilon}^p)$; on the right-hand side, from left to right, we have plastic work, thermoelastic effect, heat diffusion, heat source, heat generation due to thermomechanical defect density evolution and a coupled term between $\boldsymbol{\alpha}$ and deformation evolution. \cref{eqn:temp_evol_smallstrains} retains the same structure as \cref{eqn:temperature_evolution_2} in terms of the derivatives of $\theta$ involved, so that the analysis in \ref{sec:linear_stability} also applies to this case.

The equations of the geometrically linearised theory are grouped and presented in the set below. As in \mbox{\cref{sec:IBVP_large_def}}, we direct the reader to \mbox{\ref{sec:decomposition_Ue}} for an approach to solving for $\boldsymbol{U}^e$ using the Stokes-Helmholtz decomposition.
\begin{subequations}
    \def\myquad{\hskip10\fontdimen6\font}
    \def\myrulesize{0.9\textwidth}
    \begin{align}
        &\rule{\myrulesize}{0.4pt}\nonumber\\
        &\text{Kinematics and dislocation density evolution}\nonumber\\
        \label{eqn:setGen_SD_Ue_dot}
        &\dot{\boldsymbol{U}}^e = \nabla\dot{\boldsymbol{u}} - \boldsymbol{\alpha}\times\boldsymbol{v}^d - \boldsymbol{\gamma}\dot\theta\\
        \label{eqn:setGen_SD_alpha_dot}
        &\dot{\boldsymbol{\alpha}} = - \nabla\times\big( \boldsymbol{\alpha} \times \boldsymbol{v}^d + \boldsymbol{\gamma}\dot\theta \big)\\
        \label{eqn:setGen_SD_V_dislocation}
        &\boldsymbol{v}^d = \frac{1}{B}\left[(\boldsymbol{\sigma} + \nabla\times\rho\partial_{\boldsymbol{\alpha}}\Psi)^T\boldsymbol{\alpha}\right]:\mathbf{X}\\
        &\rule{\myrulesize}{0.4pt}\nonumber\\
        &\text{Mechanical equilibrium}\nonumber\\
        \label{eqn:setGen_SD_divsig}
        &\nabla\cdot\boldsymbol{\sigma} +\rho\boldsymbol{b}_f = \rho \ddot{\boldsymbol{u}}\\
        \label{eqn:setGen_SD_sig_constitutive}
        &\boldsymbol{\sigma} = \rho\partial_{(\boldsymbol{\varepsilon}-\boldsymbol{\varepsilon}^p)}\Psi\\
        &\rule{\myrulesize}{0.4pt}\nonumber\\
        &\text{Temperature evolution}\nonumber\\
        \label{eqn:setGen_SD_temperature_evol}
        \begin{split}
            &\left[\left(\nabla\times\rho\partial_{\boldsymbol{\alpha}}\Psi - \rho\theta\partial^2_{\theta(\boldsymbol{\varepsilon} - \boldsymbol{\varepsilon}^p)}\Psi\right):\boldsymbol{\gamma} - \rho\theta\partial^2_{\theta\theta}\Psi\right]\dot\theta + \theta\nabla\times\left(\rho\partial^2_{\boldsymbol{\alpha}\theta}\Psi\dot\theta\right):\boldsymbol{\gamma} - \rho\left(\partial_{\boldsymbol{\alpha}}\Psi + \theta\partial^2_{\theta\boldsymbol{\alpha}}\Psi\right):\nabla\times(\boldsymbol{\gamma}\dot\theta) \\
            &\hspace{.4cm}+ \theta\nabla\times\left[\rho\partial^2_{\boldsymbol{\alpha}(\boldsymbol{\varepsilon} - \boldsymbol{\varepsilon}^p)}\Psi: \boldsymbol{\gamma}\dot\theta - \rho\partial^2_{\boldsymbol{\alpha\alpha}}\Psi:\nabla\times(\boldsymbol{\gamma}\dot\theta) \right] : \boldsymbol{\gamma} = \boldsymbol{\sigma}:\left(\boldsymbol{\alpha}\times\boldsymbol{v}^d\right) + \rho\theta\partial^2_{\theta(\boldsymbol{\varepsilon} - \boldsymbol{\varepsilon}^p)}\Psi : \dot{\boldsymbol{\varepsilon}}^e \\
            &\hspace{.4cm}+ \bigg\{\rho(\partial_{\boldsymbol{\alpha}}\Psi - \theta\partial^2_{\theta\boldsymbol{\alpha}}\Psi):\nabla\times(\boldsymbol{\alpha}\times\boldsymbol{v}^d) + \theta\nabla\times\left[ \rho\partial^2_{\boldsymbol{\alpha\alpha}}\Psi:\nabla\times(\boldsymbol{\alpha}\times\boldsymbol{v}^d) \right] : \boldsymbol{\gamma}\bigg\} \\
            &\hspace{.4cm}- \theta\nabla\times\left(\rho\partial^2_{\boldsymbol{\alpha}(\boldsymbol{\varepsilon} - \boldsymbol{\varepsilon}^p)}\Psi:\dot{\boldsymbol{\varepsilon}}^e \right) : \boldsymbol{\gamma} -\nabla\cdot\boldsymbol{q} + \rho r
        \end{split}\\ 
        \label{eqn:setGen_SD_q_constitutive}
        &\boldsymbol{q} = -\boldsymbol{K}\nabla\theta.\\
        &\rule{\myrulesize}{0.4pt}\nonumber
    \end{align}
    \label{eqn:complete_eq_set_general_smalldef}
\end{subequations}


\subsection{Boundary conditions}\label{sec:BC_smalldef}

Standard displacement and traction boundary conditions on complementary parts of the boundary are necessary to solve for the equilibrium \cref{eqn:setGen_SD_divsig}.

To solve for $\boldsymbol{\alpha}$ and $\theta$, the required boundary conditions are similar to \cref{eqn:BC_alpha,eqn:BC_theta}.

\subsection{Initial conditions}\label{sec:IC_smalldef}

Similarly to \mbox{\cref{sec:IC_large_def}}, the defect transport \mbox{\cref{eqn:setGen_SD_alpha_dot}} is solved considering a given initial defect density $\boldsymbol{\alpha}_0$, i.e. $\boldsymbol{\alpha}(\boldsymbol{x}, 0) = \boldsymbol{\alpha}_0(\boldsymbol{x})$.

The balance of linear momentum \mbox{\cref{eqn:setGen_SD_divsig}} requires the specification of an initial displacement and velocity $\boldsymbol{u}_0$ and $\dot{\boldsymbol{u}}_0$, respectively, such that $\boldsymbol{u}(\boldsymbol{x}, 0) = \boldsymbol{u}_0(\boldsymbol{x})$ and $\dot{\boldsymbol{u}}(\boldsymbol{x}, 0) = \dot{\boldsymbol{u}}_0(\boldsymbol{x})$.

Furthermore, the temperature evolution \mbox{\cref{eqn:setGen_SD_temperature_evol}} requires the specification of an initial temperature profile in the body, that is, $\theta(\boldsymbol{x}, 0) = \theta_0(\boldsymbol{x})$ for a given $\theta_0$.

\section{Some examples for given Helmholtz free energy density expressions}\label{sec:example_psi}

\subsection{Finite deformation: Saint-Venant-Kirchoff model with defect core energy}

Consider the following expression that models a Saint-Venant-Kirchhoff material and allows for large deformations and temperature changes while specifying a quadratic dislocation core energy term (\cite{arora_dislocation_2020}): 


\begin{equation}
    \begin{split}
        \Psi(\boldsymbol{W}, \theta, \boldsymbol{\alpha}) = \frac{1}{2\rho_0}\boldsymbol{E}:\mathds{C}:\boldsymbol{E}
        + c_\varepsilon\left[\Delta\theta - \theta \text{ln}\left(\frac{\theta}{\theta_0}\right)\right] 
        + \frac{\xi}{2\rho_0}\boldsymbol{\alpha}:\boldsymbol{\alpha},
    \end{split}
    \label{eqn:psi_example}
\end{equation}

\noindent where $\boldsymbol{E} = \frac{1}{2}(\boldsymbol{C}^e - \mathds{1})$ is the Green-Lagrange strain tensor, $\mathds{C}$ is the fourth order isotropic stiffness tensor (which in the general case could be $\theta$ and/or $\boldsymbol{W}$-dependent), $\boldsymbol{C}^e = \boldsymbol{W}^{-T}\boldsymbol{W}^{-1}$ is the right Cauchy-Green deformation tensor, $c_\varepsilon$ is the specific heat capacity at constant strain, $\rho_0$ is the mass density for a reference state, 
$J = \text{det} (\boldsymbol{W}^{-1})$, and $\xi$ is a material constant with dimensions $\text{stress}\times\text{length}^2$. The expression in \cref{eqn:psi_example} captures the elastic stored energy in the first term, the thermal contribution in the second term, and the energy stored in dislocation cores (\cite{ACHARYA2010_newInroads}, \cite{acharya_tartar_2011}) in the third term. 

For the $\Psi$ expression in \cref{eqn:psi_example}, we show the partial derivatives with respect to its arguments in \ref{app:derivatives_psi}. Using these, the set in \cref{eqn:complete_eq_set_general} becomes:\vspace{-.7cm}

\begin{subequations}
    \def\myquad{\hskip8\fontdimen5\font}
    \def\myrulesize{0.9\textwidth}
    \begin{align}
        &\rule{\myrulesize}{0.4pt}\nonumber\\
        &\text{Kinematics and dislocation density evolution}\nonumber\\  
        \label{eqn:set_W_rate}
        &\dot{\boldsymbol{W}} + \boldsymbol{WL} = \boldsymbol{\alpha} \times \boldsymbol{v}^d + \boldsymbol{W\gamma}\dot\theta \\
        \label{eqn:set_alpha_circ}
        &\accentset{\circ}{\boldsymbol{\alpha}} = - \nabla\times\big( \boldsymbol{\alpha} \times \boldsymbol{v}^d + \boldsymbol{Y}\dot\theta \big), \quad \accentset{\circ}{\boldsymbol{\alpha}} = \text{tr}(\boldsymbol{L}) \boldsymbol{\alpha} + \dot{\boldsymbol{\alpha}} - \boldsymbol{\alpha}\boldsymbol{L}^T\\
        \label{eqn:set_V_dislocation}
        &\boldsymbol{v}^d = \frac{1}{B}\left\{ \left[ \boldsymbol{\sigma}_H  + \xi\left(\nabla\times \frac{\rho}{\rho_0}\boldsymbol{\alpha}\right)^T\boldsymbol{W}\right]\boldsymbol{W}^{-1}\boldsymbol{\alpha} \right\} : \textbf{X}\\
        &\rule{\myrulesize}{0.4pt}\nonumber\\
        &\text{Mass density evolution}\nonumber\\
        \label{eqn:set_rho_evolution}
        &\dot{\rho} + \nabla\cdot \left( \rho \boldsymbol{v} \right) = 0\\
        &\rule{\myrulesize}{0.4pt}\nonumber\\
        &\text{Mechanical equilibrium}\nonumber\\
        \label{eqn:set_divsig}
        &\nabla\cdot\boldsymbol{\sigma} +\rho\boldsymbol{b}_f = \rho \dot{\boldsymbol{v}}\\
        \label{eqn:set_sig_constitutive}
        &\boldsymbol{\sigma} = \frac{\rho}{\rho_0}\boldsymbol{W}^{-1}(\mathbb{C}:\boldsymbol{E})\boldsymbol{W}^{-T} 
        + \frac{\rho}{\rho_0}\xi\big[\boldsymbol{\alpha}^T\boldsymbol{\alpha} - \big(\boldsymbol{\alpha}:\boldsymbol{\alpha})\mathds{1}\big]\\
        &\rule{\myrulesize}{0.4pt}\nonumber\\
        &\text{Temperature evolution}\nonumber\\
        \label{eqn:set_temperature_evol}
        \begin{split}
            &\Bigg\{\bigg[ \xi\nabla\times\frac{\rho}{\rho_0}\boldsymbol{\alpha}(\mathds{1} + \theta\boldsymbol{\gamma}^T) - \rho\theta\boldsymbol{W\gamma}:\mathds{A} -\theta\boldsymbol{W}^{-T}\boldsymbol{\sigma}_H \boldsymbol{\gamma}^T - \theta(\boldsymbol{W}^{-T}\boldsymbol{\sigma}_H - \xi\nabla\times\frac{\rho}{\rho_0}\boldsymbol{\alpha}):\boldsymbol{W}\mathds{G}\bigg] :  \boldsymbol{W\gamma} +\rho c_\varepsilon \Bigg\}\dot\theta\\
            &- \xi\frac{\rho}{\rho_0}\boldsymbol{\alpha}:\nabla\times\left(\boldsymbol{W\gamma}\dot\theta\right) - \xi\theta\nabla\times\bigg[\frac{\rho}{\rho_0}\text{tr}\left(\nabla\times\left(\boldsymbol{W\gamma}\dot\theta\right)\right)\mathds{1} \bigg] : \boldsymbol{W\gamma}\\
            &\quad= - \nabla\cdot \boldsymbol{q} + \rho r +\bigg[\boldsymbol{W}^{-T}\boldsymbol{\sigma}_H\left(\mathds{1} - \theta\boldsymbol{\gamma}^T\right) -\theta\left(\boldsymbol{W}^{-T}\boldsymbol{\sigma}_H + \xi\nabla\times\frac{\rho}{\rho_0}\boldsymbol{\alpha}\right):\boldsymbol{W}\mathds{G} - \xi\theta\nabla\times\left(\frac{\rho}{\rho_0}\boldsymbol{\alpha}\right)\boldsymbol{\gamma}^T \\
            &\myquad+ \rho\theta\boldsymbol{W\gamma}:\mathds{A}\bigg]:\left(\boldsymbol{\alpha}\times\boldsymbol{v}^d\right) + \Bigg\{- \rho\theta\boldsymbol{W}^T\left( \boldsymbol{W\gamma}:\mathds{A}\right)
            + \theta\Bigg[\boldsymbol{\sigma}_H\boldsymbol{\gamma}^T +\xi\boldsymbol{W}^{T}\nabla\times\left(\frac{\rho}{\rho_0      }\boldsymbol{\alpha}\right)\boldsymbol{\gamma}^T\\
            &\myquad+ \left(\boldsymbol{\sigma}_H + \xi\boldsymbol{W}^{T}\nabla\times\frac{\rho}{\rho_0}\boldsymbol{\alpha} \right):\boldsymbol{W}\mathds{G} \Bigg]\Bigg\}:\boldsymbol{L} + \xi\frac{\rho}{\rho_0}\boldsymbol{\alpha}:\nabla\times\left(\boldsymbol{\alpha}\times\boldsymbol{v}^d\right) \\
            &\qquad+ \Bigg\{ \xi\theta\frac{\dot\rho}{\rho} \nabla\times\frac{\rho}{\rho_0}\boldsymbol{\alpha} -\theta\nabla\times\bigg[\xi\frac{\dot\rho}{\rho_0}\boldsymbol{\alpha} - \xi\frac{\rho}{\rho_0}\text{tr}\big(\text{tr}(\boldsymbol{L})\boldsymbol{\alpha} - \boldsymbol{\alpha L}^T + \nabla\times(\boldsymbol{\alpha}\times\boldsymbol{v}^d)\big)\mathds{1} \bigg] \\
            &\myquad- \xi\theta\left[\nabla\left(\frac{\rho}{\rho_0}\boldsymbol{\alpha}\right)\boldsymbol{L}\right]\, :\,\mathbf{X}\Bigg\} : \boldsymbol{W\gamma}
        \end{split}\\ 
        \label{eqn:set_q_constitutive}
        &\boldsymbol{q} = -\boldsymbol{K}\nabla\theta,\\
        &\rule{\myrulesize}{0.4pt}\nonumber
    \end{align}
\end{subequations}

\noindent where $\boldsymbol{\sigma}_H$ is given by \cref{eqn:delPsi_delW} and $\mathbb{A} := \partial^2_{\boldsymbol{W}\boldsymbol{W}}\Psi$ by \cref{eqn:del2psi_delWdelW}.

\subsection{Small deformation: Saint-Venant-Kirchoff model with defect core energy}\label{sec:smalldefexample}

Considering \cref{eqn:Fe_W_small_def}, we take the same expression for $\Psi$ from \cref{sec:example_psi}, which then becomes
\begin{equation}
    \Psi(\boldsymbol{\varepsilon}^e, \theta, \boldsymbol{\alpha}) = \frac{1}{2\rho_0}\boldsymbol{\varepsilon}^e:\mathbb{C}:\boldsymbol{\varepsilon}^e
    + c_\varepsilon\left[\Delta\theta - \theta \text{ln}\left(\frac{\theta}{\theta_0}\right)\right] 
    + \frac{\xi}{2\rho_0}\boldsymbol{\alpha}:\boldsymbol{\alpha}.
    \label{eqn:psiexample_smalldef_withEpsE}
\end{equation}
From \cref{eqn:additiv_decomp_grad_u}, we have that 
\begin{equation*}
    \boldsymbol{\varepsilon}^e = \boldsymbol{\varepsilon} - \boldsymbol{\varepsilon}^p - \boldsymbol{\varepsilon}^\theta
\end{equation*}
such that we can write \cref{eqn:psiexample_smalldef_withEpsE} as 
\begin{equation}
    \Psi(\boldsymbol{\varepsilon} - \boldsymbol{\varepsilon}^p, \theta, \boldsymbol{\alpha}) = \frac{1}{2\rho_0}(\boldsymbol{\varepsilon}-\boldsymbol{\varepsilon}^p):\mathbb{C}:(\boldsymbol{\varepsilon}-\boldsymbol{\varepsilon}^p)
    - \frac{\Delta\theta}{\rho_0}\boldsymbol{\beta}:(\boldsymbol{\varepsilon}-\boldsymbol{\varepsilon}^p) 
    + c_\varepsilon\left[\Delta\theta - \theta \text{ln}\left(\frac{\theta}{\theta_0}\right)\right] 
    + \frac{\xi}{2\rho_0}\boldsymbol{\alpha}:\boldsymbol{\alpha},
    \label{eqn:psiexample_smalldef}
\end{equation}
where $\boldsymbol{\beta} := \mathds{C}:\boldsymbol{\gamma}$, and the term in $(\Delta\theta)^2$ is neglected, the thermal contribution to $\Psi$ being considered in the third term on the right-hand side. Using \cref{eqn:constit_sig_smalldef,eqn:Stokes-helm-Ue,eqn:delPsi_delEpse}, we write the constitutive equation for the stress tensor as
\begin{equation}
    \boldsymbol{\sigma} = \mathds{C}:(\boldsymbol{\varepsilon} - \boldsymbol{\varepsilon}^p) - \Delta\theta\boldsymbol{\beta}:(\boldsymbol{\varepsilon} - \boldsymbol{\varepsilon}^p) = \mathds{C}:\boldsymbol{U}^e = \mathds{C}:\left[\nabla(\boldsymbol{u} - \boldsymbol{z}) + \boldsymbol{\zeta}\right],
    \label{eqn:sigma_small_def}
\end{equation}
where the tensors $\nabla\boldsymbol{z}$ and $\boldsymbol{\zeta}$ contains inelastic and thermal effects, as detailed in \ref{sec:decomposition_Ue}.

The constitutive expression for the dislocation velocity is given by \cref{eqn:constit_dislVel_smalldef,eqn:delPsi_delAlpha}:
\begin{equation}
    \boldsymbol{v}^d = \frac{1}{B}\left\{\left[\boldsymbol{\sigma} + \xi\left(\nabla\times\boldsymbol{\alpha}\right)^T\right]\boldsymbol{\alpha}\right\}:\textbf{X}.
    \label{eqn:disl_velocity_small_def}
\end{equation}

\vspace{-.7cm}

\begin{subequations}
    \def\myquad{\hskip13\fontdimen6\font}
    \def\myrulesize{0.9\textwidth}
    \begin{align}
        &\rule{\myrulesize}{0.4pt}\nonumber\\
        &\text{Kinematics and dislocation density evolution}\nonumber\\
        \label{eqn:setLin_Ue_dot}
        &\dot{\boldsymbol{U}}^e = \nabla\dot{\boldsymbol{u}} - \boldsymbol{\alpha}\times\boldsymbol{v}^d - \boldsymbol{\gamma}\dot\theta\\
        \label{eqn:setLin_alpha_dot}
        &\dot{\boldsymbol{\alpha}} = - \nabla\times\left( \boldsymbol{\alpha} \times \boldsymbol{v}^d + \boldsymbol{\gamma}\dot\theta \right)\\
        \label{eqn:setLin_V_dislocation}
        &\boldsymbol{v}^d = \frac{1}{B}\left\{\left[\boldsymbol{\sigma}  + \xi\left(\nabla\times\boldsymbol{\alpha}\right)^T\right]\boldsymbol{\alpha}\right\}:\textbf{X}\\
        &\rule{\myrulesize}{0.4pt}\nonumber\\
        &\text{Mechanical equilibrium}\nonumber\\
        \label{eqn:setLin_divsig}
        &\nabla\cdot\boldsymbol{\sigma} + \rho\boldsymbol{b}_f = \rho \ddot{\boldsymbol{u}}\\
        \label{eqn:setLin_sig_constitutive}
        &\boldsymbol{\sigma} = \mathds{C}:(\boldsymbol{\varepsilon} - \boldsymbol{\varepsilon}^p) - \Delta\theta\boldsymbol{\beta}:(\boldsymbol{\varepsilon} - \boldsymbol{\varepsilon}^p) \\
        &\rule{\myrulesize}{0.4pt}\nonumber\\
        &\text{Temperature evolution}\nonumber\\
        \label{eqn:setLin_temperature_evol}
        \begin{split}
            &\left[\left(\xi\nabla\times\boldsymbol{\alpha} + \theta\boldsymbol{\beta} \right):\boldsymbol{\gamma} + \rho c_\varepsilon\right]\dot\theta - \xi\boldsymbol{\alpha}:\nabla\times(\boldsymbol{\gamma}\dot\theta) - \xi\theta\nabla\times\left[\text{tr}\left(\nabla\times\big(\boldsymbol{\gamma}\dot\theta\big)\right)\mathds{1}\right]:\boldsymbol{\gamma} \\
    &\qquad= \boldsymbol{\sigma}:(\boldsymbol{\alpha}\times\boldsymbol{v}^d) - \theta\boldsymbol{\beta}:\dot{\boldsymbol{\varepsilon}}^e - \nabla\cdot\boldsymbol{q} + \rho r + \xi\boldsymbol{\alpha}:\nabla\times\left(\boldsymbol{\alpha}\times\boldsymbol{v}^d\right) \\
    &\qquad\quad + \xi\theta \nabla\times\left[\text{tr}\left(\nabla\times\left(\boldsymbol{\alpha}\times\boldsymbol{v}^d\right)\right)\mathds{1}\right]:\boldsymbol{\gamma}
        \end{split}\\ 
        \label{eqn:setLin_q_constitutive}
        &\boldsymbol{q} = -\boldsymbol{K}\nabla\theta.\\
        &\rule{\myrulesize}{0.4pt}\nonumber
    \end{align}
    \label{eqn:complete_eq_set_linearized}
\end{subequations}

\textcolor{black}{In \cite{berdichevskiiDynamicTheoryContinuously1967}, a thermomechanical theory including the mechanics of dislocations was proposed. The theory has some variational attributes but its governing equations are not the Euler-Lagrange equations of a given functional. A small deformation model of dislocation mechanics is defined (in section 6 of that work) but, to our knowledge, it has not been exercised to explore the capabilities of the model and whether basic dislocation thermomechanical behavior can be recovered within the setting (cf. \cite{2024_limachaves_upadhyay}). Focusing on essentials, the dislocation density tensor $\boldsymbol{S}$ in that work  satisfies, up to terms not depending on it (and some constants), an equation of the type (taken from equations 6.7, 12, 13, 14 in that work):
    \[
    curl\left( (\boldsymbol{M} - \boldsymbol{L}) \dot{\boldsymbol{S}} + \boldsymbol{E} \boldsymbol{S} \right)  + \boldsymbol{C} \boldsymbol{S} + \ldots = 0 
    \] 
where $\boldsymbol{M, L, E, C}$ are tensors of material constants. Clearly, the structure of these equations is  different from the dislocation transport equation of our theory, even when restricted to linear transport."}

\subsection{Comparison with the model proposed by \cite{upadhyay_thermomechanical_2020}}\label{sec:comparison_upadhyay}

In \cref{eqn:alpha_dot_smallstrains}, the term $\dot{\boldsymbol{\alpha}}^p = -\nabla\times(\boldsymbol{\alpha}\times\boldsymbol{v}^d)$ represents the evolution of the line-character of the thermomechanical defect and is directly associated with the evolution of the dislocation ensemble with velocity $\boldsymbol{v}^d$. On the other hand, $\boldsymbol{S}^\theta = -\nabla\times(\boldsymbol{\gamma}\dot\theta)$ contains the contribution of the incompatibility of the transient temperature field to the evolution of the defect character (Burgers vector). The evolution statement in \cref{eqn:alpha_dot_smallstrains} is similar to the one derived in \cite{upadhyay_thermomechanical_2020} (Eq. $3.25_3$). However, a notable difference lies in the equation for $\dot{\boldsymbol{\alpha}}^p$. In \cite{upadhyay_thermomechanical_2020}, $\boldsymbol{\alpha}^p$ was introduced as being the density of dislocation lines in the body, independent of the areal density $\boldsymbol{S}^\theta$ (defined there as $\boldsymbol{\alpha}^\theta$), and thus its evolution was given by $\dot{\boldsymbol{\alpha}}^p =-\nabla\times(\boldsymbol{\alpha}^p\times\boldsymbol{v}^d)$. In this work, in $\dot{\boldsymbol{\alpha}}^p$ the whole thermomechanical defect $\boldsymbol{\alpha}$ is transported with velocity $\boldsymbol{v}^d$, while $\boldsymbol{S}^\theta$ acts as a source term for $\boldsymbol{\alpha}$, i.e., there is no clear separation between a dislocation line and a \say{thermal} line-type defect. Such a description seems to be better suited to describe the dislocation density state in a body involving transient thermal gradients since an experimental observation of this state could not allow for a clear distinction of the contribution due to thermal effects.

In \cite{upadhyay_thermomechanical_2020}, the total strain tensor was additively decomposed into a sum of elastic, plastic, and thermal parts, as $\boldsymbol{\varepsilon} = \boldsymbol{\varepsilon}^e + \boldsymbol{\varepsilon}^p + \boldsymbol{\varepsilon}^\theta$, with $\boldsymbol{\varepsilon}^\theta = \boldsymbol{\gamma}\Delta\theta$. The Helmholtz free energy was taken as $\hat\Psi = \hat\Psi(\boldsymbol{\varepsilon}^e) \equiv \hat\Psi(\boldsymbol{\varepsilon} - \boldsymbol{\varepsilon}^p, \theta)$, such that the dependence of $\Psi$ on internal variables, such as $\boldsymbol{\alpha}$, was not studied. To establish a comparison, we consider the same expression for $\hat\Psi$, i.e.
\begin{equation}
    \hat\Psi(\boldsymbol{\varepsilon}-\boldsymbol{\varepsilon}^p, \theta) = \frac{1}{2\rho_0}(\boldsymbol{\varepsilon}-\boldsymbol{\varepsilon}^p):\mathbb{C}:(\boldsymbol{\varepsilon}-\boldsymbol{\varepsilon}^p)
    - \frac{\Delta\theta}{\rho_0}\boldsymbol{\beta}:(\boldsymbol{\varepsilon}-\boldsymbol{\varepsilon}^p) 
    + c_\varepsilon\left[\Delta\theta - \theta \text{ln}\left(\frac{\theta}{\theta_0}\right)\right],
    \label{eqn:Psi_example_comparison}
\end{equation}
such that the dissipation can be written from \cref{eqn:localsecondlaw2}
\begin{equation}
    \begin{aligned}
        D := \int_{\Omega_t} \left[ - \rho \left( \partial_{(\boldsymbol{\varepsilon}-\boldsymbol{\varepsilon}^p)}\hat\Psi:(\dot{\boldsymbol{\varepsilon}} - \dot{\boldsymbol{\varepsilon}}^p) + \partial_{\theta}\hat\Psi:\dot\theta + \eta \dot\theta \right) - \frac{1}{\theta} \boldsymbol{q} \cdot \nabla \theta + \boldsymbol{\sigma}:\boldsymbol{L} \right] \text{d}v \geq 0.
    \end{aligned}
    \label{eqn:global_D_comparison}
\end{equation}  
Noting that $\boldsymbol{\sigma}:\boldsymbol{L} = \boldsymbol{\sigma}:\dot{\boldsymbol{\varepsilon}}$, we arrive at the following constitutive relations based on \cref{eqn:global_D_comparison} 
\begin{subequations}
    \begin{align}
        \label{eqn:constit_sig_comp}
        &\boldsymbol{\sigma} = \rho\partial_{(\boldsymbol{\varepsilon}-\boldsymbol{\varepsilon}^p)}\hat\Psi\\
        \label{eqn:constit_eta_comp}
        &\eta = -\partial_\theta\hat\Psi\\
        \label{eqn:constit_v_comp}
        & \boldsymbol{v}^d = \frac{1}{B}(\boldsymbol{\sigma\alpha}):\mathbf{X}\\
        \label{eqn:constit_q_comp}
        &\boldsymbol{q} = -\boldsymbol{K}\nabla\theta,
    \end{align}
    \label{eqn:constitutive_eqs_comparison}
\end{subequations}
which corresponds to what was obtained in \cite{upadhyay_thermomechanical_2020}. Note that, in our model, we do not explicitly introduce a plastic distortion tensor. Instead, plasticity arises from the motion and generation of dislocations, which could be expressed in terms of a plastic (slip) distortion rate in the form $\dot{\boldsymbol{U}}^p = \boldsymbol{\alpha}\times\boldsymbol{v}^d$ from \cref{eqn:velGradient_smallStrain}. If we ignore the dislocation source term $\boldsymbol{S}$ we have that $\boldsymbol{\sigma}:\dot{\boldsymbol{U}}^p = \boldsymbol{\sigma}:\dot{\boldsymbol{\varepsilon}}^p = \boldsymbol{\sigma} : (\boldsymbol{\alpha}\times\boldsymbol{v}^d)$, which leads to the definition in \cref{eqn:constit_v_comp}.

From \cref{eqn:constit_eta_comp}, we have
\begin{equation*}
    \dot\eta = -\partial^2_{\theta(\boldsymbol{\varepsilon} - \boldsymbol{\varepsilon}^p)}\hat\Psi : (\dot{\boldsymbol{\varepsilon}} - \dot{\boldsymbol{\varepsilon}}^p) - \partial^2_{\theta\theta}\hat\Psi\dot\theta,
\end{equation*}
with which, by following a similar procedure from \cref{sec:temperature_evolution_LD}, the temperature evolution equation is obtained as 
\begin{equation}
    -\rho\theta\partial^2_{\theta\theta}\hat\Psi\dot\theta = -\nabla\cdot\boldsymbol{q} + \rho\theta\partial^2_{\theta(\boldsymbol{\varepsilon} - \boldsymbol{\varepsilon}^p)}\hat\Psi:(\dot{\boldsymbol{\varepsilon}} - \dot{\boldsymbol{\varepsilon}}^p) + \boldsymbol{\sigma}:\dot{\boldsymbol{\varepsilon}}^p + \rho r.
\end{equation}
Considering the expression in \cref{eqn:Psi_example_comparison}, we get 
\begin{equation}
    \rho c_\varepsilon\dot\theta = -\nabla\cdot\boldsymbol{q} - \theta\boldsymbol{\beta} : (\dot{\boldsymbol{\varepsilon}} - \dot{\boldsymbol{\varepsilon}}^p) + \boldsymbol{\sigma}:\dot{\boldsymbol{\varepsilon}}^p + \rho r.
\end{equation}
The equation set of the model in this case is
\begin{subequations}
    \begin{align}
        \label{eqn:setComp_Ue}
        &\boldsymbol{\varepsilon}^e = \boldsymbol{\varepsilon} - \boldsymbol{\varepsilon}^p - \boldsymbol{\varepsilon}^\theta  \\
        \label{eqn:setComp_alpha_dot}
        &\dot{\boldsymbol{\alpha}} = - \nabla\times\left( \boldsymbol{\alpha} \times \boldsymbol{v}^d + \boldsymbol{\gamma}\dot\theta \right)\\
        \label{eqn:setComp_V_dislocation}
        &\boldsymbol{v}^d = \frac{1}{B}\left(\boldsymbol{\sigma}\boldsymbol{\alpha}\right):\textbf{X}\\
        \label{eqn:setComp_divsig}
        &\nabla\cdot\boldsymbol{\sigma} = \rho \ddot{\boldsymbol{u}}\\
        \label{eqn:setComp_sig_constitutive}
        &\boldsymbol{\sigma} = \mathds{C}:(\boldsymbol{\varepsilon} - \boldsymbol{\varepsilon}^p) - \boldsymbol{\beta}\Delta\theta\\
        \label{eqn:setComp_temp_evol}
        &\rho c_\varepsilon\dot\theta = -\nabla\cdot\boldsymbol{q} - \theta\boldsymbol{\beta} : (\dot{\boldsymbol{\varepsilon}} - \dot{\boldsymbol{\varepsilon}}^p) + \boldsymbol{\sigma}:\dot{\boldsymbol{\varepsilon}}^p + \rho r,
    \end{align}
    \label{eqn:set_comparison}
\end{subequations}
and is completed with the boundary and initial conditions in \cref{sec:BC_smalldef,sec:IC_smalldef}. Hence, under a similar assumption of the form of $\Psi$, the present model reduces to the theory proposed in \cite{upadhyay_thermomechanical_2020}. The main difference stems from the additive decomposition of the strain tensor into elastic, plastic and thermal parts in the latter, leading to the introduction of the areal densities $\boldsymbol{\alpha}^p$ and $\boldsymbol{\alpha}^\theta$ which are independent of each other, as mentioned at the beginning of this section. From this definition, each density has its evolution statement, with $\dot{\boldsymbol{\alpha}}^\theta = -\nabla\times(\dot{\overline{\boldsymbol{\gamma}\theta}})$. In the present model, however, without considering a multiplicative decomposition of the deformation gradient, which lies in the introduction of thermal and plastic distortion tensors, the evolution of the thermomechanical defect density (\cref{eqn:setComp_alpha_dot}) comprises both thermal and dislocation line slip effects, without distinction, due to an indistinguishable contribution of thermal gradients and dislocation lines to the incompatibility in $\boldsymbol{W}$ (\cref{eqn:burgersdef}). 

\section{Conclusion}

In this work, a fully nonlinear model of field dislocations thermomechanics is proposed. The motivation behind its development arose from the need for a continuum framework capable of computing the evolution of dislocations based on a physical conservation argument under any thermomechanical boundary conditions while allowing for large deformations. It expands on previous work on the isothermal field dislocation mechanics model (\cite{ACHARYA2001761}, \cite{acharya2004constitutive}, \cite{ARORA2020_FEM_FDM}) and the small-strain thermal field dislocation mechanics with its numerical implementation (\cite{upadhyay_thermomechanical_2020}, \cite{2024_limachaves_upadhyay}).

Although not adopting the standard multiplicative decomposition of the deformation gradient into multiple components, as is customary in elastoplasticity models, the additive decomposition of the velocity gradient into elastic, plastic, and thermal parts is recovered, purely as a result of the conservation of the Burgers vector (\cite{acharya2015dislocation}), from which the evolution statement for the dislocation density tensor is also derived. Based on the standard conservation laws from continuum mechanics (mass, linear and angular momenta, and energy), the remaining governing equations are presented.

A central point of the model is the kinematical assumption of the contribution of a transient heterogeneous temperature field as a source term to the evolution of the thermomechanical defect character (which comprises dislocations and a line-type defect that arises through the incompatibility in thermal strains). Considering a Helmholtz free energy density $\Psi$ dependent on the inverse elastic distortion tensor $\boldsymbol{W}$, temperature $\theta$ and thermomechanical defect density $\boldsymbol{\alpha}$, it is shown that the resulting expression of the fraction of plastic work converted into heat (\cref{eqn:beta_rate}) is a function of the material type and strain rates. This fraction was introduced in the experimental work of \cite{taylorquinney_latent_1934}, and has been shown to depend on the material type and the loading conditions (\cite{rittel2017dependence}), with such dependence emerging in the present model as a consequence of the evolution of the thermomechanical defect density obtained from the conservation of the Burgers vector. Moreover, the temperature evolution equation is shown to allow for solutions in the form of dispersive waves with finite propagation speed, despite using Fourier's law of heat conduction as the constitutive assumption of the heat flux vector. Well-posedness of the equation is shown in a linearised setting. Along with the expected decay for a range of wavenumbers, the solution also admits well-posed growth, which can be the source of spatial patterning resulting from the growth of some Fourier components.

Considering $\Psi$ for a Saint-Venant-Kirchhoff material, the set of fully coupled equations of the model is presented. A linearisation of the model is also shown, which allows a comparison with the small deformation theory presented in \cite{upadhyay_thermomechanical_2020}. The main difference is shown to come from the additive decomposition of the strain tensor in that work into elastic, plastic, and thermal parts, which results in separate evolution statements for dislocations $\dot{\boldsymbol{\alpha}}^p$ and \say{thermal} defects $\dot{\boldsymbol{\alpha}}^\theta$. The present framework, however, by not explicitly introducing a plastic distortion tensor, upon linearization results in a single expression for the thermomechanical defect density $\dot{\boldsymbol{\alpha}}$ (\cref{eqn:setComp_alpha_dot}) that comprises both dislocation line evolution and thermal effects.

Prospective work includes the numerical implementation of the proposed model in a finite element framework to (i) study the model capabilities, (ii) verify the approach through comparison with experimentally-obtained conversion rates of plastic work into heat (e.g. in \cite{nieto-fuentes_dislocation-based_2018}, \cite{rittel2017dependence}) during deformation, and (iii) apply the theory to study dislocation thermomechanics under extreme processing conditions, as in the context of metal additive manufacturing.

\appendix

\section{Notation}\label{sec:notation}
Scalars are denoted with an italic font (e.g., $r$ or $\theta$). Vectors are denoted by a lowercase bold and italic Latin font (e.g., $\boldsymbol{q}$). Considering fixed {Cartesian reference frames}, the orthonormal basis vectors in $\Omega_r$ are denoted $\{\hat{\boldsymbol{e}}^r_I\}$, $I = 1,2,3$, while the orthonormal basis vectors in $\Omega_t$ are denoted $\{\hat{\boldsymbol{e}}_i\}$, $i = 1,2,3$. Lowercase (uppercase) indices refer to quantities in $\Omega_t$ ($\Omega_r$). Points in $\Omega_t$ ($\Omega_r$) are denoted $\boldsymbol{x}$ ($\boldsymbol{x}_r$). The second-order identity tensor is denoted $\mathds{1}$, whose components are $\delta_{ij}$ (the Kronecker delta). The third-order Levi-Civita permutation tensor is denoted $\mathbf{X}$, with components $e_{ijk}$ (the permutation symbol). Fourth-order tensors are denoted by double-stroke letters (e.g., $\mathds{C}$). The null tensor is denoted 0 for any tensor order. Summation of repeated indices is implied unless otherwise stated. Consider the vectors $\boldsymbol{u},\boldsymbol{v} \in \Omega_t$, as well as the tensors $\boldsymbol{\alpha},\boldsymbol{B} \in \Omega_t$. Then, we define the following operations:

\begin{multicols}{2}
    \noindent\textit{Tensor product}:\\
    \indent$\boldsymbol{u}\otimes\boldsymbol{v} = u_i v_j \, \hat{\boldsymbol{e}}_i \otimes \hat{\boldsymbol{e}}_j$
    
    \noindent\textit{Inner product}:\\
    \indent$\boldsymbol{u} \cdot \boldsymbol{v} = u_iv_i$\\
    \indent $\boldsymbol{\alpha}:\boldsymbol{B} = \alpha_{ij}B_{ij}$
    
    \noindent\textit{Cross product}:\\
    \indent$\boldsymbol{u} \times \boldsymbol{v} = e_{ijk} u_j v_k \, \hat{\boldsymbol{e}}_i$\\
    \indent $\boldsymbol{\alpha} \times \boldsymbol{u} = e_{jkl} \alpha_{ik} u_l \, \hat{\boldsymbol{e}}_i \otimes \hat{\boldsymbol{e}}_j$
    
    \noindent\textit{Tensor multiplication and action on vectors}:\\
    \indent$\boldsymbol{\alpha} \boldsymbol{B} = \alpha_{ij} B_{jk} \, \hat{\boldsymbol{e}}_i \otimes \hat{\boldsymbol{e}}_k$\\
    \indent $\boldsymbol{\alpha} \boldsymbol{u} = \alpha_{ij} u_{j} \, \hat{\boldsymbol{e}}_i$\\
    \indent $\boldsymbol{v} \boldsymbol{B} = v_{i} B_{ij} \, \hat{\boldsymbol{e}}_j$
    
    \noindent\textit{Double-dot product}:\\
    \indent $\mathbf{X} : \boldsymbol{\alpha} = e_{ijk} \alpha_{jk} \, \hat{\boldsymbol{e}}_i$\\
    \indent $\mathds{C} : \boldsymbol{\alpha} = \mathds{C}_{ijkl} \alpha_{kl} \, \hat{\boldsymbol{e}}_i \otimes \hat{\boldsymbol{e}}_j$
    
    \noindent\textit{Differential operators on $\Omega_t$} (comma indicates differentiation with respect to a given coordinate):\\
    \indent $\nabla \boldsymbol{u} = \text{grad}\, \boldsymbol{u} = u_{i,j}\, \hat{\boldsymbol{e}}_i \otimes \hat{\boldsymbol{e}}_j$\\
    \indent $\nabla \boldsymbol{\alpha} = \text{grad}\, \boldsymbol{\alpha} = \alpha_{ij,k}\, \hat{\boldsymbol{e}}_i \otimes \hat{\boldsymbol{e}}_j \otimes \hat{\boldsymbol{e}}_k$\\
    \indent $\nabla \cdot \boldsymbol{u} = \text{div}\, \boldsymbol{u} = u_{i,i}$\\
    \indent $\nabla \cdot \boldsymbol{\alpha} = \text{div}\, \boldsymbol{\alpha} = \alpha_{ij,j} \, \hat{\boldsymbol{e}}_i$ \\
    \indent $\nabla \times \boldsymbol{u} = \text{curl}\, \boldsymbol{u} = e_{ijk} u_{k,j}\, \hat{\boldsymbol{e}}_i$ \\
    \indent $\nabla \times \boldsymbol{\alpha} = \text{curl}\, \boldsymbol{\alpha} = e_{jkl} \alpha_{il,k}\, \hat{\boldsymbol{e}}_i \otimes \hat{\boldsymbol{e}}_j\, ,$
    
\end{multicols}

\noindent A \textit{two-point tensor} is defined as $\boldsymbol{W} = W_{Ij} \hat{\boldsymbol{e}}^r_I \otimes \hat{\boldsymbol{e}}_j$ or $\boldsymbol{F} = F_{iJ} \hat{\boldsymbol{e}}_i \otimes \hat{\boldsymbol{e}}^r_J$.

\noindent The \textit{material time derivative} in $\Omega_t$ is denoted by a superposed dot $\dot{\dottedsquare}$. $\text{det}(\dottedsquare)$ and $\text{tr}(\dottedsquare)$ indicate the \textit{determinant} and the \textit{trace} of a tensor, respectively. The \textit{symmetric} and \textit{skew-symmetric} parts of a tensor are denoted by $\text{sym}(\dottedsquare)$ and $\text{skew}(\dottedsquare)$, respectively. The Frobenius norm of a second-order tensor is denoted by $\lVert \boldsymbol{B} \rVert := (\boldsymbol{B}:\boldsymbol{B})^{1/2}$.

\section{Invariance requirements of $\boldsymbol{Y}$}\label{sec:invariance_Y}

Consider a point in $\Omega_t$ expressed as $\boldsymbol{x}(\boldsymbol{X}, t)$, with $\boldsymbol{X} \in \Omega_{ref}$, an arbitrarily fixed reference configuration. A rigid body motion of $\Omega_t$ can be expressed as 
\begin{equation}
    \boldsymbol{x}^*(\boldsymbol{x}, t) = \boldsymbol{Q}(t)\boldsymbol{x}(\boldsymbol{X}, t) + \boldsymbol{c}(t)
    \label{eqn:app_RBM}
\end{equation}
for any proper rotation tensor $\boldsymbol{Q}$ and translation vector $c$. 
Considering a rotated configuration $\Omega_t^*$, this allows us to write:
\begin{subequations}
    \begin{align}
        \label{eqn:Fe_RBM}
        &\boldsymbol{F}^{e*} = \boldsymbol{QF}^e\\
        \label{eqn:W_RBM}
        &\boldsymbol{W}^{*} = \boldsymbol{WQ}^T.
    \end{align}
\end{subequations}
For the transformation of $\boldsymbol{\alpha}$, we have the requirement that 
\begin{equation}
    \begin{split}
        \boldsymbol{\alpha}\hat{\boldsymbol{n}} = \boldsymbol{\alpha}^*\hat{\boldsymbol{n}}^* \quad \forall \hat{\boldsymbol{n}}, \hat{\boldsymbol{n}}^*,
    \end{split}
\end{equation}
with $\hat{\boldsymbol{n}}^* = \boldsymbol{Q}\hat{\boldsymbol{n}}$, which leads to
\begin{equation}
    \begin{split}
        \boldsymbol{\alpha}\hat{\boldsymbol{n}} = \boldsymbol{\alpha}^*\boldsymbol{Q}\hat{\boldsymbol{n}} \implies (\boldsymbol{\alpha} - \boldsymbol{\alpha}^*\boldsymbol{Q})\hat{\boldsymbol{n}} = 0\quad \forall \hat{\boldsymbol{n}} \iff \boldsymbol{\alpha}^* = \boldsymbol{\alpha Q}^T.
    \end{split}
    \label{eqn:alpha_star_RBM}
\end{equation}
Consistency with the evolution statement in \cref{eqn:Wrate} requires that 
\begin{equation}
    \begin{aligned}
        &\dot{\boldsymbol{W}}^* + \boldsymbol{W}^*\boldsymbol{L}^* = \boldsymbol{\alpha}^* \times \boldsymbol{v}^{d*} + \boldsymbol{Y}^*\dot\theta + \boldsymbol{S}^*\\
        \implies &\dot{\overline{\boldsymbol{WQ}^T}} + \boldsymbol{WQ}^T\left(\dot{\boldsymbol{Q}}\boldsymbol{Q}^T + \boldsymbol{Q}\boldsymbol{L}\boldsymbol{Q}^T\right) = (\boldsymbol{\alpha}\boldsymbol{Q}^T) \times \boldsymbol{v}^{d*} + \boldsymbol{Y}^*\dot\theta + \boldsymbol{S}^*\\
        \implies &\dot{\boldsymbol{W}}\boldsymbol{Q}^T + \boldsymbol{W}\dot{\boldsymbol{Q}}^T + \boldsymbol{WQ}^T\dot{\boldsymbol{Q}}\boldsymbol{Q}^T + \boldsymbol{W}\boldsymbol{L}\boldsymbol{Q}^T = (\boldsymbol{\alpha}\boldsymbol{Q}^T) \times \boldsymbol{v}^{d*} + \boldsymbol{Y}^*\dot\theta + \boldsymbol{S}^*\\
        \implies &\dot{\boldsymbol{W}} + \boldsymbol{W}\boldsymbol{L} + \boldsymbol{W}\dot{\boldsymbol{Q}}^T\boldsymbol{Q} + \boldsymbol{WQ}^T\dot{\boldsymbol{Q}} = [(\boldsymbol{\alpha}\boldsymbol{Q}^T) \times \boldsymbol{v}^{d*}]\boldsymbol{Q} + \boldsymbol{Y}^*\boldsymbol{Q}\dot\theta + \boldsymbol{S}^*\boldsymbol{Q}\\
        \implies &\dot{\boldsymbol{W}} + \boldsymbol{W}\boldsymbol{L} = [(\boldsymbol{\alpha}\boldsymbol{Q}^T) \times \boldsymbol{v}^{d*}]\boldsymbol{Q} + \boldsymbol{Y}^*\boldsymbol{Q}\dot\theta + \boldsymbol{S}^*\boldsymbol{Q}.
    \end{aligned}
    \label{eqn:wrate_star}
\end{equation}
Assuming $\boldsymbol{v}^{d*} = \boldsymbol{Qv}^d$, $\boldsymbol{Y}^* = \boldsymbol{YQ}^T$ and $\boldsymbol{S}^* = \boldsymbol{SQ}^T$, it can be shown that \cref{eqn:wrate_star} leads to
\begin{equation}
    \dot{\boldsymbol{W}} + \boldsymbol{W}\boldsymbol{L} = \boldsymbol{\alpha} \times \boldsymbol{v}^{d} + \boldsymbol{Y}\dot\theta + \boldsymbol{S}
\end{equation}
Hence, to comply with invariance requirements, under a rigid body motion $\boldsymbol{v}^d$ must transform as an objective vector, and $\boldsymbol{Y}$ and $\boldsymbol{S}$ must transform as second-order two-point tensors as in \cref{eqn:W_RBM} (\cite{acharya2004constitutive}). To satisfy this invariance requirement of $\boldsymbol{Y}$, the simplest candidate is 
\begin{equation}
    \boldsymbol{Y} = \boldsymbol{W}\boldsymbol{\gamma},
\end{equation}
with $\boldsymbol{\gamma}$ being a tensor of thermal expansion coefficients defined in $\Omega_t$ that transforms under a rigid body motion as $\boldsymbol{\gamma}^* = \boldsymbol{Q}\boldsymbol{\gamma}\boldsymbol{Q}^T$, which would give $\boldsymbol{Y}^* = \boldsymbol{W}^*\boldsymbol{\gamma}^* = \boldsymbol{W}\boldsymbol{Q}^T\boldsymbol{Q}\boldsymbol{\gamma}\boldsymbol{Q}^T = \boldsymbol{W\gamma Q}^T = \boldsymbol{Y}\boldsymbol{Q}^T$.

\section{Ericksen's identity}\label{sec:ericksen_equality}

The balance of angular momentum requires the symmetry of the Cauchy stress tensor $\boldsymbol{\sigma}$ (\cref{eqn:angularmomentum}). To analyze the consistency of the right-hand side of \cref{eqn:elasticlaw} as an expression for $\boldsymbol{\sigma}$, we require that $\Psi$ be invariant under any rigid body motion (\cite{acharya_fressengeas_2015}, \cite{ericksen_conservation_1961}), i.e.
\begin{equation}
    \Psi^*(\boldsymbol{W}^*,\theta^*,\boldsymbol{\alpha}^*) = \Psi(\boldsymbol{W},\theta,\boldsymbol{\alpha})
    \label{eqn:psiStar_eq_psi}
\end{equation}
for a motion given by \cref{eqn:app_RBM}.

Consider $\Psi$ in an arbitrarily fixed state $(\boldsymbol{W}, \theta, \boldsymbol{\alpha})$, at a given instant of time $t$, and a specific rigid body motion for which $\boldsymbol{Q}(t) = \mathds{1}$ and $\dot{\boldsymbol{Q}}(t) = \boldsymbol{P}$, where $\boldsymbol{P}$ is an arbitrarily fixed skew tensor. Next, noting that, under a rigid body motion, $\boldsymbol{W}$ and $\boldsymbol{\alpha}$ transform as in \cref{eqn:W_RBM} and \cref{eqn:alpha_star_RBM}, respectively, we compute the rate $\dot\Psi^*$ as
\begin{equation}
    \begin{split}
        \dot\Psi^* &= \partial_{\boldsymbol{W}^*}\Psi^* : \dot{\overline{\boldsymbol{W}\boldsymbol{Q}^T}} + \partial_{\theta^*}\Psi^*\dot\theta^* + \partial_{\boldsymbol{\alpha}^*}\Psi^*:\dot{\overline{\boldsymbol{\alpha}\boldsymbol{Q}^T}} \\
        &= \partial_{\boldsymbol{W}}\Psi : (\dot{\boldsymbol{W}} - \boldsymbol{W}\boldsymbol{P}) + \partial_{\theta}\Psi\dot\theta - \partial_{\boldsymbol{\alpha}}\Psi:(\dot{\boldsymbol{\alpha}} - \boldsymbol{\alpha}\boldsymbol{P}).
    \end{split}
    \label{eqn:psi_start_rate}
\end{equation}
\cref{eqn:psiStar_eq_psi} implies $\dot\Psi^* = \dot\Psi$, such that 
\begin{equation}
    \begin{split}
        &\big(\boldsymbol{W}^T\partial_{\boldsymbol{W}}\Psi + \boldsymbol{\alpha}^T\partial_{\boldsymbol{\alpha}}\Psi\big) : \boldsymbol{P} = 0 \\
        \implies &\frac{1}{2}\left[\boldsymbol{W}^T\partial_{\boldsymbol{W}}\Psi - (\partial_{\boldsymbol{W}}\Psi)^T\boldsymbol{W} + \boldsymbol{\alpha}^T\partial_{\boldsymbol{\alpha}}\Psi - (\partial_{\boldsymbol{\alpha}}\Psi)^T\boldsymbol{\alpha}\right] = 0
    \end{split}
    \label{eqn:skew_sig_eq_0}
\end{equation}
due to the arbitrariness of $\boldsymbol{P}$. The left-hand side of \cref{eqn:skew_sig_eq_0}$_2$ is equal to $-\text{skew}(\boldsymbol{\sigma}) = 0$, with $\boldsymbol{\sigma}$ given by \cref{eqn:elasticlaw}, which is thus shown to be symmetric. Hence, \cref{eqn:elasticlaw} is consistent with the balance of the angular momentum \cref{eqn:angularmomentum}, and also with the requirement of no dissipation due to material spin, given by $\text{skew}(\boldsymbol{L})$.

\section{Linear stability analysis of the temperature evolution equation}\label{sec:linear_stability}
\setcounter{figure}{0}

In this section, a 1-d, constant coefficient linear PDE is used to clarify the implications of the temperature evolution statement in \cref{eqn:temperature_evolution_2}. For that, we take into account the same temporal and spatial derivatives of $\theta$ present and consider
\begin{equation}
    a\theta_t + b\theta_{tx} + c\theta_{txx} = d\theta_{xx} + g\theta
    \label{eqn:1Dlinear_theta_equation}
\end{equation}
for $a,b,c,d, g\in\mathds{R}$, and assume $a,d \geq 0$. The subscripts $t$ and $x$ indicate partial differentiation with respect to time and space, respectively, and the last term on the right-hand side incorporates the presence of source terms in \cref{eqn:temperature_evolution_2} that depend linearly on $\theta$.  We take the ansatz of a plane-wave solution 
\begin{equation}
    \theta = \exp\big(i(kx + \omega t)\big)
    \label{eqn:theta_ansatz}
\end{equation}
considering $k \in \mathds{R}^+$ and insert it into \cref{eqn:1Dlinear_theta_equation} to get 
\begin{equation}
    \begin{split}
        &ia\omega - b\omega k - ic\omega k^2 = -dk^2 + g\\
        \implies& \omega(k) = \frac{dk^2 - g}{i(ck^2 - a) + bk } \frac{\big[bk - i(ck^2 - a)\big]}{\big[bk - i(ck^2 - a)\big]}\\
        \implies& \omega(k) = \frac{(bdk^3 - bgk) - i(cdk^4 - cgk^2 - adk^2 + ag)}{(ck^2 - a)^2 + b^2k^2}
    \end{split}
    \label{eqn:wave_dispersion_1Dtheta}
\end{equation}
which corresponds to the dispersion relation of the plane wave in \cref{eqn:theta_ansatz}. Denoting $\omega = \omega_R + i\omega_I$, with $\omega_R$ and $\omega_I$ the real and complex parts of $\omega$, respectively, \cref{eqn:theta_ansatz} becomes
\begin{equation}
    \theta = \exp(-\omega_I t)\exp\big(i(kx + \omega_R t)\big),
    \label{eqn:theta_wr_wi}
\end{equation}
hence showing that the stability of the solution is dependent on the behaviour of $\omega_I(k)$. Consider an initial condition as the superposition of waves with different $k$:
\begin{equation}
    \theta_0(x) = \sum_k A_k \exp(ikx).
\end{equation}
Small perturbations in the initial condition, $\delta\theta_0(x) = \sum_k \delta A_k \exp(ikx)$, may contain components with arbitrarily large $k$, such that, after some time $t$, the perturbed solution would be 
\begin{equation}
    \theta(x,t) + \delta\theta(x,t) = \sum_k (A_k + \delta A_k)\exp(-\omega_I t)\exp\big(i(kx + \omega_R t)\big).
    \label{eqn:perturbed_theta}
\end{equation}
If $\omega_I(k)\to -\infty$ when $k\to\infty$, then the arbitrarily large $k$ in $\delta\theta_0$ would result in unbounded growth of \cref{eqn:perturbed_theta} for arbitrarily small $t$, such that continuous dependence on initial data is not verified and thus the problem is ill-posed. More generally, for a given $M \gg 1$ and $\epsilon \ll 1$ arbitrarily fixed, ill-posed growth implies $\exp(|\omega(k)|\epsilon) \geq M$ for some $k$. In that case, $\omega(k)$ is an unbounded function of $k$. Hence, we define \say{well-posed growth} as growth in the solution (\cref{eqn:theta_wr_wi}) with time that does not attain arbitrary magnitudes in arbitrarily small time intervals with $\omega(k)$ thus being bounded as a function of $k$. To assess the boundedness of $\omega_I(k)$, we evaluate it on the limits $k \to \infty$ and $k\to 0$; for $k \in (0, \infty)$, we solve $\omega_I'(k) = 0$ to determine the critical values $k^*$, and show that $\omega(k^*)$ is bounded.

\paragraph{Case $d = 0$} From \cref{eqn:wave_dispersion_1Dtheta}, we have that 
\begin{equation}
    \begin{split}
        \omega(k) &= \frac{- bgk + i(cgk^2 - ag)}{(ck^2 - a)^2 + b^2k^2}
    \end{split}
    \label{eqn:wave_dispersion_adiabatic}
\end{equation}
which corresponds to the dispersion relation in the adiabatic case. From \cref{eqn:wave_dispersion_adiabatic}, we see that the presence of the term in $\theta_{tx}$ in \cref{eqn:1Dlinear_theta_equation} leads to a wave-like solution for temperature evolution, with a non-zero real part of $\omega$. In this case, from \cref{eqn:wave_dispersion_adiabatic} we have that 
\begin{equation}
    \quad \lim_{k\to    \infty} \omega(k) = 0, \quad \lim_{k\to 0} \omega(k) = -i\frac{g}{a},
    \label{eqn:omega_I_limits_adiabatic}
\end{equation}
which shows the boundedness of $\omega_I(k)$, thus guaranteeing well-posed growth for the limiting values of $k$. Now, taking the derivative of \cref{eqn:wave_dispersion_adiabatic}, we have 
\begin{equation}
    \omega_I'(k) = \frac{2 g k \left[c \left(b^{2} k^{2} + \left(a - c k^{2}\right)^{2}\right) + \left(a - c k^{2}\right) \left(b^{2} - 2 c \left(a - c k^{2}\right)\right)\right]}{\left[b^{2} k^{2} + \left(a - c k^{2}\right)^{2}\right]^{2}}.
\end{equation}
By solving $\omega_I'(k) = 0$, we find each $k^*_i$, $i=1,..,5$, where $\omega_I(k)$ attains a maximum or a minimum, which gives
\begin{equation}
    \begin{split}
        &k^*_1 = 0;\quad k^*_2 = - \sqrt{\frac{ac^2 - b \sqrt{ac^{3}}}{c^3}}; \quad k^*_3 = -k^*_2;\quad k^*_4 = - \sqrt{\frac{ac^2 + b \sqrt{ac^{3}}}{c^3}};\quad k^*_5 = -k^*_4\\
        &\omega_I\left(k^*_1\right) = -\frac{g}{a}; \quad \omega_I\left(k^*_2\right) = \frac{gc^{5/2}\sqrt{a}}{b \left(b\sqrt{ac^{3}} - 2 a c^{2}\right)}; \quad \omega_I\left(k^*_3\right) = \omega_I\left(k^*_2\right);\\&\omega_I\left(k^*_4\right) = \frac{gc^{5/2}\sqrt{a}}{b \left(b\sqrt{ac^{3}} + 2 a c^{2}\right)};\quad \omega_I\left(k^*_5\right) = \omega_I\left(k^*_4\right).
    \end{split}
    \label{eqn:omega_I_critical_adiabatic}
\end{equation}
Hence, since all the extrema of $\omega_I$ are bounded, we conclude that the solution admits well-posed growth or decay. The values of $k$ that will result in the growth or decay of the solution depend on the sign of the coefficients $g$ and $c$ and can be analysed from \cref{eqn:wave_dispersion_adiabatic} by solving
\begin{equation}
    \omega_I = \frac{cgk^2 - ag}{(ck^2 - a)^2 + b^2k^2} >0 \implies g(ck^2 - a) > 0.
\end{equation}
From this equation, the \textit{decay} of the solution occurs for
\begin{equation}
    \begin{split}
        g > 0 &,\, c > 0,\, k > \sqrt{\frac{a}{c} }\\
        g < 0 &, 
            \begin{cases}
                c < 0,\, k \in \mathds{R}^+,\\
                c > 0,\, k < \sqrt{\frac{a}{c}}\\
            \end{cases},    
    \end{split}
\end{equation}
whereas \textit{well-posed growth} occurs for
\begin{equation}
    \begin{split}
        g > 0 &, 
            \begin{cases}
                c < 0,\, k\in\mathds{R}^+,\\
                c > 0,\, k < \sqrt{\frac{a}{c} } \\
            \end{cases}\\
        g < 0 &,\,c > 0 ,\, k > \sqrt{\frac{a}{c}}.
    \end{split}
\end{equation}
We highlight the fact that, considering the solution as a superposition of plane waves with different $k$, the wave components whose $k$ lie in a growth region could give rise to a spatial patterning of the temperature profile. 

The phase velocity of the adiabatic temperature wave is given by 
\begin{equation}
    v_p(k) = \frac{\omega_R}{k} = \frac{- bg}{(ck^2 - a)^2 + b^2k^2}
\end{equation}
and is a function of the wavenumber $k$, such that the solution for $\theta$ is in the form of dispersive waves.

\paragraph{Case $d \neq 0$} From \cref{eqn:wave_dispersion_1Dtheta}, we have 
\begin{equation}
    \omega_R(k) = \frac{bdk^3 - bgk }{(ck^2 - a)^2 + b^2k^2} \quad \text{and} \quad \omega_I(k) = \frac{(a - ck^2)(dk^2 - g)}{(ck^2 - a)^2 + b^2k^2}.
    \label{eqn:wr_wi_general_case}
\end{equation}
Similar to the previous case, we are interested in analysing the boundedness of $\omega_I(k)$, and we have
\begin{equation}
    \lim_{k\to -\infty} \omega_I(k) = -\frac{d}{c}, \quad \lim_{k\to 0} \omega_I(k) = -\frac{g}{a},
\end{equation}
which once again ensures well-posed growth of the solution for the limiting values of $k$. Deriving $\omega_I$ in \cref{eqn:wr_wi_general_case} with respect to $k$ gives
\begin{equation}
    \omega_I'(k) = \frac{2 k \left[\left(c k^{2} - a\right) \left(b^{2} - 2 c \left(a - c k^{2}\right)\right) \left(d k^{2} - g\right) + \left(b^{2} k^{2} + \left(a - c k^{2}\right)^{2}\right) \left(c \left(g - d k^{2}\right) + d \left(a - c k^{2}\right)\right)\right]}{\left[\left(a - c k^{2}\right)^{2} + b^{2} k^{2}\right]^{2}}
\end{equation}
As before, solving $\omega_I'(k) = 0$ for $k$ yields five critical points $k^*_i$, which results in bounded extrema $\omega_I\left(k^*_i\right)$, $i = 1,...,5$, therefore also implying well-posed growth or decay of the solution (the solutions are not shown here due to their considerable size). From \cref{eqn:wr_wi_general_case}, the conditions for growth or decay can be established by solving
\begin{equation}
    (ck^2 - a)(dk^2 - g) < 0,
\end{equation}
from which the \textit{decay} of the solution occurs for 
\begin{equation}
    \begin{split}
        g > 0 &,\,
            \begin{cases}
                c < 0 ,\, k > \sqrt{\frac{g}{d}}\\
                c > 0 ,\, \begin{cases}
                        \sqrt{\frac{g}{d}} < k < \sqrt{\frac{a}{c}}\quad \text{if} \quad \frac{a}{c} > \frac{g}{d}\\
                        \sqrt{\frac{a}{c}} < k < \sqrt{\frac{g}{d}}\quad \text{if} \quad \frac{a}{c} < \frac{g}{d}\\
                    \end{cases}
            \end{cases}\\
        g < 0 &,\,
            \begin{cases}
                c < 0 ,\, k \in \mathds{R}^+\\
                c > 0 ,\, k > \sqrt{\frac{g}{d}}\\
            \end{cases}    
    \end{split}
\end{equation}
whereas \textit{well-posed growth} occurs for
\begin{equation}
    \begin{split}
        g > 0 &,\, 
            \begin{cases}
                c < 0 ,\, k < \sqrt{\frac{g}{d}}\\
                c > 0 ,\, \begin{cases}
                        0 < k < \sqrt{\frac{g}{d}} \quad \text{or} \quad  k > \sqrt{\frac{a}{c}}\quad \text{if} \quad \frac{a}{c} > \frac{g}{d}\\
                        0 < k < \sqrt{\frac{a}{c}} \quad \text{or} \quad k > \sqrt{\frac{g}{d}}\quad \text{if} \quad \frac{a}{c} < \frac{g}{d}\\
                    \end{cases}
            \end{cases}\\
        g < 0 &,\, c > 0 ,\, k < \sqrt{\frac{g}{d}}
    \end{split}
\end{equation}
The phase velocity in this case is 
\begin{equation}
    v_p(k) = \frac{\omega_R}{k} = \frac{b(dk^2 - g)}{(ck^2 - a)^2 + b^2k^2},
\end{equation}
and the solution is again in the form of dispersive waves.

\section{The Stokes-Helmholtz decomposition of \texorpdfstring{$\boldsymbol{W}$}{The Stokes-Helmholtz decomposition of the inverse elastic distortion tensor}}

\subsection{Large deformations}\label{sec:decompose_W}

It is sometimes convenient to adopt a Stokes-Helmholtz decomposition of $\boldsymbol{W}$ into incompatible (i.e., divergence-free) and compatible (i.e., curl-free) parts for solving problems, e.g. in dislocation statics, the equations of equilibrium and the incompatibility equation pose 12 equations in 9 variables, while having solutions (unique, in the linear case) despite its overdetermined appearance. Thus, we use

\begin{equation}
    \boldsymbol{W} = \boldsymbol{\chi} + \nabla\boldsymbol{f}
    \label{eqn:W_decomp_1}
\end{equation}
where $\boldsymbol{f}$ is to be considered as the plastic position vector (\cite{acharya_size_2006}). 
Now, given $\boldsymbol{\alpha}$, the following equation set allows for the unique determination of $\boldsymbol{\chi}$:

\begin{subequations}
    \begin{alignat}{2}
        \nabla\times\boldsymbol{\chi} &= -\boldsymbol{\alpha} \quad &&\text{in} \quad \Omega_t\\
        \nabla\cdot\boldsymbol{\chi} &= 0 \quad &&\text{in} \quad \Omega_t\\
        \boldsymbol{\chi}\hat{\boldsymbol{n}} &= 0 \quad &&\text{on} \quad \partial\Omega_t
    \end{alignat}
    \label{eqn:chi_given_alpha_LD}
\end{subequations}
Note that \cref{eqn:chi_given_alpha_LD} also ensures that $\boldsymbol{\chi} = 0$ whenever $\boldsymbol{\alpha} = 0$.

To compute the rate of the inverse elastic distortion gradient tensor $\dot{\boldsymbol{W}}$, we perform the analysis in a \say{relative} description (\cite{acharya2004constitutive}), in which we fix the body in a given configuration $\Omega_{t'}$ at time $t = t'$, and consider a motion from this configuration onwards, parametrised by a time-like variable $\tau$. By denoting $\boldsymbol{x}(t')$ the points in $\Omega_{t'}$, we have that $\boldsymbol{x}_\tau(\tau = 0) = \boldsymbol{x}(t')$, $\boldsymbol{x}_\tau \in \Omega_\tau$. Let $\boldsymbol{F}_\tau$ be the deformation gradient associated with this motion, and $\nabla_{\tau}$ the nabla operator in $\Omega_\tau$. Then, we can rewrite the decomposition in \cref{eqn:W_decomp_1} as

\begin{equation}
    \begin{aligned}
    &\boldsymbol{W} = \boldsymbol{\chi} + (\nabla_{\tau}\boldsymbol{f}) \boldsymbol{F}^{-1}_{\tau}  
    \implies \boldsymbol{W}\boldsymbol{F}_{\tau} = \boldsymbol{\chi}\boldsymbol{F}_{\tau} + (\nabla_{\tau}\boldsymbol{f}) \\
    \implies& \dot{\boldsymbol{W}}\boldsymbol{F}_{\tau} + \boldsymbol{W}\dot{\boldsymbol{F}_{\tau}} = \dot{\overline{\boldsymbol{\chi}\boldsymbol{F}_{\tau}}} + (\nabla_{\tau}\dot{\boldsymbol{f}}) 
    \implies \dot{\boldsymbol{W}} + \boldsymbol{W}\dot{\boldsymbol{F}_{\tau}}\boldsymbol{F}^{-1}_{\tau} = \dot{\overline{\boldsymbol{\chi}\boldsymbol{F}_{\tau}}}\boldsymbol{F}^{-1}_{\tau} + (\nabla_{\tau}\dot{\boldsymbol{f}})\boldsymbol{F}^{-1}_{\tau} \\
    \implies&\dot{\boldsymbol{W}} = \dot{\overline{\boldsymbol{\chi}\boldsymbol{F}_{\tau}}}\boldsymbol{F}^{-1}_{\tau} + (\nabla_{\tau}\dot{\boldsymbol{f}})\boldsymbol{F}^{-1}_{\tau} - \boldsymbol{WL}.
    \end{aligned}
    \label{eqn:W_decomp_dot}
\end{equation}
Now, we can write 
\begin{equation}
    \dot{\overline{\boldsymbol{\chi}\boldsymbol{F}_{\tau}}} = \dot{\boldsymbol{\chi}}\boldsymbol{F}_\tau + \boldsymbol{\chi}\dot{\boldsymbol{F}}_\tau \implies \dot{\overline{\boldsymbol{\chi}\boldsymbol{F}_{\tau}}}\boldsymbol{F}_\tau^{-1} = \dot{\boldsymbol{\chi}} + \boldsymbol{\chi}\boldsymbol{L}
    \label{eqn:dotChiFtau}
\end{equation}
such that \cref{eqn:W_decomp_dot} becomes

\begin{equation}
    \begin{aligned}
        \dot{\boldsymbol{W}} = \dot{\boldsymbol{\chi}} + \boldsymbol{\chi}\boldsymbol{L} + (\nabla_{\tau}\dot{\boldsymbol{f}})\boldsymbol{F}^{-1}_{\tau} - \boldsymbol{WL}
    \end{aligned}
    \label{eqn:dotW_chiFtau_replaced}
\end{equation}
At $\tau = 0$, \cref{eqn:dotW_chiFtau_replaced} evaluates to 

\begin{equation}
    \dot{\boldsymbol{W}} = \dot{\boldsymbol{\chi}} + \boldsymbol{\chi}\boldsymbol{L} + \nabla\dot{\boldsymbol{f}} - \boldsymbol{WL}.
    \label{eqn:dotW_final}
\end{equation}
which remains valid for any $t$, since the choice of $t'$ is arbitrary.

As established in \cite{acharya2004constitutive}, $\nabla\boldsymbol{f}$ is related to the permanent deformation of the body. Thus, we would like $\nabla\dot{\boldsymbol{f}}$ to bear a dependence on the general defect evolution in the body, given by $\boldsymbol{\alpha}\times\boldsymbol{v}^d + \boldsymbol{\phi}^\theta$ (\cref{eqn:transport}). Considering \cref{eqn:Wrate}, we can write \cref{eqn:dotW_final} as

\begin{equation}
    \begin{aligned}
        &\dot{\boldsymbol{W}} + \boldsymbol{WL} = \dot{\boldsymbol{\chi}} + \boldsymbol{\chi}\boldsymbol{L} + \nabla\dot{\boldsymbol{f}} = \boldsymbol{\alpha}\times\boldsymbol{v}^d + \boldsymbol{\phi}^\theta  \\
        \implies& \nabla\cdot\nabla\dot{\boldsymbol{f}} = \nabla\cdot(\boldsymbol{\alpha}\times\boldsymbol{v}^d + \boldsymbol{\phi}^\theta - \dot{\boldsymbol{\chi}} - \boldsymbol{\chi}\boldsymbol{L}) \quad \text{in}\quad \Omega_t
    \end{aligned}\label{eqn:introducing_divgradDotf}
\end{equation}
To obtain a unique solution for $\dot{\boldsymbol{f}}$, \cref{eqn:introducing_divgradDotf} requires the following boundary condition

\begin{equation}
    (\nabla\dot{\boldsymbol{f}})\hat{\boldsymbol{n}} = (\boldsymbol{\alpha}\times\boldsymbol{v}^d + \boldsymbol{\phi}^\theta - \dot{\boldsymbol{\chi}} - \boldsymbol{\chi}\boldsymbol{L})\hat{\boldsymbol{n}}\quad\text{on}\quad \partial\Omega_t
\end{equation}
where $\hat{\boldsymbol{n}}$ is the outward normal field to $\partial\Omega_t$. It can be shown that the evolution statement in \cref{eqn:introducing_divgradDotf} is compatible with dissipation requirements imposed by the second law of thermodynamics (\cite{acharya2004constitutive}, \cite{acharya_microcanonical_2011}).

\subsection{Small deformations}\label{sec:decomposition_Ue}

Considering \cref{eqn:additiv_decomp_grad_u}, we write

\begin{equation}
    \boldsymbol{U}^e = \nabla\boldsymbol{u} - \boldsymbol{U}^p - \boldsymbol{\varepsilon}^\theta.
    \label{eqn:Ue_gradu_Up_epstheta}
\end{equation}
By denoting $\boldsymbol{A}^\parallel$ the compatible part of a tensor $\boldsymbol{A}$, we have that

\begin{equation}
    \boldsymbol{U}^{e\parallel} = \nabla\boldsymbol{u} - \boldsymbol{U}^{p\parallel} - \boldsymbol{\varepsilon}^{\theta\parallel}.
    \label{eqn:UeParallel_gradu_Up_epstheta}
\end{equation}
Hence, we decompose $\boldsymbol{U}^e$, $\boldsymbol{U}^p$ and $\boldsymbol{\varepsilon}^\theta$ into compatible and incompatible parts as

\begin{equation}
    \boldsymbol{U}^e = \nabla(\boldsymbol{u} - \boldsymbol{z}) + \zeta;\quad \boldsymbol{U}^p = \nabla\boldsymbol{z}^p + \zeta^p;\quad \boldsymbol{\varepsilon}^\theta = \nabla\boldsymbol{z}^\theta + \zeta^\theta,
    \label{eqn:Stokes-helm-Ue}
\end{equation}
such that, considering \cref{eqn:Ue_gradu_Up_epstheta,eqn:UeParallel_gradu_Up_epstheta}, we have the following:

\begin{align}
    &\nabla\boldsymbol{z} = \nabla\boldsymbol{z}^p + \nabla\boldsymbol{z}^\theta\\
    &\boldsymbol{\zeta} = -\boldsymbol{\zeta}^p -\boldsymbol{\zeta}^\theta.
\end{align}
The vector $\boldsymbol{z}$ is the \say{plastic displacement} (\cite{acharya_size_2006}) and  $\boldsymbol{\zeta}$ is a divergence-free tensor, determined by solving the system 

\begin{equation}
    \def\myquad{\hskip7.6\fontdimen5\font}
    \begin{aligned}
        \left.\begin{aligned}
            &\nabla\times\boldsymbol{\zeta} = \boldsymbol{\alpha} \\
            &\nabla\cdot\boldsymbol{\zeta} = 0
        \end{aligned}\right\rbrace\quad&\text{in}\quad\Omega\\
        \boldsymbol{\zeta}\hat{\boldsymbol{n}} = 0 \myquad &\text{on}\quad \partial\Omega,    
    \end{aligned}
\end{equation}
similarly to \cref{eqn:chi_given_alpha_LD}. In the small strains approximation, we have that $\boldsymbol{L} = \nabla\boldsymbol{v} = \nabla\dot{\boldsymbol{u}}$, such that, by using \cref{eqn:velGradient_smallStrain,eqn:Stokes-helm-Ue}, we can write

\begin{equation}
    \begin{aligned}
        &\nabla\dot{\boldsymbol{u}} = \nabla(\dot{\boldsymbol{u}} - \dot{\boldsymbol{z}}) + \dot{\boldsymbol{\zeta}} + \boldsymbol{\alpha}\times\boldsymbol{v}^d + \boldsymbol{\gamma}\dot\theta\\
        \implies& \nabla\cdot\nabla\dot{\boldsymbol{z}} = \nabla\cdot(\boldsymbol{\alpha}\times\boldsymbol{v}^d 
        + \boldsymbol{\gamma}\dot\theta) \quad \text{in}\quad \Omega,
    \end{aligned}
    \label{eqn:divGradDotZ_smallstrains}
\end{equation}
since, in the geometrically linear case, $\dot{\boldsymbol{\zeta}}$ is incompatible (i.e., $\nabla\cdot\dot{\boldsymbol{\zeta}} = 0$). \cref{eqn:divGradDotZ_smallstrains} also requires the following boundary condition to ensure the uniqueness of the solution:

\begin{equation}
    (\nabla\dot{\boldsymbol{z}})\hat{\boldsymbol{n}} = ( \boldsymbol{\alpha}\times\boldsymbol{v}^d + \boldsymbol{\gamma}\dot\theta)\hat{\boldsymbol{n}} \quad \text{on} \quad \partial\Omega.
\end{equation}

\section{Derivation of global dissipation rate $D$}
\label{app:derivation_dissipation}

The following identities are used in the derivation:

\begin{equation}
    \partial_{\boldsymbol{W}} \Psi : (\boldsymbol{W}\boldsymbol{L}) = \frac{\partial\Psi}{\partial W_{mn}} W_{mi} L_{in} = W_{mi} \frac{\partial\Psi}{\partial W_{mn}} L_{in} = \left(\boldsymbol{W}^T  \partial_{\boldsymbol{W}}\Psi\right) : \boldsymbol{L};
    \label{eqn:appB_delW_delPsi_W_L}
\end{equation}

\begin{equation}
    \partial_{\boldsymbol{\alpha}} \Psi : \left(\boldsymbol{\alpha}\boldsymbol{L}^T\right) = \frac{\partial\Psi}{\partial \alpha_{ij}} \alpha_{ip} L_{jp} =  \left[(\partial_{\boldsymbol{\alpha}}\Psi)^T \boldsymbol{\alpha}\right] : \boldsymbol{L};
    \label{eqn:appB_delAlpha_delPsi_alpha_L^T}
\end{equation}
\vspace{-.2cm}
\begin{equation}
    \begin{split}
        \int_\Omega \boldsymbol{A} : \nabla\times\boldsymbol{B}\, \text{d}v &= \int_\Omega A_{ij} \epsilon_{jkl} B_{il,k}\, \text{d}v = \int_\Omega (A_{ij} \epsilon_{jkl} B_{il})_{,k}\, \text{d}v - \int_\Omega A_{ij,k} \epsilon_{jkl} B_{il}\, \text{d}v \\
        &= -\int_{\partial\Omega} A_{ij} \epsilon_{jlk} B_{il}\hat{n}_k\, \text{d}s + \int_\Omega \epsilon_{lkj} A_{ij,k} B_{il}\, \text{d}v 
        = -\int_{\partial\Omega} \boldsymbol{A}:(\boldsymbol{B}\times\hat{\boldsymbol{n}})\, \text{d}s + \int_\Omega (\nabla\times\boldsymbol{A}):\boldsymbol{B}\, \text{d}v; 
    \end{split}
    \label{eqn:appB_int_A_curlB}
\end{equation}

\begin{equation}
    \boldsymbol{A} : \left(\boldsymbol{\alpha} \times \boldsymbol{v}^d\right) = A_{ij} \epsilon_{jkl}\alpha_{ik}v^d_l = A_{ij} \alpha_{ik} \epsilon_{jkl} v^d_l  = \big[(\boldsymbol{A}^T\boldsymbol{\alpha}) : \textbf{X}\big]\cdot \boldsymbol{v}^d.
    \label{eqn:appB_A_alphaTimesV}
\end{equation}

For completeness, we recall the global dissipation inequality \cref{eqn:D_after_replacing_dotf}:

\begin{equation}
    \begin{aligned}
        D = &\int_{\Omega_t} \left(\boldsymbol{\sigma} : \boldsymbol{L} - \frac{1}{\theta}\boldsymbol{q}\cdot\nabla\theta\right)\, \text{d}v + \int_{\Omega_t} \rho\boldsymbol{W}^T\partial_{\boldsymbol{W}}\Psi : \boldsymbol{L}\, \text{d}v - \int_{\Omega_t} \rho\partial_{\boldsymbol{W}}\Psi : \left(\boldsymbol{\alpha}\times\boldsymbol{v}^d + \boldsymbol{W\gamma}\dot\theta \right)\, \text{d}v \\
        &- \int_{\Omega_t} \rho\partial_{\boldsymbol{\alpha}}\Psi : \left[-\text{tr}(\boldsymbol{L})\boldsymbol{\alpha} + \boldsymbol{\alpha L}^T - \nabla\times\left( \boldsymbol{\alpha} \times \boldsymbol{v}^d + \boldsymbol{W\gamma}\dot\theta \right)\right]\, \text{d}v - \int_{\Omega_t}\rho \big(\partial_\theta\Psi + \eta \big)\dot\theta\, \text{d}v \geq 0,
    \end{aligned}
    \label{eqn:appB_dissip_1}
\end{equation}

After separating terms, we get

\begin{equation}
    \begin{split}
        D = \int_{\Omega_t}\Big[
            &\rho\boldsymbol{W}^T\partial_{\boldsymbol{W}}\Psi:\boldsymbol{W}\boldsymbol{L} - \rho\boldsymbol{W}^T\partial_{\boldsymbol{W}}\Psi:(\boldsymbol{\alpha}\times\boldsymbol{v}^d) - \rho\boldsymbol{W}^T\partial_{\boldsymbol{W}}\Psi : \boldsymbol{W\gamma}\dot\theta - \rho\partial_\theta\Psi\dot\theta \\ 
            &-\rho\partial_{\boldsymbol{\alpha}}\Psi : (\boldsymbol{\alpha}\boldsymbol{L}^T) + \rho\text{tr}(\boldsymbol{L})\partial_{\boldsymbol{\alpha}}\Psi : \boldsymbol{\alpha} + \rho\partial_{\boldsymbol{\alpha}}\Psi : \nabla\times(\boldsymbol{\alpha} \times \boldsymbol{v}^d + \boldsymbol{W\gamma}\dot{\theta}) - \rho\eta\dot\theta
        \Big] \, \text{d}v \\
        + \int_{\Omega_t} &\left(\boldsymbol{\sigma}:\boldsymbol{L} -\frac{1}{\theta} \boldsymbol{q}\cdot\nabla\theta\right)\, \text{d}v \geq 0.
    \end{split}
    \label{eqn:appB_dissip_2}
\end{equation}

By using \cref{eqn:appB_delW_delPsi_W_L,eqn:appB_delAlpha_delPsi_alpha_L^T,eqn:appB_int_A_curlB,eqn:appB_A_alphaTimesV} in \cref{eqn:appB_dissip_2}, and regrouping the terms in $\boldsymbol{L}$, $\dot\theta$, $\boldsymbol{v}^d$ and $\boldsymbol{S}$, we obtain \cref{eqn:dissipation}.

\section{Evaluation of the derivatives of $\Psi$ for a Saint-Venant-Kirchhoff material}\label{app:derivatives_psi}

\subsection{Large deformations}\label{sec:PsiDerivativesLargeDef}

We make use of the following identity:

\begin{equation}
    \frac{\partial W^{-1}_{il}}{\partial W_{mn}} = -W^{-1}_{im}W^{-1}_{nl} \big(\hat{\boldsymbol{e}}_i \otimes \hat{\boldsymbol{e}}_l \otimes \hat{\boldsymbol{e}}_m \otimes \hat{\boldsymbol{e}}_n\big),
\end{equation}

\noindent which can be readily obtained from partial derivation with respect to $\boldsymbol{W}$ of $\boldsymbol{W}^{-1}\boldsymbol{W}=\mathds{1}$. With this in hand, consider the following derivatives of $\Psi$ (\cref{eqn:psi_example}): 

\begin{flalign}
        \frac{1}{\rho}\boldsymbol{W}^{-T}\boldsymbol{\sigma}_H = \partial_{\boldsymbol{W}}\Psi = \frac{1}{2\rho_0}\partial_{\boldsymbol{W}}(\boldsymbol{E}:\mathbb{C}:\boldsymbol{E}) = \frac{1}{\rho_0}\frac{\partial E_{ij}}{\partial F^e_{pq}}\frac{\partial W^{-1}_{pq}}{\partial W_{mn}}C_{ijkl}E_{kl} = -\frac{1}{\rho_0}(\boldsymbol{W}^{-T}\boldsymbol{W}^{-1})(\mathbb{C}:\boldsymbol{E})\boldsymbol{W}^{-T};&&
        \label{eqn:delPsi_delW}
\end{flalign}

\begin{flalign}
    \partial^2_{\boldsymbol{W}\boldsymbol{W}}\Psi &= \frac{\partial^2\Psi}{\partial W_{mn}\partial W_{rs}} = \frac{\partial}{\partial W_{rs}}\left(-\frac{1}{\rho_0}W^{-1}_{im}W^{-1}_{ik}\mathbb{C}_{kjpq}E_{pq}W^{-1}_{nj}\right)\nonumber\\
    &= -\frac{1}{\rho_0}\mathbb{C}_{kjpq}\left(\frac{\partial W^{-1}_{im}}{\partial W_{rs}}W^{-1}_{ik}W^{-1}_{nj}E_{pq}
    + W^{-1}_{im}\frac{\partial W^{-1}_{ik}}{\partial W_{rs}}W^{-1}_{nj}E_{pq}
    + W^{-1}_{im}W^{-1}_{ik}\right.\frac{\partial W^{-1}_{nj}}{\partial W_{rs}}E_{pq}
    \nonumber\\&\quad+ W^{-1}_{im}W^{-1}_{ik}W^{-1}_{nj}\left.\frac{\partial E_{pq}}{\partial F^e_{ab}}\frac{\partial W^{-1}_{ab}}{\partial W_{rs}}\right)\nonumber\\
    &= \frac{1}{\rho_0}\mathbb{C}_{kjpq}\left(W^{-1}_{ir}W^{-1}_{sm}W^{-1}_{ik}W^{-1}_{nj}E_{pq}
    + W^{-1}_{im}W^{-1}_{ir}W^{-1}_{sk}W^{-1}_{nj}E_{pq}
    + W^{-1}_{im}W^{-1}_{ik}\right.W^{-1}_{nr}W^{-1}_{sj}E_{pq}
    \nonumber\\&\quad+ W^{-1}_{im}W^{-1}_{ik}W^{-1}_{nj}\left.W^{-1}_{aq}W^{-1}_{ar}W^{-1}_{sp} + W^{-1}_{im}W^{-1}_{ik}W^{-1}_{nj}W^{-1}_{ap}W^{-1}_{ar}W^{-1}_{sq}\right);&&
    \label{eqn:del2psi_delWdelW}
\end{flalign}

\begin{flalign}
    \partial_{\boldsymbol{\alpha}}\Psi = \frac{\xi}{\rho_0}\boldsymbol{\alpha} \implies \partial^2_{\boldsymbol{\alpha}\boldsymbol{\alpha}}\Psi = \frac{\xi}{\rho_0}\mathds{1}\otimes\mathds{1};&&
    \label{eqn:delPsi_delAlpha}
\end{flalign}

\begin{flalign}
    \partial_{\theta}\Psi = - \frac{1}{\rho_0}\boldsymbol{\beta}:\boldsymbol{E} - c_\varepsilon\text{ln}\frac{\theta}{\theta_0} \implies \partial^2_{\theta\theta}\Psi = -\frac{c_\varepsilon}{\theta};&&
    \label{eqn:delPsi_delTheta}
\end{flalign}

\subsection{Small deformations}

The following derivatives are given for the $\Psi$ expression in \cref{eqn:psiexample_smalldef}:\vspace{.2cm}

\noindent For compactness, denote $\boldsymbol{A} \equiv (\boldsymbol{\varepsilon} - \boldsymbol{\varepsilon}^p)$. Then,
\begin{flalign}
    \partial_{\boldsymbol{A}}\Psi &= \frac{1}{2\rho_0} \frac{\partial}{\partial A_{mn}}\big(A_{ij}\mathds{C}_{ijkl}A_{kl}\big) - \frac{\Delta\theta}{\rho_0}\beta_{ij} \frac{\partial A_{ij}}{\partial A_{mn}} = \frac{1}{2\rho_0}\big(\delta_{im}\delta_{jn}\mathds{C}_{ijkl}A_{kl} + A_{ij}\mathds{C}_{ijkl}\delta_{km}\delta_{ln}\big) - \frac{\Delta\theta}{\rho_0}\beta_{ij} \delta_{im}\delta_{jn}\nonumber\\
    &= \frac{1}{\rho_0}\mathds{C}:\boldsymbol{A} - \frac{\Delta\theta}{\rho_0}\boldsymbol{\beta}\quad \text{since $\mathds{C}_{ijkl} = \mathds{C}_{klij}$}&&
    \label{eqn:delPsi_delEpse}
\end{flalign}

\begin{flalign}
    \partial^2_{\boldsymbol{A}\theta} = \partial^2_{\theta\boldsymbol{A}} = -\frac{1}{\rho_0}\boldsymbol{\beta}&&
    \label{eqn:del2Psi_delepsEdelTheta}
\end{flalign}
The other partial derivatives of $\Psi$ are the same as in \ref{sec:PsiDerivativesLargeDef}.

\section*{Acknowledgments}
GDLC and MVU are grateful to the European Research Council (ERC) for their support through the European Union's Horizon 2020 research and innovation program for project GAMMA (Grant agreement No. 946959). AA's work was supported by the Center for Extreme Events in Structurally Evolving Materials, Army Research Laboratory Contract No. W911NF2320073.

\section*{Declaration of interests}

The authors declare that they have no known competing financial interests or personal relationships that could have appeared to influence the work reported in this paper.

\bibliographystyle{model5-names}\biboptions{authoryear}
\bibliography{mybibfile}

\end{document}